%% file: Main.tex
\pdfoutput=1
\documentclass{aa}  

\usepackage{graphicx}
\usepackage{float}
\usepackage{lscape}
\usepackage{rotating}
\usepackage{caption}
\usepackage{subcaption} 
\usepackage{morefloats}
\usepackage{threeparttable} 
\usepackage{txfonts}
\usepackage{hyperref}
\hypersetup{colorlinks=true, linkcolor=blue, urlcolor=blue, citecolor=blue}
\defcitealias{Lombardi2011_Orion}{L11}
\defcitealias{Lombardi2010_Perseus_Taurus}{L10}
\defcitealias{Schlafly2015_OrionDust}{S15}

\begin{document}

\title{Three-dimensional dust density structure of the Orion, Cygnus X, Taurus, and Perseus star-forming regions \thanks{Our code is available at \url{https://github.com/Thavisha/Dustribution} and the latest results are available from \url{www.mwdust.com}}}


\author{T. E. Dharmawardena
    \inst{1}\thanks{dharmawardena@mpia.de},
    C. A. L. Bailer-Jones
    \inst{1},
    M. Fouesneau
    \inst{1},
    \and
    D. Foreman-Mackey
    \inst{2}
    }

\institute{Max Plank Institute for Astronomy (MPIA), Königstuhl 17, 69117 Heidelberg, Germany
\and
Center for Computational Astrophysics, Flatiron Institute, 162 5th Ave, New York, NY 10010, USA
    }

\date{Received Month Date, Year; accepted Month Date, Year}

 
\abstract
{

Interstellar dust affects many astronomical observations through absorption and reddening, yet this extinction is also a powerful tool for studying interstellar matter in galaxies.
Three-dimensional reconstructions of dust extinction and density in the Milky Way have suffered from artefacts such as the fingers-of-god effect and negative densities, and have been limited by large computational costs.
Here we aim to overcome these issues with a novel algorithm that derives the three-dimensional (3D) extinction density of dust in the Milky Way using a latent variable Gaussian Process in combination with variational inference. Our model maintains non-negative density and hence monotonically  non-decreasing extinction along all lines-of-sight, while performing the inference within a reasonable computational time.

Using extinctions for hundreds of thousands of stars computed from optical and near infrared photometry, together with distances based on Gaia parallaxes, we use our algorithm to infer the structure of the Orion, Taurus, Perseus, and Cygnus X star-forming regions.  
A number of features that are
superimposed in 2D extinction maps are clearly deblended in 3D dust extinction density maps.
For example, we find a large filament on the edge of Orion that may host a number of star clusters. We also identify a coherent structure that may link the Taurus and Perseus regions, and show that Cygnus X is located at 1300--1500~pc, in line with VLBI measurements.

We compute dust masses of the regions and find these to be slightly higher than previous estimates, likely a consequence of our input data recovering the highest column densities more effectively. By comparing our predicted extinctions to Planck data, we find that known relationships between density and dust processing, where high-extinction lines of sight have the most processed grains, hold up in resolved observations when density is included, and that they exist at smaller scales than previously suggested. This can be used to study the changes in size or composition of dust as they are processed in molecular clouds.

}

\titlerunning{3D dust density}
\authorrunning{Dharmawardena et al.,}

\maketitle

%

\section{Introduction}

Interstellar dust affects how we view the Universe by absorbing and scattering starlight, leading to interstellar reddening and extinction.
Although interstellar dust encompasses only a small fraction ($\sim 1
\%$) of the baryonic matter in our Universe, it is crucial to understand the properties of dust in order to interpret observations correctly. The absorption of starlight also plays a key role in the thermal balance of the interstellar medium, driving chemical reactions and enabling gas clouds to collapse. Knowing the spatial distribution of interstellar dust is therefore key both to understanding the interstellar medium as whole and
to being able to account for the effects of interstellar extinction. 

Past studies have made great strides in mapping the spatial distribution of the interstellar medium in both extinction and dust density. One of the most comprehensive works was by \citet{Schlegel1998}, who used IRAS and COBE far-IR emission to produce an all-sky 2D dust extinction map. This was followed by two decades of improvements fuelled by advances in machine learning techniques and the application of Bayesian statistics leading to both new 2D and 3D dust extinction and extinction density maps \citep[e.g.,][]{Marshall2006, Schlafly2011, Sale2014_IPHAS, Sale2014,  Lallement2014, Green2015, Hanson2016, Rezaei2017, Sale2018, Lallement2019, Green2019_Bayestars19, Babusiaux2020, Leike2020}.

Gaussian Processes (GPs) have been widely use to study the 3D structure of interstellar dust. \citet{Sale2014} used them to predict extinction in simulated data on individual lines-of-sight and then applied it to IPHAS photometry in a slice of the Galactic plane \citep{Sale2014_IPHAS}. This was followed by \citet{Sale2018} who improved the method to account for correlations between lines-of-sight and applied it to simulated data along individual line-of-sight.

\citet{Green2019_Bayestars19} took a similar approach and produced a 3D extinction map covering the entire sky north of $-30^{\circ}$ up to 2~kpc. The authors used 2MASS, Gaia and Pan-STARRS data to carry out important sampling on a parameter grid assuming a GP prior on the logarithm of the dust reddening density. These maps have become a go-to resource for estimating interstellar extinction along lines-of-sight as they are easily accessible. The works by \citet{Rezaei2018B, Rezaei2018A} used GP techniques to predict the 3D dust density of the Orion region and the Milky Way disk using Gaia DR2, 2MASS and WISE and APOGEE DR14 data for 30\,000 stars respectively. \citet{Leike2019} and \citet{Leike2020} modelled the dust density as a GP and simultaneously inferred the power-spectrum of dust density. Using this method they modelled the inner 400~pc molecular clouds coverage using 30 million sources from multiple surveys including Gaia.  

Modelling the 3D dust distribution of the Milky Way with GPs is not without its challenges and issues, however. One is the so-called fingers-of-god effect, in which the density distribution is elongated along the line-of-sight due to higher tangential than radial accuracy. It can be avoided if high-accuracy distances are available and/or if correlations between points in 3D space are incorporated explicitly rather than as individual lines-of-sight coupled together in the plane of the sky. Another key feature which must be considered is the physical requirement that densities be positive and hence extinction must be monotonically non-decreasing along any line-of-sight. One way of achieving this is to model the logarithm of the density instead of density or extinction itself. 
GPs are computationally intensive, naively scaling with $\mathcal{O}(N^{3})$ in number of sources. Methods with improved scaling are therefore important to ensure that a wide range of problems are feasible with minimal trade-offs between, for example, resolution, map size, and number of sources. Finally, the reproducibility of the results may be hampered by the lack of code availability or non-use of public libraries.

In this paper we present a technique that directly accounts for the 3D correlations between density points and enforces positive density throughout. 
To accelerate the processing, we use GP Latent Variable Models in combination with Variational Inference to infer the logarithm of the dust density. We describe our method in section \ref{sec:Method}, validate it using simulated data in section \ref{sec:SimData}, then using input data from section \ref{sec:Data} apply it to the four  well-known star-forming regions, Orion, Cygnus X, Taurus and Perseus in sections \ref{sec:Orion} and \ref{sec:Cyx_Per_Taurus} to map their dust in 3D. From this we can compute their total masses (section \ref{sec:Mass}), and in section \ref{sec:Planck} we compare our results to Planck dust emission measures.


\section{3D Extinction and Density Mapping Model}
\label{sec:Method}

\subsection{Overview}

To ensure that density is always positive, we model the logarithm of the density, and impose a GP prior to account for correlations between the densities at points in three dimensions. While we will use the term "dust density" or simply "density" for ease throughout this work, in reality we are actually mapping \textit{dust extinction density} in $\mathrm{mag \ pc^{-1}}$ units. For it to be a traditional density in $\mathrm{g \ cm^{-3}}$ units the dust extinction density needs to be converted using the dust optical depth and dust opacity.

We optimise the hyperparameters for this GP along with several other parameters which approximately condition the GP as described in Sect.~\ref{sec:VI} and \ref{sec:Implement}, to infer the logarithm of the density. At the end we draw sample maps from the final conditioned GP to explore the dust density distribution and its uncertainty; these are integrated to predict extinction and its uncertainty.    

Our approach requires measurements and uncertainties of the extinctions towards a sample of stars, as well as their positions in three dimensions. Uncertainty on distance is included following the same approach as \citet{RezaeiKh.2020}, which involves incorporating them into the observed extinctions by inflating their uncertainties (see their sec.~2 and eqns.~1 and 2). We utilise  GP package \emph{GPyTorch} \citep{Gardner2018_GpyTorch} and probabilistic programming package \emph{Pyro} \citep{Bingham2018_Pyro1, Phan2019_Pyro2} in our implementation. Both packages are built upon \emph{PyTorch} \citep{Paszke2019_PyTorch}, an open-source machine learning framework.



\subsection{Latent Gaussian Process Function}\label{sec:LatentGP}

Mathematically, our quantity of interest, dust density ($\rho$), can be described as a latent quantity or function -- a quantity that we do not directly observe. This latent quantity can be transformed by integrating along lines-of-sight to forward model the observed quantity, which is extinction. 
Dust density is expected to vary in a complex way spatially, which cannot be accurately captured by a parametric function. As a result, the model for the latent function should ideally be flexible and non-parametric, which makes GPs ideal for this purpose.



We model the measured extinctions with a Gaussian likelihood. Given that $\mathbf{A_{mod}} = \int^{d}_{0} \rho\ d\vec{s}_i$, then if we adopted a GP prior on $\rho$, then the posterior would also be Gaussian in $\rho$, with analytic solutions for its mean and covariance. This is the approach of \cite{Rezaei2017} (their appendix A). However, to enforce non-negative densities, we define our GP prior on $\log_{10}$($\rho$), with the consequence that the exact posterior no longer has a simple analytic form (see appendix \ref{sec:Appendix:ApproxPosterior}).
We handle this using variational inference (see section~\ref{sec:VI}) and approximate the posterior in  $\log_{10}$($\rho$) as normally distributed.
The prior on $\log_{10}$($\rho$) -- which we now denote $\phi$ for brevity -- is modelled using a Gaussian Process with a constant mean \citep[see chapter 2.2 and eqn.~2.14 in ][]{Rasmussen2006_GPbook},  
\begin{equation}
   P\left(\phi\left(x,y,z\right)\right) = \mathcal{GP}\left(c, \alpha K(\vec{x}_{1}-\vec{x}_{2})\right),
   \label{eqn:GPfunc}
\end{equation}
and elements of the covariance matrix given by a 3D radial basis function (RBF) kernel
\begin{equation}
    K(\vec{x}_{1},\vec{x}_{2}) = \exp \left( -\frac{1}{2}(\vec{x}_{1} - \vec{x}_{2})^{\mathrm{T}}\vec{\gamma}^{-2}(\vec{x}_{1} - \vec{x}_{2}) \right),
    \label{eqn:rbfKernel}
\end{equation}
where $\vec{x}_{1}$ and $\vec{x}_{2}$ are two position vectors in 3D space, $\vec{\gamma}$ is a diagonal matrix of scale lengths, $\alpha$ is a scale factor, and $c$ is a constant mean \citep[see chapter 4.2 eqn.~4.9 and chapter 2.2 eqn.~2.14 in ][]{Rasmussen2006_GPbook}\footnote{available online at \url{http://www.gaussianprocess.org/gpml/}} \footnote{see the \emph{Gpytorch} documentation for the n-dimensional generalisation of the RBF kernel: \url{https://docs.gpytorch.ai/en/stable/kernels.html}}. We abbreviate the right hand side of equation~\ref{eqn:GPfunc} to $\mathcal{GP}$($\Theta$) throughout this paper, where  $\Theta$ represents the hyperparameters of the GP. 

Our GP has five  hyperparameters: three physical scale lengths in the three Heliocentric Cartesian coordinates ($x,y,z$) of physical space; one exponential scale factor; the mean density. These hyperparameters are used to generate the GP prior from which sets of $\phi$ are predicted. In effect, treating $\phi$ as a latent variable introduces one free parameter at each point in $x,y,z$, all coupled together through the GP prior. 
We evaluate the GP on a grid, which we take to be regular in Galactic (spherical) coordinates ($l,b,d$) as this eases the implementation of line-of-sight integration. We then convert the coordinates of the centres of the cells to Heliocentric Cartesian coordinates for use in eq.~\ref{eqn:GPfunc}.

To compare to observed extinction ($A_{obs}$) to what our model density distribution predicts, we must integrate this density distribution along the line-of-sight. We first exponentiate the distribution of $\phi$ to obtain $\rho$ and then numerically integrate along the lines-of-sight to all stars in our observed data set. The numerical integration is described in Sect.~\ref{sec:LosInt}. These integrated densities, i.e.\ the model extinctions ($A_{mod}$), are compared to our observed extinctions (via the likelihood) in order to optimise the model. The posterior for our model is therefore
\begin{equation}
     P(\phi, \Theta \,\vert\, A_{obs}) \propto P(A_{obs} \,\vert\, \phi) \,P(\phi \,\vert\, \Theta) \,P\left(\Theta\right)
     \label{eqn:posterior}
\end{equation}
where the first term on the right does not depend on $\Theta$ because $A_{\rm obs}$ is independent of $\Theta$ once conditioned on $\phi$. 
As mentioned above, as we have a prior on log density rather than density, we no longer have close form solutions for the integrals over the model densities. We therefore use numerical methods to compute the integrals to achieve a Gaussian approximation of the posterior in $\phi$. We do this via variational inference, as explained in the next section.
When doing this, we are free to infer the hyperparameters of the GP prior at the same time. As shown in eq.~\ref{eqn:posterior}, our model is implicitly hierarchical: the GP is a prior on (the logarithm of) dust density, the plausible values of which are determined by the hyperparameters, and we optimise the dust density as a function of position.



\subsection{Variational Inference}
\label{sec:VI}

The above approach to modelling the dust density effectively makes the problem one of inferring the joint distribution of densities at all points, and as a result the problem has a very high-dimensional parameter space. To handle this effectively we use two techniques. The first is to reduce the dimensionality of the problem by conditioning the GP only on a subset of points, known as the inducing points. The locations of these inducing points are parameters which can be optimised along with the other free parameters. Although the use of inducing points lowers the dimensionality, it is still high, because there are three free parameters per inducing point. There is no "correct" choice of the number of inducing points. A rule of thumb is to incorporate as many inducing points as computationally feasible given the available resources \citep[e.g.,][]{Wang2019_IndPs, Wu2021HierarchicalIP, TowardsDataSci_IPs}. Results in the literature show that there is minimal gain once the number of inducing points rises above several percent of the full data set \citep{Wang2019_IndPs}. We typically use 500--1000 of inducing points in this work, which is as much our computational setup allowed at the time. 

To circumvent the still large number of free parameters, we employ variational inference to approximate the target marginalisation integrals. Variational inference replaces the target posterior with an approximate posterior that is easier to work with, and finds the parameters for this approximation that best reproduce the true posterior \citep{Bishop2006_VIandELBObook, Blei2017:VI}. This allows the direct computation of the approximate posterior and its gradient with respect to the free parameters, thereby enabling the use of gradient-descent optimisers, as described below in Sect.~\ref{sec:Implement}. 

This approximate posterior is referred to as the variational distribution.  As we are using a GP, we use a multivariate Gaussian distribution for this approximate posterior (as described above in Sect.~\ref{sec:LatentGP}) to give the joint distribution of $\phi$ at the inducing points. The mean and covariance of this Gaussian are what we need to infer. In our case, the variational distribution approximates the true (non-Gaussian) posterior of $\phi$ at the inducing points, analogous to conditioning a GP on a function which is approximating the real (non-Gaussian) distribution, and only conditioning at a subset of the available points \citep{Bishop2006_VIandELBObook, Blei2017:VI}. 



We want to find that approximate posterior that is most similar to the true one. We achieve this by finding the parameters of the (variational) distribution that minimizes the difference between this distribution and $\mathrm{P(\phi \, \vert \, A_{obs}})$. 
To minimise this difference, we maximise the \textit{Evidence Lower Bound (ELBO)}. ELBO is a variational bound, i.e.\ it bounds a quantity by a functional over some set of functions \citep[chapter 10]{Bishop2006_VIandELBObook}. It computes a minimum value for the evidence (the integral of the likelihood over all possible parameter values) given a set of likelihood evaluations.

The motivation for the ELBO is that it is equal to the logarithm of the evidence (which is equal to the expectation of the likelihood over the prior) plus the negative of the Kullback-Leibler (KL) divergence {\citep{Bishop2006_VIandELBObook, Blei2017:VI}}
\begin{equation}
    ELBO\left(V\right) = \mathbb{E}\left(\log P(A_{obs}\,\vert\, \phi)\right) -KL\left(V\left(\phi\right) \vert\vert P\left(\phi\,\vert\,A_{obs} \right)\right),
\end{equation}
where the KL divergence is a measure of the dissimilarity of two distributions, and $V$ is the variational distribution. In our case the two distributions whose KL divergence is measured are the exact posterior $P\left(\phi\,\vert\,A_{obs}\right)$ and the approximation, the variational distribution. This is calculated as:
\begin{equation}
    KL\left(V\left(\phi\right) \vert\vert P\left(A_{obs}, \phi\right)\right) = 
     \int V\left(\phi\right) \ln \left(\frac{V\left(\phi\right)}{P\left(\phi \vert A_{obs}\right)}\right) d\phi,
     \label{eqn:KL}
\end{equation}
following \citet[chapter 10 eq.~10.2--10.4]{Bishop2006_VIandELBObook} and \citet[sec.~2.2 eq.~ 11--14]{Blei2017:VI}.
This can be efficiently minimised using stochastic gradient descent algorithms. The full set of free parameters to be inferred in our model therefore consists of the GP hyperparameters, the locations of the inducing points and the parameters of variational distribution i.e. the means and covariances.

Once the inference is complete, we have our conditioned GP. From this we can draw samples of $\phi$ which can be exponentiated to then determine the 16th, 50th and 84th percentiles of $\rho$. We can also numerically integrate (as described in Sect.~\ref{sec:LosInt}) the samples of $\rho$ over distance to determine the 16th, 50th, and 84th percentiles of extinction. We use samples to determine confidence intervals as our distributions of $\rho$ and extinction are not Gaussian; if they were, the resulting integration (or sum) would itself be Gaussian, however, we end up needing to sum over log-normal random variates which does not necessarily have similar convenient properties. Sampling, on the other hand, is computationally efficient.

\subsection{Implementation}
\label{sec:Implement}

The GP is implemented using \emph{GPytorch} \citep{Gardner2018_GpyTorch}, and \emph{Pyro} \citep{Bingham2018_Pyro1, Phan2019_Pyro2} is used for variational inference. The inputs to our algorithm are the positions and extinctions (including uncertainties) of a sample of stars within a region of interest in the Galaxy. We define a \textit{training} grid on which the GP can infer densities. The grid is composed of cells, and every cell is assumed to be uniform, that is, the value of the density is the same at every point within a given cell. The boundaries of this grid are defined in galactic coordinates to enclose the region of interest. The number of cells along each axis of the grid is independent, and can be spaced arbitrarily. The number of cells define the resolution of our training grid and hence the minimum size of the recovered features in the final reconstruction. 

We have three sets of free parameters in our model which are optimised. The GP hyperparameters, the locations of the inducing points within our training grid and the means and covariances of the variational distribution. The hyperparameters of the GP require starting values to begin the optimisation. For the mean of the RBF kernel we estimate the mean density of our region of interest based on prior knowledge and the specific values are given in sec.~\ref{sec:Orion} and \ref{sec:Cyx_Per_Taurus}. The scale factor of the kernel is initiated from the expected scale (amplitude) of deviations from the mean density in dex (since our model is logarithmic); these values must be below $1\,\mathrm{dex}$ because we do not expect variations in dust density of an order of magnitude on scales of 1\,pc in the ISM. Finally, we choose the starting scale length based on the typical size of clumps to be recovered. Table~\ref{tab:HPs} below presents the input hyperparameters as well as their conditioned counterparts.  We randomly select a subset of the positions of stars in our input sample for the initial locations of the inducing points (however, a user is free to place them in any initial configuration of their choice and can leave these positions fixed throughout the optimisation process if they wish). The variational distribution is initiated by \emph{GPyTorch} with means of zero and variances of 1. 

As seen in Eqn. \ref{eqn:KL}, estimating ELBO requires integration; this is calculated using 32 Monte Carlo samples at each iteration, and Adaptive Moment Estimation with Weight decay \citep[AdamW;][]{Loshchilov2017_AdamW}, implemented in \emph{Pyro}, is used to maximise ELBO. AdamW has several tunable parameters, including the learning rate, which determines the step size at each iteration. We choose relatively large learning rates to avoid the optimisation becoming stuck in a local minimum or taking too long to optimise, while also avoiding moving around too much and so potentially missing the optimum values while maintaining numerical stability. AdamW is set to terminate training after a fixed number of iterations chosen based on when the variation of the ELBO is less than $1\%$ across the last 10 iterations and the model has converged. 

For our training grid, we choose a set of boundaries in all three dimensions that closely encompass our region of interest with padding in $l,b,d$. This speeds up the run time of the model compared to using a larger region, but requires that we account for the effects of the foreground and background dust that have been omitted. The padding added in the $d$ dimension accounts for the fore and background dust. 

In the case of foreground dust, which would directly affect all observed extinctions, the model would increase the density of the closer cells in order to account for the missing foreground cells (i.e.\ from zero to the lower boundary). To accommodate this we add a set of \textit{ghost cells} to hold this additional dust, similar to an approach used in grid-based hydrodynamics models \citep[e.g.][]{OBrien2018_GhostCells}. No stars are given as input to these ghost cells, giving the model the freedom to insert foreground dust to explain the extinctions of the  stars closest to the lower distance boundary. This avoids biasing the final scale factor and mean density. Fewer ghost cells are required when the foreground extinction is low (for example nearby regions, or at high galactic latitude) while high-extinction lines-of-sight, particularly for distant star-forming regions in the galactic plane, 
will require more cells. The exact amount must be tuned on a case-by-case basis. 

The case of background regions is slightly more complex. Because our model is conditioned on extinction, and the density at a point is informed by the extinctions of stars at larger distances on nearby lines-of-sight, we require a sufficient number of sources in the background to the star formation region in order to infer the density at a point within the star formation region unambiguously. Restricting the input sources exclusively to the star formation region in which we want to infer the dust density would discard vital information, so sources must be included at greater distances such that they sample the density at a sufficient number of lines-of-sight inside the region of interest. Hence, we pad the model volume at greater distances. Unlike with the foreground ghost cells, these background cells contain input sources, and so the padding must be increased until a sufficient number of sources are available to encapsulate the underlying structure of the region. Once again, this was determined for each case individually by iteratively increasing the distance until the changes in the structure are no longer significant by eye. Beyond the range probed by the input set of stars, the predicted density will rapidly converge to the mean of the GP prior, as this is the maximum \emph{a posteriori} value in the absence of data on which to condition. This does not preclude low-extinction lines-of-sight from introducing low dust-density regions in this mean density if the source is sufficiently distant and if there is little dust in front of it, as there is information available here.

Once the GP is conditioned, we are free to predict the density at arbitrary points. We exploit this to produce visualisations of the reconstructed density at higher resolution, making them easier to interpret. We therefore define a grid with 2--3 times higher resolution than the training grid but the same boundaries, which we refer to as the \textit{prediction} grid hereafter. Although it does not provide any new information compared to the training grid, it is especially beneficial for visualising structures that are small and only a few pixels in size in the training grid.

As our optimisation routine AdamW requires the gradient of the ELBO all quantities are stored as \emph{Torch} tensors or distributions to exploit \emph{PyTorch}'s autograd  module. These tools have been scaled to large datasets and give us good performance for our problem.

\subsection{Integration along lines-of-sight}
\label{sec:LosInt}

The integration of density is approximated as a summation. This is carried out in spherical coordinates with the Sun at the origin to optimise the computations by minimising the number of cell boundary crossings. 

We want to compute 
\begin{equation}
    \mathbf{A_{mod}} = \int^{d_{\ast, i}}_{0} 10^{\phi} d\vec{s}_i
\end{equation}
for each star, where $\phi$ is the logarithm of extinction density, $\vec{s}_i$ is the path to the $i$th star and $d_{\ast, i}$ is its distance.
As $\phi$ is discretised on a grid, we approximate this integral as the sum
\begin{equation}
    \mathbf{A_{mod}} = \sum\limits^{j_\ast}_{j=0} 10^{\phi_j} \Delta d_j
\end{equation}
where the index $j$ iterates over the grid cells along the line-of-sight, $j_\ast$ is the index of the cell that contains the $i$th source, $\phi_j$ is the density in the $j$th cell and $\Delta d_j$ is the length of the path crossing the $j$th cell.
This summation is refined slightly to account for the partial crossing of the final cell on the path, i.e.
\begin{equation}
    \mathbf{A_{mod}} = \sum\limits^{j_\ast - 1}_{j=0} \left( 10^{\phi_j} \Delta d_j \right) + 10^{\phi_{j_\ast}} \left(d_{\ast} - d_{j_\ast - 1}\right) 
\end{equation}
where $d_{j_\ast - 1}$ is the distance to the boundary between the $j_\ast - 1$th and the $j_\ast$th cell.

\section{Testing Model on Simulated Data}
\label{sec:SimData}

\begin{figure}
    \centering
    \includegraphics[width=0.5\textwidth, trim=3cm 3cm 3.5cm 3cm, clip]{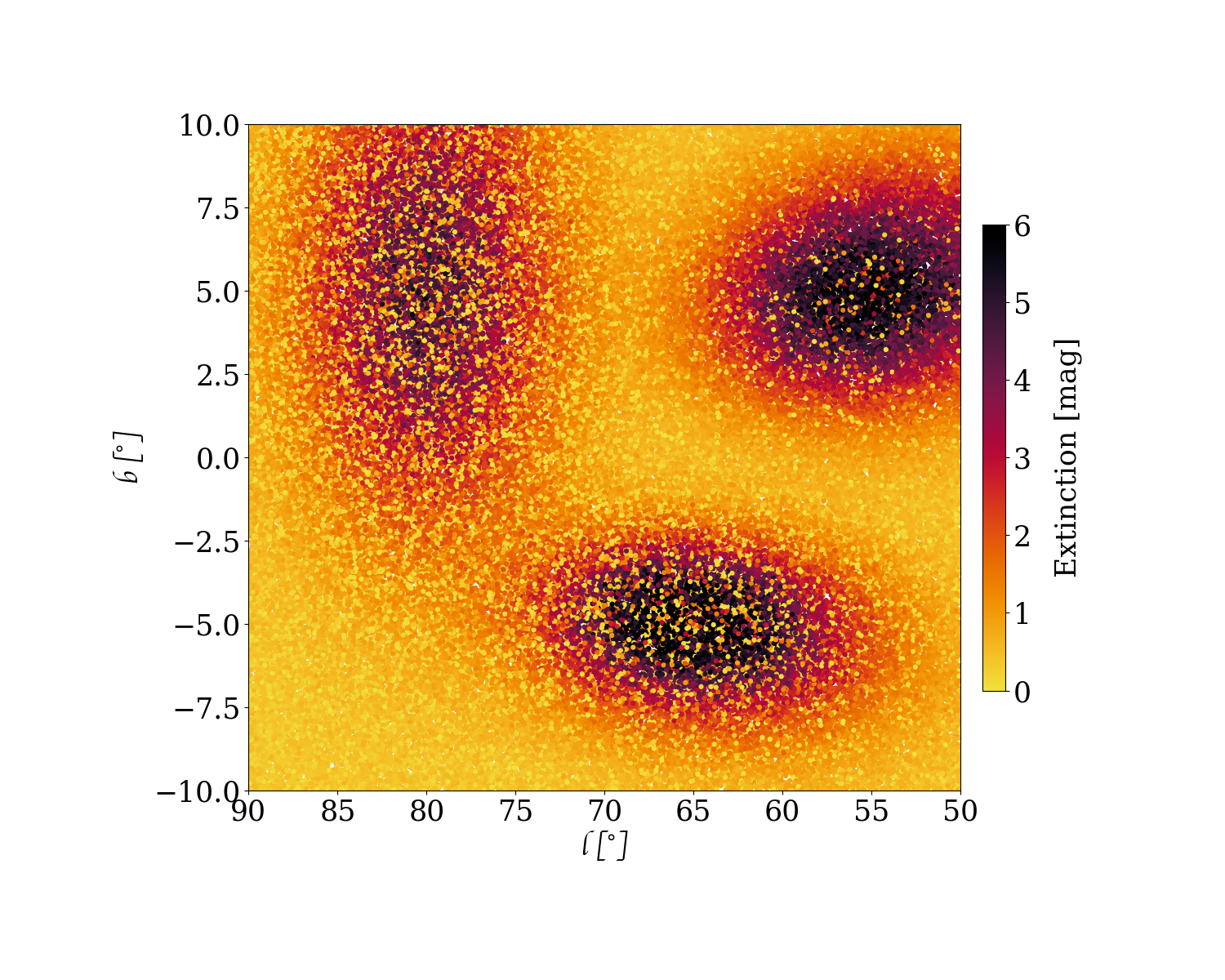}
    \caption{Synthetic extinction estimates used as input for the model, drawn from the simulated density distribution. Each point in the plot corresponds to one simulated star, with a randomly-generated $\left(l, b, d\right)$ and its corresponding extinction.}
    \label{fig:SimTabDist}
\end{figure}

As an initial test of our model we run it with a set of simulated data with known densities. We generate a 3D density field comprising of a constant background overlaid by three Gaussian overdensities randomly located within a cube. This superposition of clumps on a background provides a simple analogue to  the real distribution of matter that is easy to analyse.
The background of the density field is set to $1.6\times10^{-3} \ {\rm mag} \ {\rm pc}^{-1}$, which gives a total background extinction of 0.5 mag when integrating over the model distance range (see below) providing sufficient extinction to be visible in the model. We generate extinctions toward 120\,000 simulated sources within the model volume by integrating along lines-of-sight to a representative sample of coordinates spread randomly throughout the 3D density field. The coordinates are distributed in $l$, $b$ and $d^3$, assuming a uniform space-density of sources. The integration is performed on  a grid with boundaries $l=50^{\circ}$ to $90^{\circ}$, $b=-10^{\circ}$ to $+10^{\circ}$, and $d=180-500 \ \mathrm{ pc}$ and consists of $l\times b\times d = 100 \times 100 \times 115$ cells. We add zero-mean Gaussian noise to these extinctions with a standard deviation equal to $10\%$ of the true line-of-sight extinction. These noisy extinctions, along with their noise-free 3D positions, serve as the input to the model and are shown in Fig.~\ref{fig:SimTabDist}. We only apply the distance uncertainty method of \citep{RezaeiKh.2020} to our real star formation data and it is not applied here. 

\begin{figure}
    \centering
    \includegraphics[width=0.5\textwidth, trim=0cm 0cm 0cm 0cm, clip]{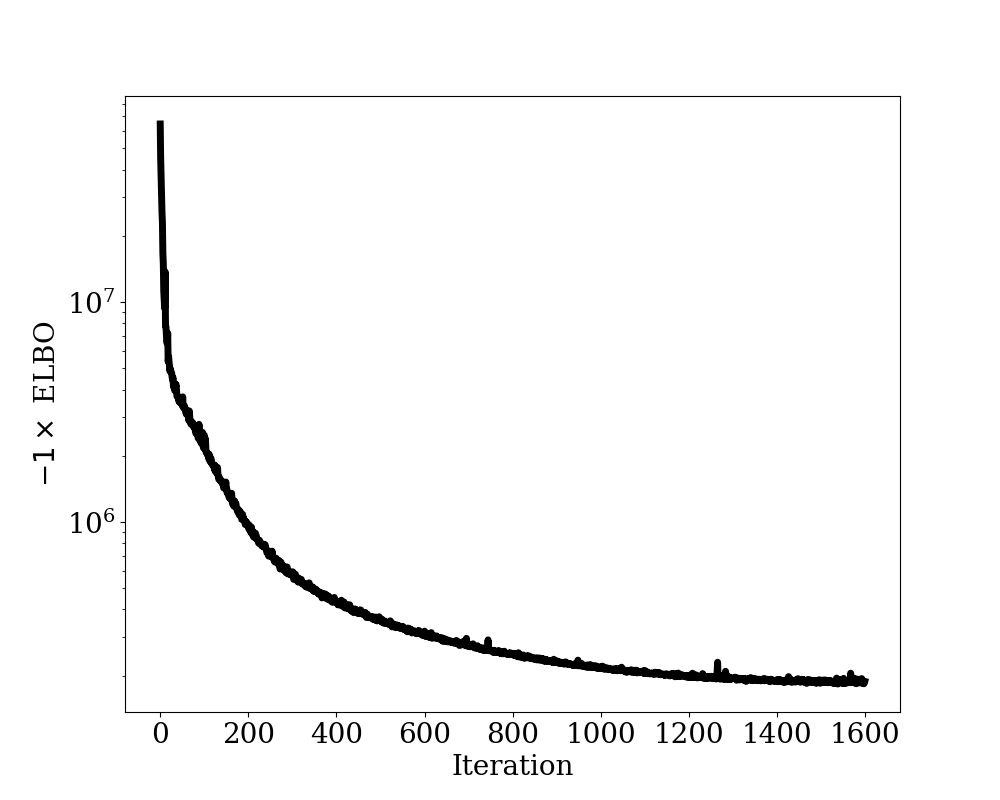}
    \caption{Variation of ELBO through model iterations. \emph{Pyro} reports ELBO as a loss function and therefore seeks to minimise $-1 \times$~ELBO instead of directly maximising ELBO, and the y-axis on this plot is therefore positive instead of negative. Minimising $-1\times$~ELBO is therefore analogous to minimising the KL divergence.}
    \label{fig:SimELBO}
\end{figure}

Our algorithm was trained on a grid of $n_l\times n_b\times n_d = 20 \times 20 \times 35$ cells\footnote{The choice of these values affect the final reconstruction; we recommend using the largest grid that fits in memory during integration on the machine used for training.} with 1000 inducing points and the output predictions were made on a grid of $n_l\times n_b\times n_d = 100 \times 100 \times 105$ cells to be able to directly compare to the ground truth maps. 
The grid boundary coordinates were the same
during training and for the prediction. Figure \ref{fig:SimELBO} shows the variation and convergence of the ELBO for our model run. The variations in ELBO over the last few iterations are below the $1\%$ level, demonstrating that the model has converged.

\begin{figure}
\begin{subfigure}{0.5\textwidth}
  \centering
 \includegraphics[width=\textwidth, trim=0cm 0cm 0cm 0cm, clip]{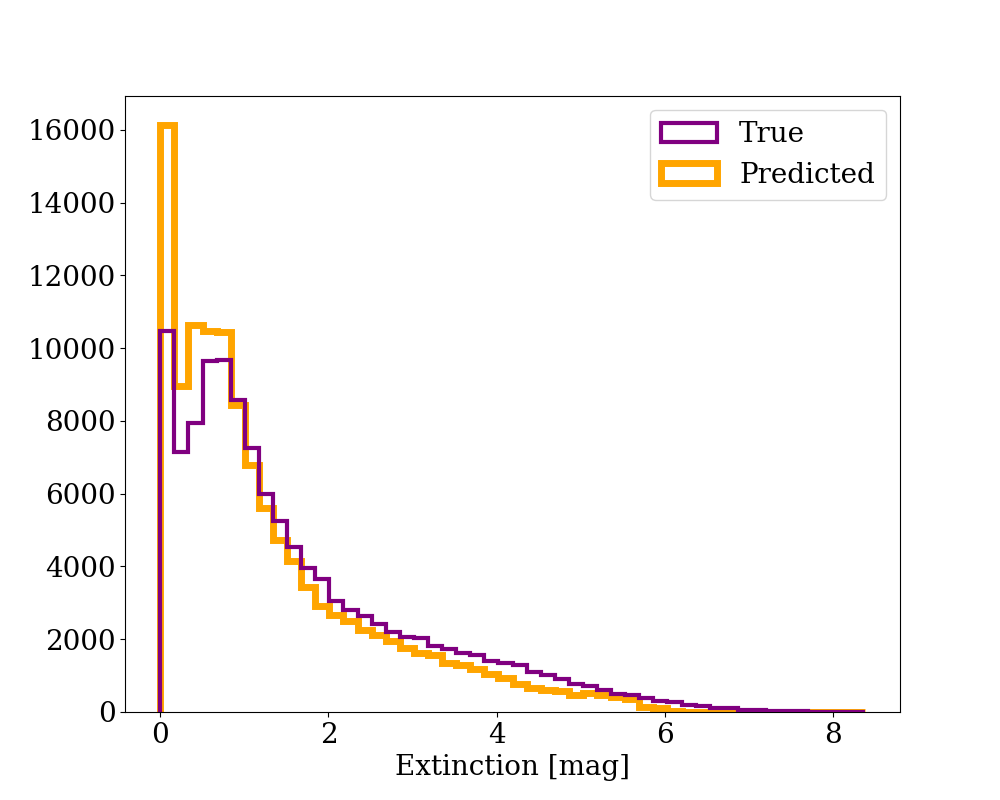}
  \end{subfigure}
\begin{subfigure}{0.5\textwidth}
  \centering
  \includegraphics[width=\textwidth, trim=0cm 0cm 0cm 0cm, clip]{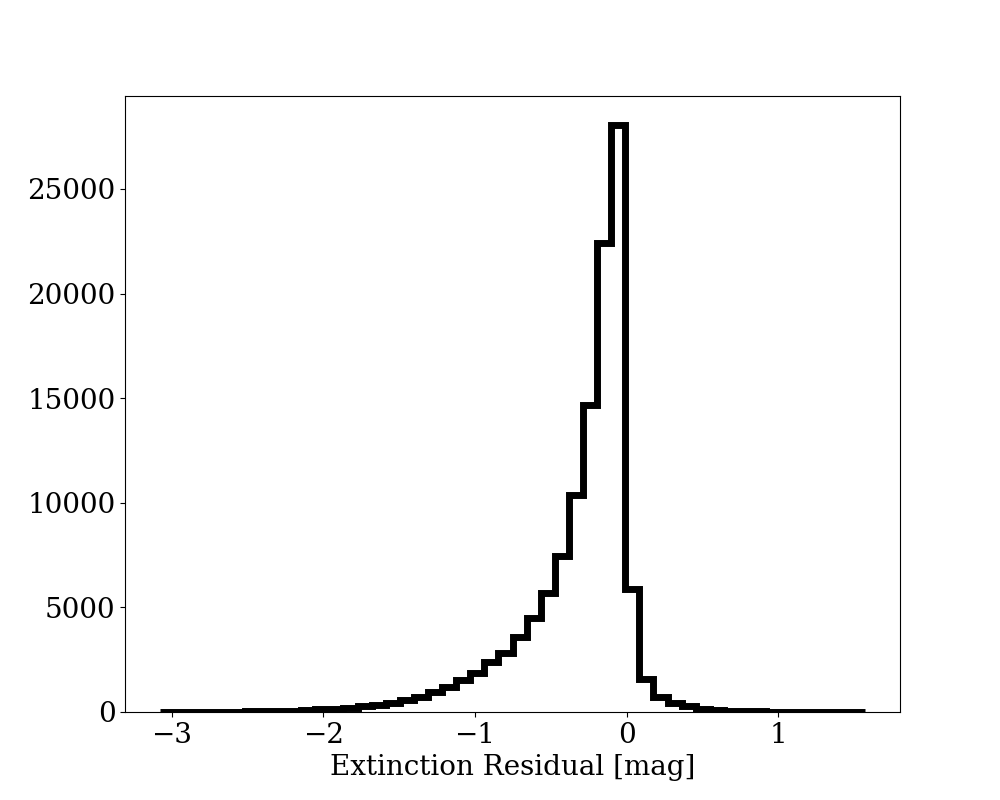}
  \end{subfigure}
  \caption{\emph{Top}: Histograms of true extinctions and predicted extinctions; \emph{Bottom}: Residual, i.e.\ predicted extinctions -  true extinctions, for the set of simulated sources.}
  \label{fig:SimExt_Hist_All}
\end{figure}

\begin{figure}
\centering
\begin{subfigure}{0.45\textwidth}
  \centering
  \includegraphics[width=\textwidth, trim=3cm 3cm 3.5cm 3cm, clip]{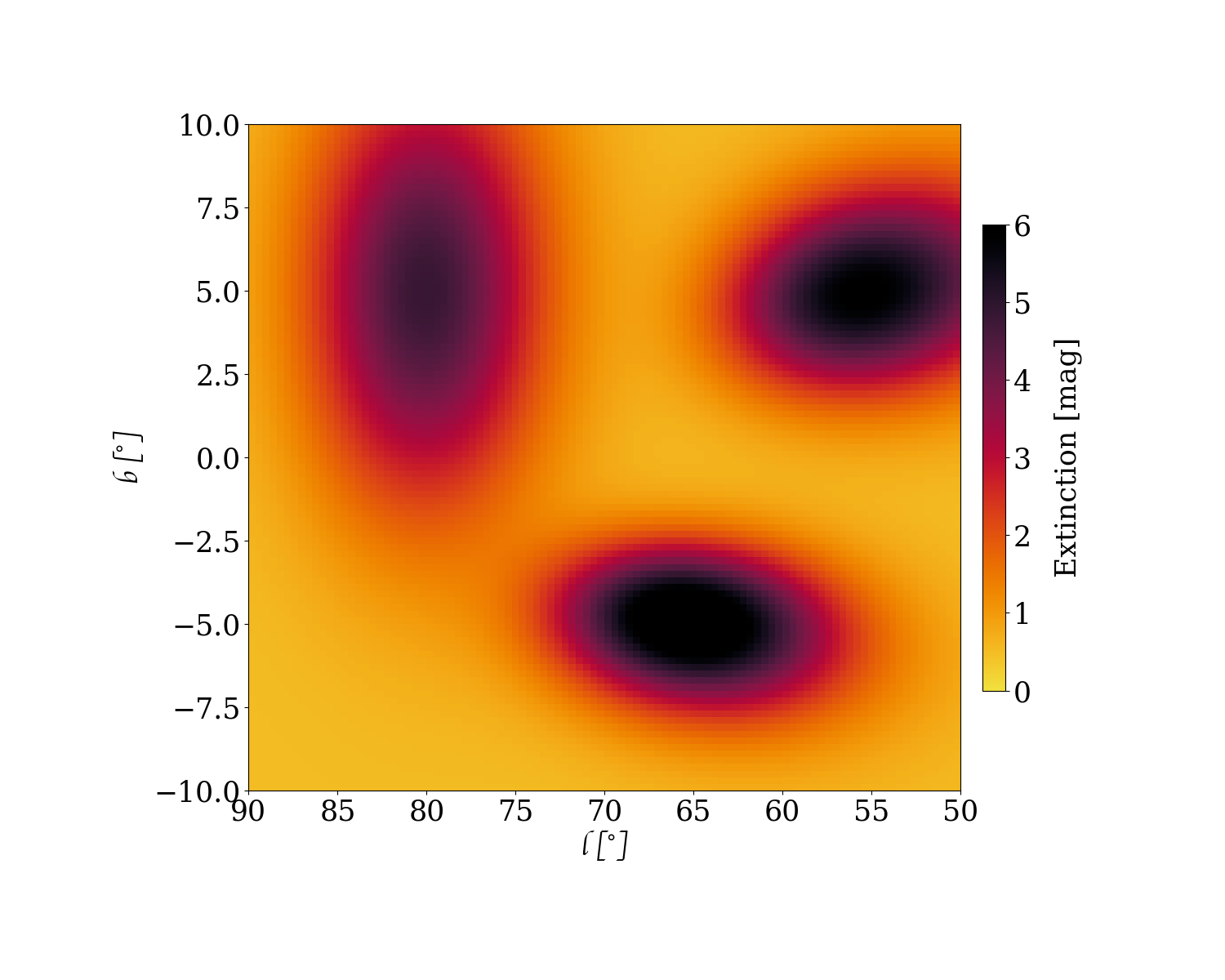}
  \end{subfigure}
\begin{subfigure}{0.45\textwidth}
  \centering
  \includegraphics[width=\textwidth, trim=3cm 3cm 3.5cm 3cm, clip]{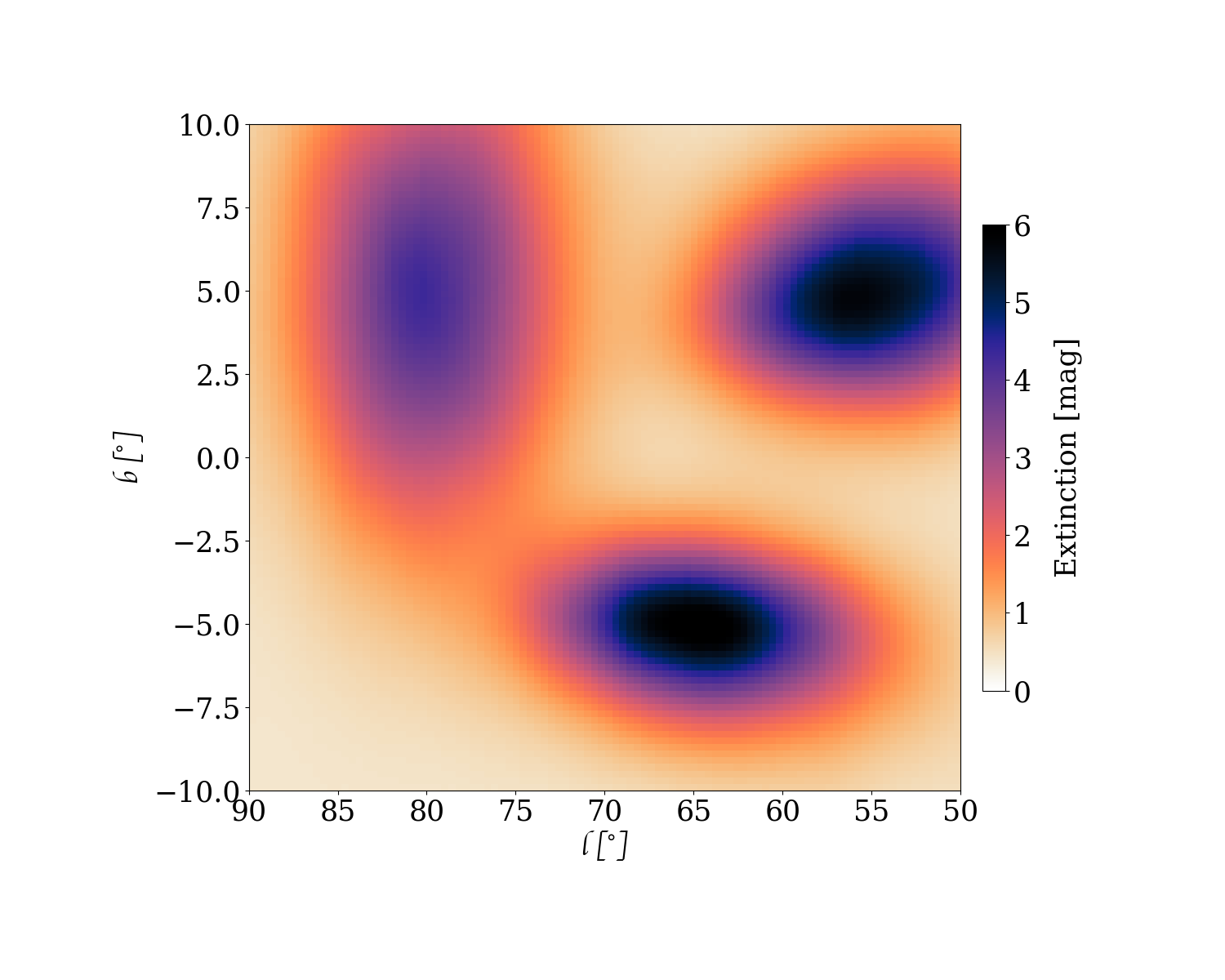}
  \end{subfigure}
\begin{subfigure}{0.45\textwidth}
  \centering
  \includegraphics[width=\textwidth, trim=3cm 3cm 3.5cm 3cm, clip]{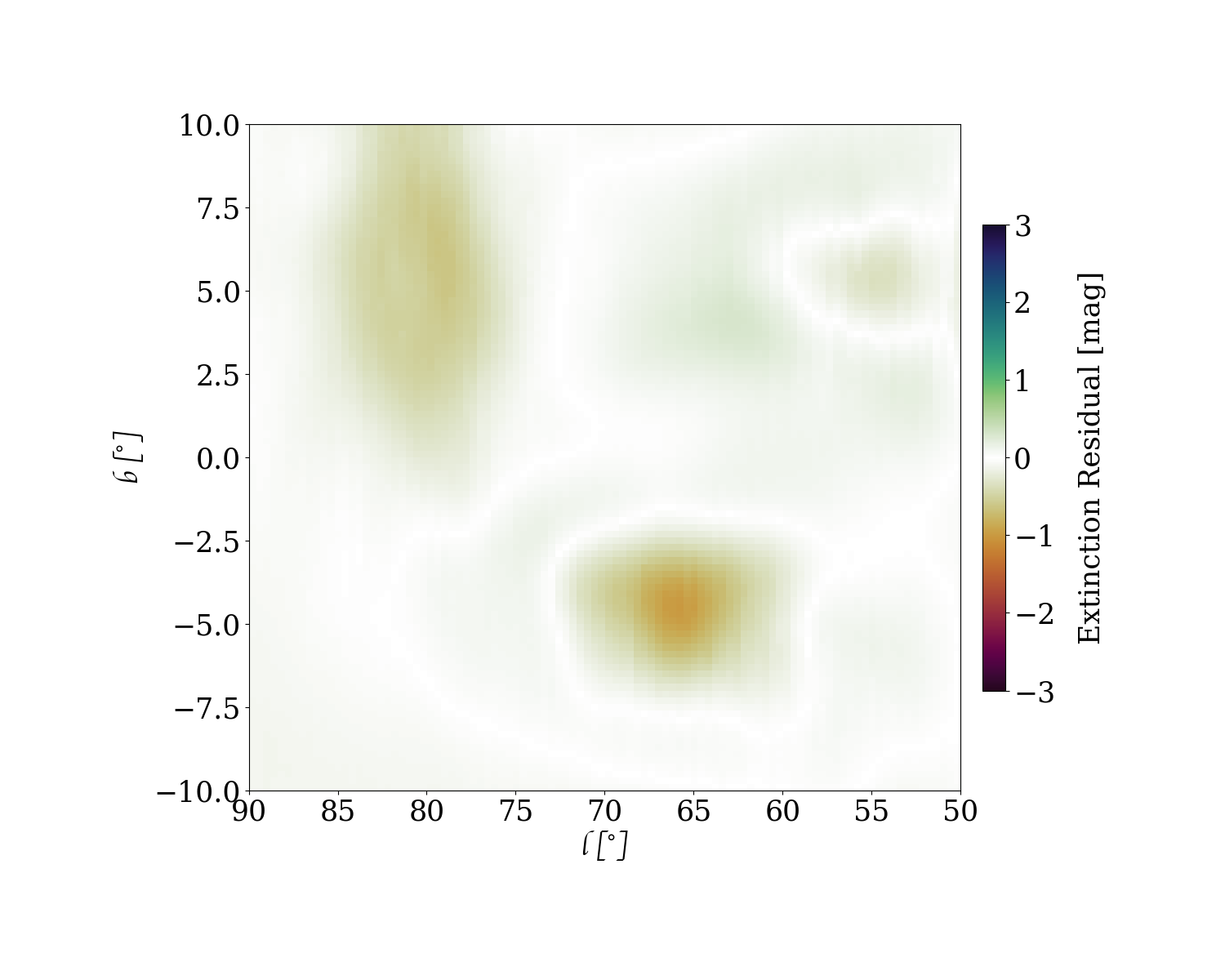}
  \end{subfigure}  
  \caption{\emph{Top}: Simulated extinction map up to 500~pc; \emph{Middle}: Our model reconstructed Extinction map up to 500~pc; \emph{Bottom}: Residual of Model Predicted - Simulated Extinction.}
  \label{fig:SimCumExtAll}
\end{figure}

\begin{figure*}
\centering
\begin{subfigure}{\textwidth}
  \centering
  \includegraphics[width=\textwidth]{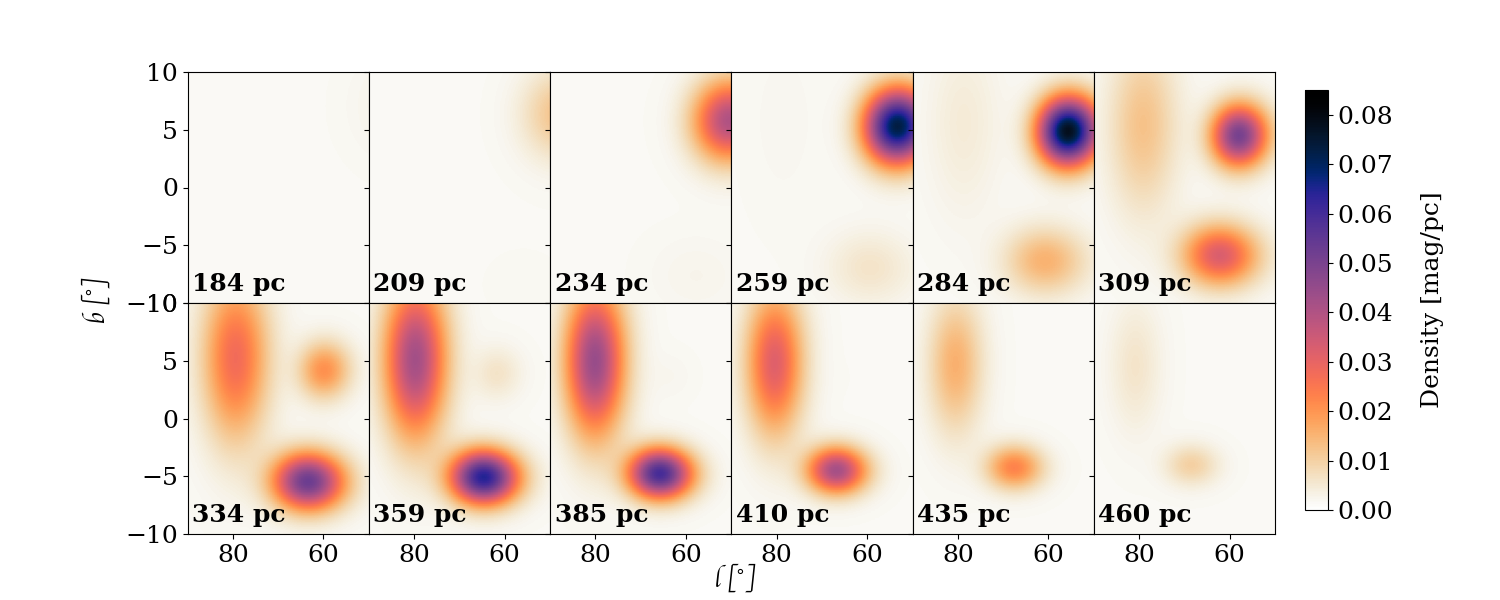}
  \end{subfigure}
\begin{subfigure}{\textwidth}
  \centering
  \includegraphics[width=\textwidth]{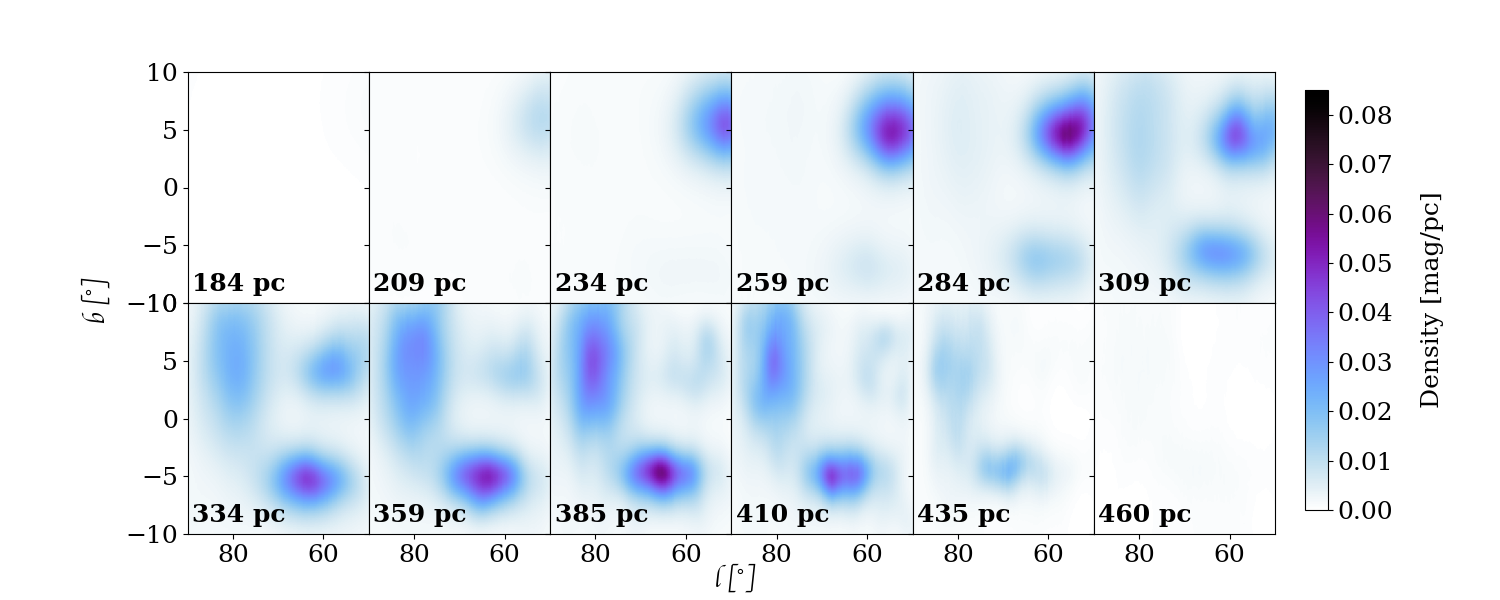}
  \end{subfigure}
   \begin{subfigure}{\textwidth}
  \centering
  \includegraphics[width=\textwidth]{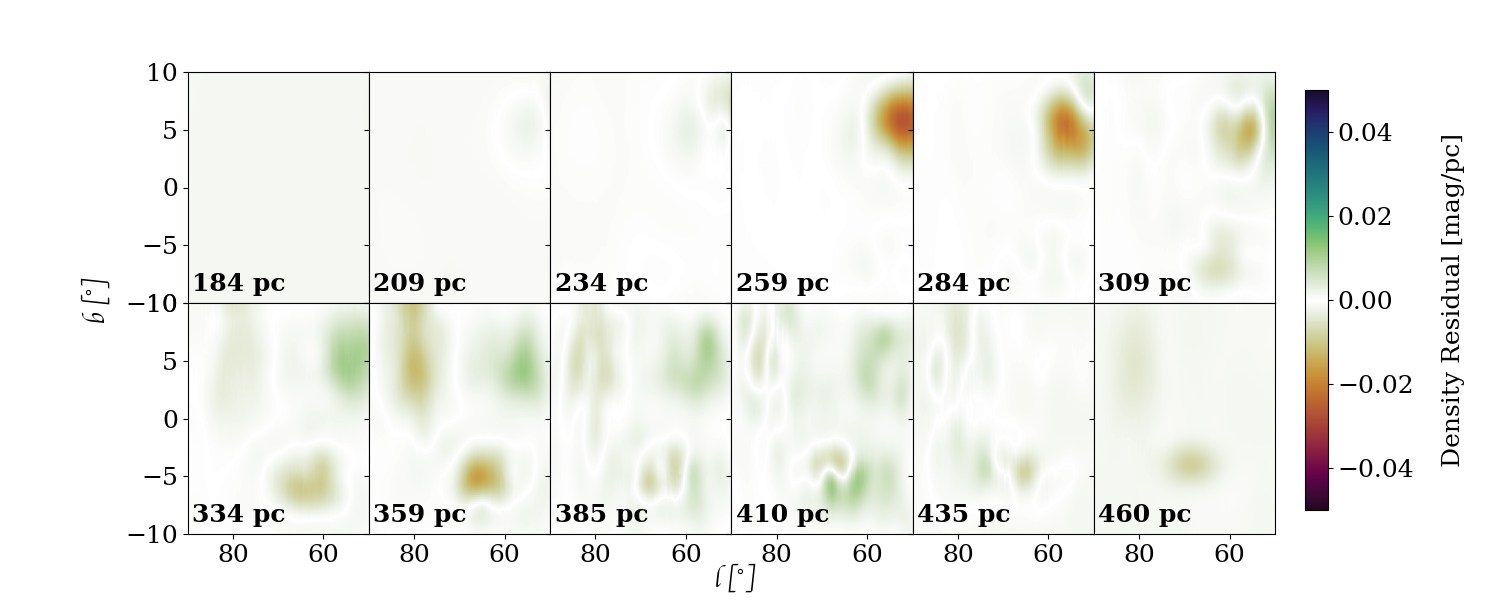}
  \end{subfigure}
  \caption{\emph{Top}:Simulated Density map sampled at the indicated distances; \emph{Middle}:Our model reconstructed Density map sampled at the same distances as the simulated map; \emph{Bottom}:Residual of Model Predicted density - Simulated Density}
    \label{fig:SimDenseAll}
\end{figure*}

As we can see from our comparison of the simulated and reconstructed data, our model reproduces the overall shape of the extinction and density distributions well. In the histogram presented in Fig~\ref{fig:SimExt_Hist_All} the predicted source-by-source extinction follows the same trend as true extinction with only a small fraction ($<10\%$) of the sources being underestimated. We do see that the model reconstructions of extinction and density are smoothed out compared to the ground truth, as is expected from a GP method. This smoothing is more evident in the plots of the residuals (Figs.~\ref{fig:SimCumExtAll} and \ref{fig:SimDenseAll}). In the total extinction we see an average underprediction of less than 1 magnitude and in density it's slightly larger, with a typical (maximal) variation of $\sim 12\%$ ($\sim 20\%$).

The underestimation of extinction, particularly the highest extinctions, is driven by a combination of two effects; the density is smoothed, spreading the same extinction over a larger range of distances, and the vanishingly-small probability of having sources that sample the very highest extinctions makes it difficult to measure them. On the other hand, densities are both over-estimated in some regions and underestimated in others, which also arises because of smoothing - when some material is moved along the line-of-sight, the clump becomes larger and hence the core will be underestimated and the wings slightly overestimated.

The smoothing performed by the GP inevitably introduces some error into the reconstructed positions of structures in density, particularly along the line-of-sight. However, from the predicted and residual plots of density shown in Fig.~\ref{fig:SimDenseAll} we can see that this is typically less than one scale length (i.e. features may be offset from their true positions by up to $20$~pc in the final map in this case). 

For our density reconstructions we also need to consider the typical uncertainty scales on the values of the density. 
In appendix~\ref{sec:Appendix:SimAlongLBDFigs} we present the density and extinction as a function distance for several lines of sight, along with cuts along $l$ and $b$ as a function distance. From these figures we infer a typical uncertainty of $\lesssim 20\%$ in density. 

Overall, the size of the uncertainties and errors of the output of our model are small, demonstrating that our algorithm provides good performance in both localising structures and estimating their magnitude, provided that the scale length is smaller than the typical molecular cloud size. 

The time taken to train our model depends on several factors. (The run times of our model for our star formation region are given in the upcoming table~\ref{tab:HPs}). The number of cells in the training grid is the largest factor, with the run time empirically scaling slightly faster than linear ($\sim \mathcal{O}\left(n^{1.4}\right)$) with the number of cells $n$. The run time relative to the number of inducing points, however, scales approximately $ \mathcal{O}\left(n^{0.95}\right)$ but consistent with linear scaling. Finally, as the number of sources changes, the run time varies $ \mathcal{O}\left(n^{0.8}\right)$. The prediction run time makes up a negligible fraction of the run time taking only several hundred seconds compared to the hours -- days required for training depending on the described factors above.  

\section{Data}
\label{sec:Data}

In the rest if this work, we use
data from the stellar parameter catalogue of Fouesneau et al.\ (submitted)\footnote{available at \url{http://dc.zah.uni-heidelberg.de/}}. 
Specifically we use their inferred extinction at 547~nm ($A_{0}$) as $A_{obs}$ and their distances as $d$. Our code is wavelength agnostic, however, and could be used with extinctions at any wavelength from any catalogue.

The Fouesneau et al., parameters are inferred from photometry from Gaia DR2, 2MASS, and WISE together with Gaia parallaxes. Their model simultaneously estimates $A_{0}$, distances, $R_{0}$, effective temperature, luminosity, surface gravity, mass and age for each star independently. The models predict the reddened spectral energy distributions (SEDs) from the data using PARSEC isochrone models \citep{Chen2014_PARSEC1, Marigo2013_PARSEC2, Rosenfield2016_PARSEC3}, the ATLAS9 atmospheric library \citep{Castelli2004_ATLAS9}, and Fitzpatrick extinction law \citep{Fitzpatrick1999}. Their estimates result from Markov chain Monte Carlo (MCMC) samples of the posterior parameter distribution. During this procedure, the models are interpolated using a neural network. In the regions we study in this present work, the typical (median) uncertainties in extinction are 0.34 mag, and the fractional parallax uncertainties are typically between 0.13 and 0.19.

\begin{figure*}
    \centering
    \includegraphics[width=\textwidth]{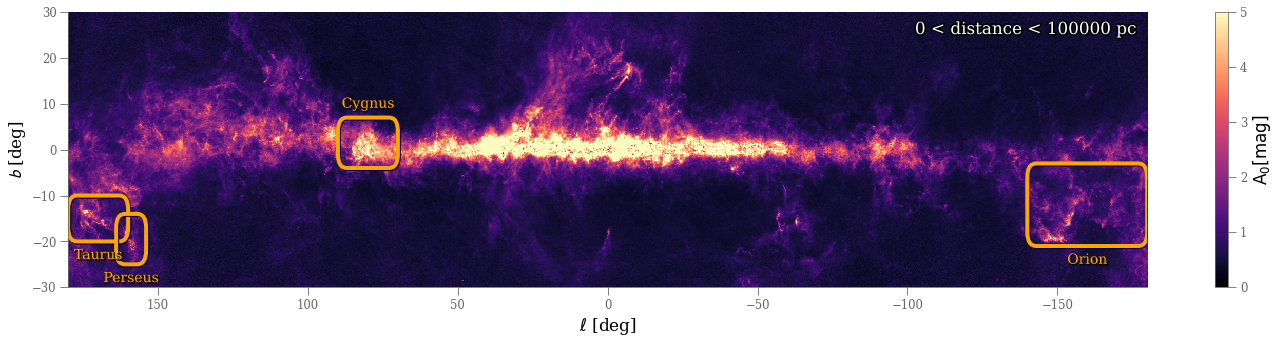}\\
    \includegraphics[width=\textwidth]{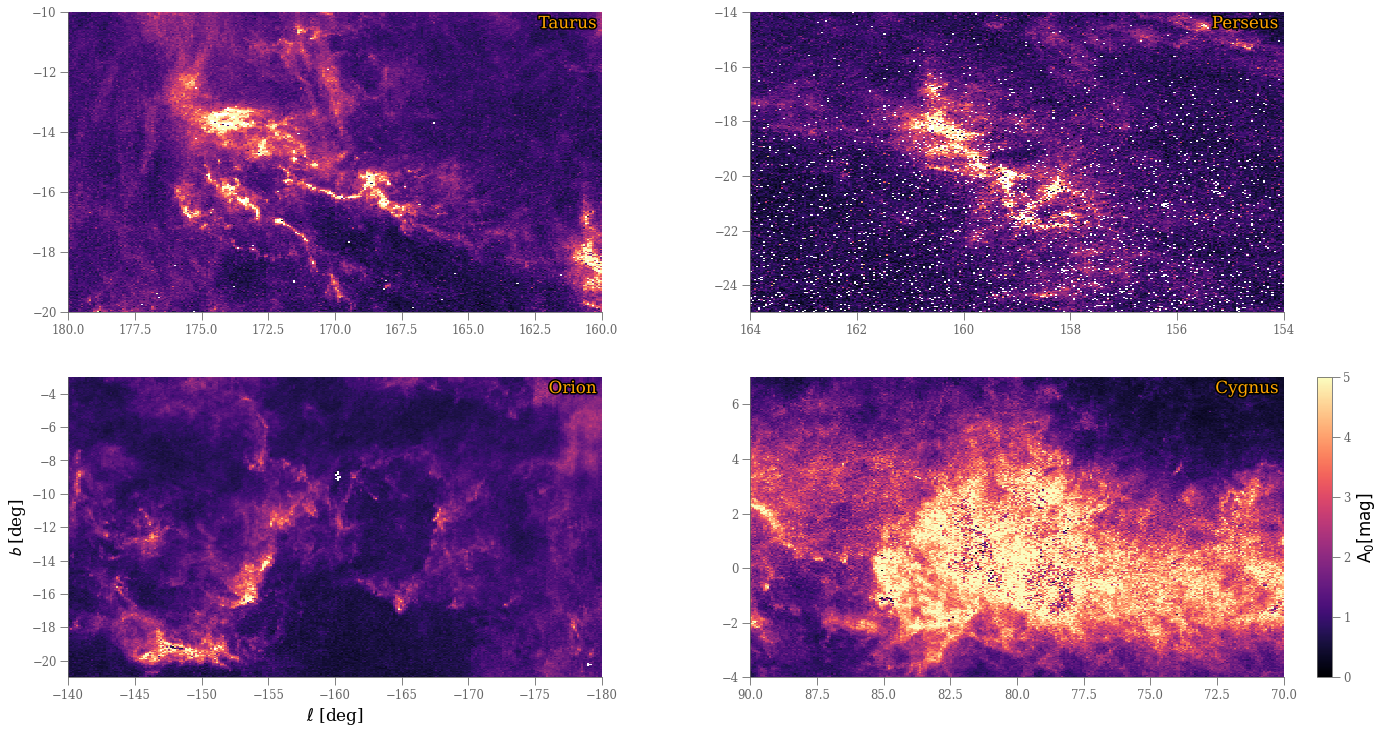}
    \caption{Extinctions as a function of Galactic coordinates from the catalogue of Fouesneau et al., subm. with the star formation regions analysed in this paper highlighted. \emph{Top}: Full Galactic extinction map; \emph{Middle left}: Taurus; \emph{Middle right}: Perseus; \emph{Bottom left}: Orion; \emph{Bottom right}: Cygnus X}
    \label{fig:lbol_inputdata}
\end{figure*}

We show their Galactic $A_0$ map and the regions we study in this work in Fig. \ref{fig:lbol_inputdata}. We do not make any cuts on the set of sources based on stellar type or stellar parameters; we simply include all sources that fall within our region boundaries. Young Stellar Objects (YSOs) often have circumstellar dust that may be incorrectly attributed to interstellar extinction when that is determined in the Fouesneau et al., catalogue. We did not filter out such objects. However, we expect these objects to be rare, because they are often very red and so would lack a BP measurement, and thus be excluded from the catalogue.

Using this catalogue as our input data provides us with advantages compared with previous Studies. First, 
one can achieve more reliable estimates of stellar parameters by combining multiple spectroscopic and photometric surveys. This catalogue relies on more than Gaia-only data instead of using GDR2 $A_G$ from \citet{Andrae2018}.  \citet{Leike2019} used those in their analysis and were limited to the systematics of the GDR2 extinctions. Second, infrared indicators such as RJCE \citep{Majewski2011_RJCEmethod}, optimized for particular applications, are less sensitive to low column density than optical bands. These $A_k$ estimates need careful considerations to extrapolate their usage on non-giant stars \citep[e.g.][]{RezaeiKh.2020}. Finally, the method in Fouesneau et al., subm. jointly estimates distance with the extinction (and other properties). As a result, we do not rely on the inverse parallax as distance measurements \citep{Bailer-Jones2015}, and we obtain a coherent set of input data.


\section{Orion}
\label{sec:Orion}

The Orion star-forming region (SFR) is the nearest cluster of young, massive stars ($> 8 \ \mathrm{M_{\odot}}$) at an approximate distance of 400~pc \citep{Menten2007_OrionDist}. The stellar ages in the Orion SFR range from 2--12 Myr and the SFR is separated into three main components, the Orion A and Orion B molecular clouds, and the $\lambda$ Orionis molecular ring \citep{Bally2008_Orion}. Orion has been studied extensively in literature, making it an ideal real-world test case for our model. Its 3D dust density distribution has been explored by \citet{Rezaei2018A, RezaeiKh.2020}, who found a foreground cloud in front of the Orion A molecular cloud at 345~pc. The dust extinction distribution was investigated by \citet{Lombardi2011_Orion} (hereafter \citetalias{Lombardi2011_Orion}) and \citet{Schlafly2015_OrionDust} (hereafter \citetalias{Schlafly2015_OrionDust}). \citetalias{Lombardi2011_Orion} explores the distances to several clouds within the Orion SFR using foreground stellar densities and infers masses for them, while  \citetalias{Schlafly2015_OrionDust} explore the 3D structure using dust reddening to study dust ring structures in detail. Attempts to understand the structure of Orion are particularly motivated by the need to study the feedback of massive stars on their environment and the impact this has on the star-forming process \citep{Bally2008_Orion}.

\subsection{Orion Model setup}

The model set up including the total number sources from Sec.~\ref{sec:Data}, chosen region size along galactic longitude $l$, galactic latitude $b$ and distance $d$ along line-of-sight ($l$ bounds, $b$ bounds, $d$ bounds), spacing of cells along $l,b,d$ ($n_l,n_b,n_d$) and initial input and final conditioned Gaussian kernel hyperparameters are presented in Table~\ref{tab:HPs}. We begin testing our model for the Orion SFR by setting the scale lengths of the  Gaussian kernel to 10~pc, a value we arrive at based on the distance to the complex and the projected sizes of the molecular clouds. We initialised the mean density to that used by \citet{Rezaei2017}. We set the initial scale factor based on the size of the perturbations expected from the mean background. We then ran the model a number of times, starting the next run from the end point of the previous one, until we found a satisfactory region of parameter space for the scale factor and mean density for the final model run. This was an intuitive process, and resulted in the initial values listed in Table~\ref{tab:HPs}. 

\begin{figure*}
    \centering
    \includegraphics[width=\textwidth, trim=4cm 4cm 5.5cm 4cm, clip]{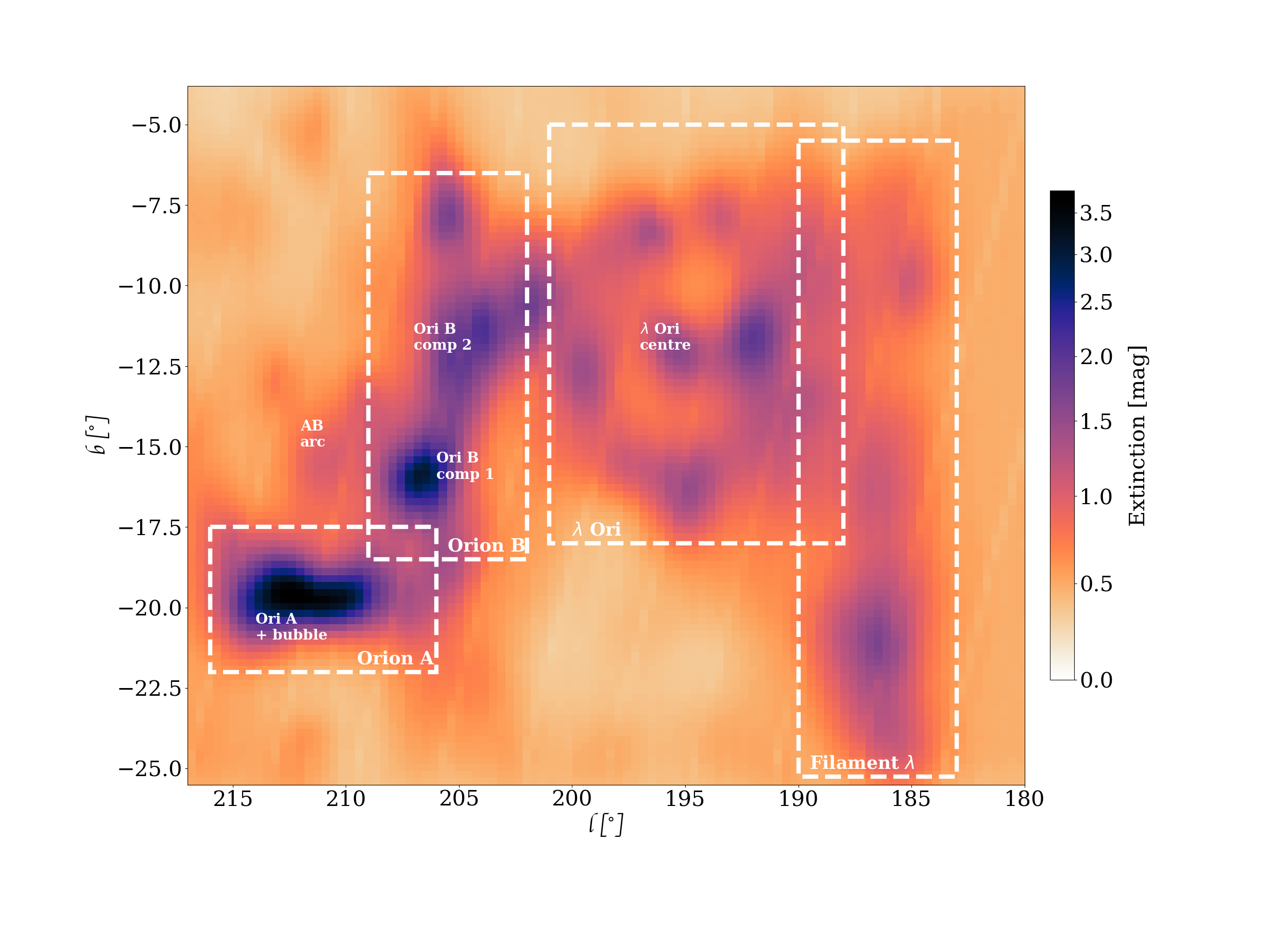}
    \caption{Total Extinction of Orion reconstructed by our model at a distance of 550~pc.}
    \label{fig:OrionCumExt}
\end{figure*}

\begin{figure*}
    \centering
    \includegraphics[width=1\textwidth, trim=1cm 3.5cm 1cm 7cm, clip]{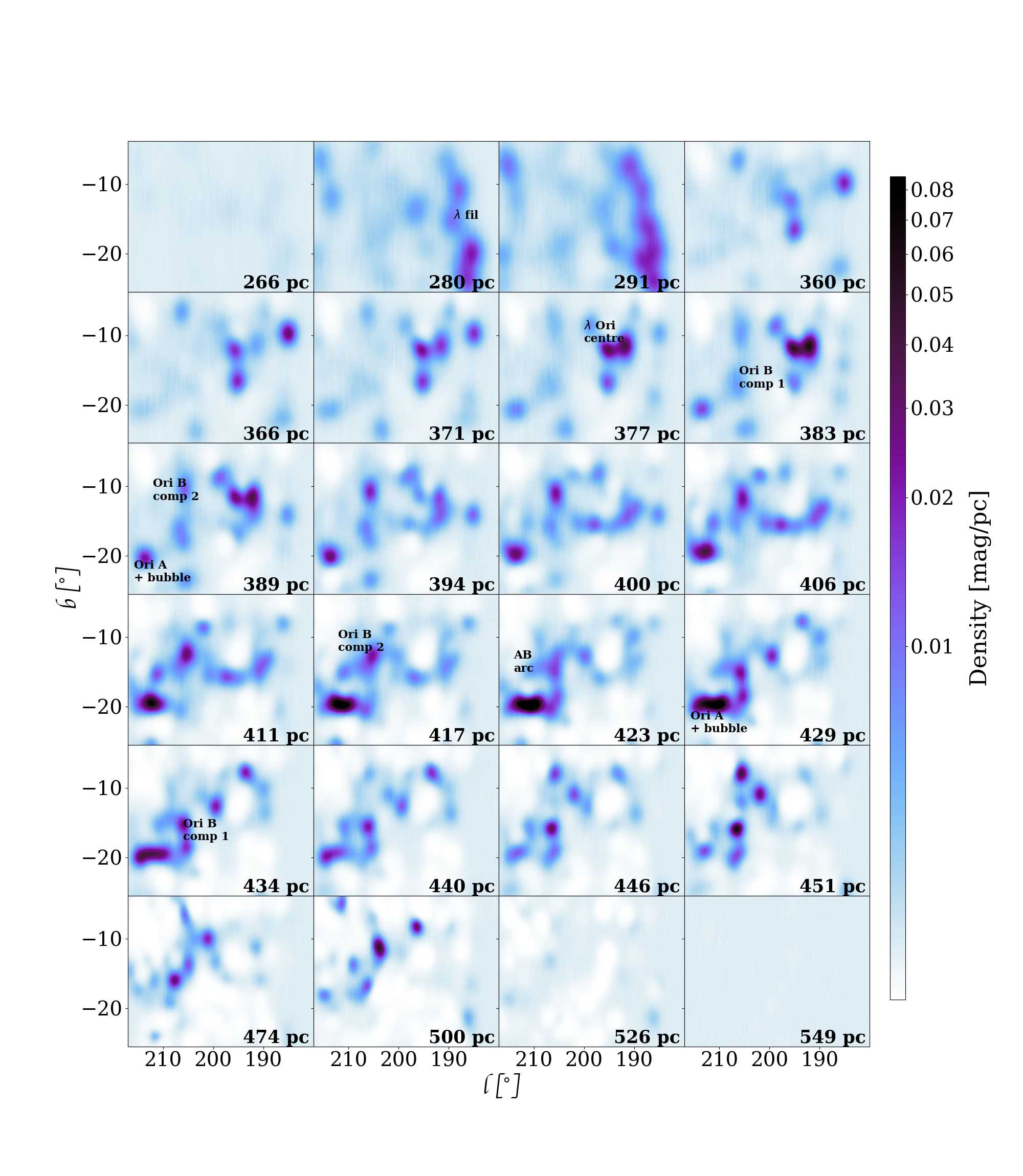}
    \caption{Three-dimensional dust density distribution of Orion sampled at the indicated distances.}
    \label{fig:OrionDensSlices}
\end{figure*}

\input{Tables/HPs}

\subsection{Inferred structure of Orion}

We summarise the coordinates and the sizes of the main components of Orion identified here in Table~\ref{tab:ClumpSizes}. Figures \ref{fig:OrionCumExt}, \ref{fig:OrionDensSlices}, and \ref{fig:OrionExtSlices} show our inferred dust densities and the line-of-sight extinctions computed from these. These maps recover the structure of the Orion region well when compared to other literature results, such as \citetalias{Lombardi2011_Orion}, \citetalias{Schlafly2015_OrionDust} and \citet{Zucker2020}, as well as the Planck dust emission \citep[shown in Fig.~\ref{fig:Planck_DustFlux_All};][]{Planck2016}. We do not see evidence for a separate foreground cloud along the line-of-sight to Orion A as suggested by \citet{RezaeiKh.2020}. We do, however, see that Orion A is quite extended, stretching from 350 to 450 \ $\mathrm{pc}$ as first suggested by \citep{Groschedl2018_OrionA} using YSO data from Gaia DR2 and later by \citep{RezaeiKh.2020} using 3D dust densities.
The Planck $850 \ \mathrm{\mu m}$ and dust intensity maps (see Figs. \ref{fig:Planck_850micronflux_All} and \ref{fig:Planck_DustFlux_All}) show a bubble at $l=215^{\circ}$ and $b=-20^{\circ}$ at the edge of Orion A in the direction of the Galactic center. This bubble is also seen in the infrared extinction by \citetalias{Lombardi2011_Orion}, and \citetalias{Schlafly2015_OrionDust} suggests it lies beyond 550~pc. However, our maps show this bubble is closer to us along the line-of-sight, appearing as an over-density in all three of our maps. It is densest at 405~pc while spanning  $380 \leq d \leq 430 \ \mathrm{ pc}$.  

We recover Orion B and find it to be composed of two segments and to be elongated along the line-of-sight (seen in Fig.~\ref{fig:OrionDensSlices}). The component adjacent to Orion A (Ori B comp 1 in Fig.~\ref{fig:OrionDensSlices}) appears at around 380~pc and disappears at around 400~pc before reappearing at larger distances with the densest region around 430~pc \citep[consistent with ][]{RezaeiKh.2020}, while the component closer to $\lambda$ Orionis (Ori B comp 2 in Fig.~\ref{fig:OrionDensSlices}) appears around 380~pc and extends up to 420~pc. 

$\lambda$ Orionis itself is the bubble-like structure whose nearside is around 360~pc and the far side is around 110~pc further along the line-of-sight, placing it closer than previous measurements such as \citetalias{Lombardi2011_Orion} who infer a distance $445\pm50 \mathrm{pc}$. 
Additionally, we recover a peak in density at the centre of the ring (on sky) at a distance of  360--390~pc; this was not seen by \citetalias{Lombardi2011_Orion} or \citetalias{Schlafly2015_OrionDust} but is clearly visible in the Planck dust intensity map shown in Fig.~\ref{fig:Planck_DustFlux_All}.

Another arc-like feature linking Orion A and Orion B can be seen at $205^{\circ} \leq l \leq 210^{\circ}$ and $-19^{\circ} \leq b \leq -15^{\circ}$ between 400 and 450~pc, possibly corresponding to NGC2170 and the dust surrounding the Barnard Loop {\sc hii} region. The large cloud seen at $205^{\circ} \leq l \leq 215^{\circ}$ and $b=-5^{\circ}$ in the Planck $850 \ \mathrm{\mu m}$ map and in extinction by \citetalias{Lombardi2011_Orion} is not visible in our results. This may imply the cloud is located beyond 550~pc, possibly indicating its association with the Canis Major star-forming region instead of Orion.

Finally, we recover a large filament at $l \simeq 186^{\circ}$ visible at 270--300~pc, even though it is not clearly visible as a coherent structure in extinction. This emphasises the power of 3D reconstructions recovering structure lost in integration to 2D (however it must be noted that if differentiated, 2D extinction maps may also have the potential to show similar localisation). We refer to this filament as the \textit{$\lambda$ filament} in Fig.~\ref{fig:OrionCumExt} as it is closest to $\lambda$ Orionis.  A similar structure is also visible in the Planck maps (figures \ref{fig:Planck_850micronflux_All} and \ref{fig:Planck_DustFlux_All}). This filament may be host to several small clouds such as LDN 1558 and TGU L156 \citepalias{Lombardi2011_Orion} as they overlap on sky. However without confirmed distances to these condensations we cannot corroborate this.

\input{Tables/ClumpSizes}

\section{Cygnus X, Perseus, and Taurus}
\label{sec:Cyx_Per_Taurus}

We select three more star forming regions to demonstrate the capabilities of our algorithm. These were chosen on the grounds that their 3D dust densities have been less well studied, and because they have different properties from Orion. Once again, we summarise the coordinates and the sizes of the main components of these regions identified through this work in Table~\ref{tab:ClumpSizes}.

Cygnus X is the most massive star forming region within 2~kpc of the Sun, with a total mass of $3\times10^{6} \ \mathrm{M}_{\odot}$ \citep{Schneider2006_CygX} and cluster ages of up to 18 Myr \citep{Maia2016_CygAge}. It is home to the largest number of massive protostars and the largest OB stellar association Cygnus OB2 \citep{Guarcello2013_CygX}. While the Orion SFR is located south of the Galactic plane, Cygnus X is in the Galactic plane, so shows higher foreground extinction and crowding. Of the four regions studied here, Cygnus X is the furthest away from us, with estimated distances in the range 1300-2000~pc \citep{Rygl2010_CygXDist1, Rygl2012_CygXDist2, Schneider2006_CygX}. 

The Taurus star formation region is the nearest to us at a distance of 145~pc \citep{Yan2019_Taurus_Perseus} and an age below 5 Myr. Deviating from the common embedded cluster mode star formation seen in other SFRs such as Orion and Perseus, Taurus is the prototype low-mass star formation region ($0.7-1.0 \mathrm{M_{\odot}}$) where stars appear to form in relative isolation \citep{Kraus2017_Taurus}. The proximity of Taurus to us also makes it one of the largest SFRs on sky. This, combined with its large population of young stars, makes it an interesting star formation region for analysis. 

Finally, we study the Perseus star formation region which neighbours Taurus on sky and is at a distance of 310~pc \citep{Yan2019_Taurus_Perseus}. Perseus is home to several regions of active or recent star formation with typical ages of 1--5 Myr \citep{Pavlidou2021_Perseus}. It is the smallest of the four regions in angular size presented in this work. There are several filamentary structures spreading towards it from the other nearby star formation regions including Taurus and California. With a mass of around 100 $\mathrm{M_{\odot}}$, this SFR is the closest region which is still actively forming low--intermediate-mass stars \citep{Bally2008_Perseus}.

The model-setup for these three regions is similar to the one we used for Orion, bar a few modifications to match properties of these regions as given in table \ref{tab:HPs}.

\subsection{Inferred structure of the regions}

\subsubsection{Cygnus X}
\begin{figure*}
    \centering
    \includegraphics[width=\textwidth, trim=1cm 3.5cm 1cm 7cm, clip]{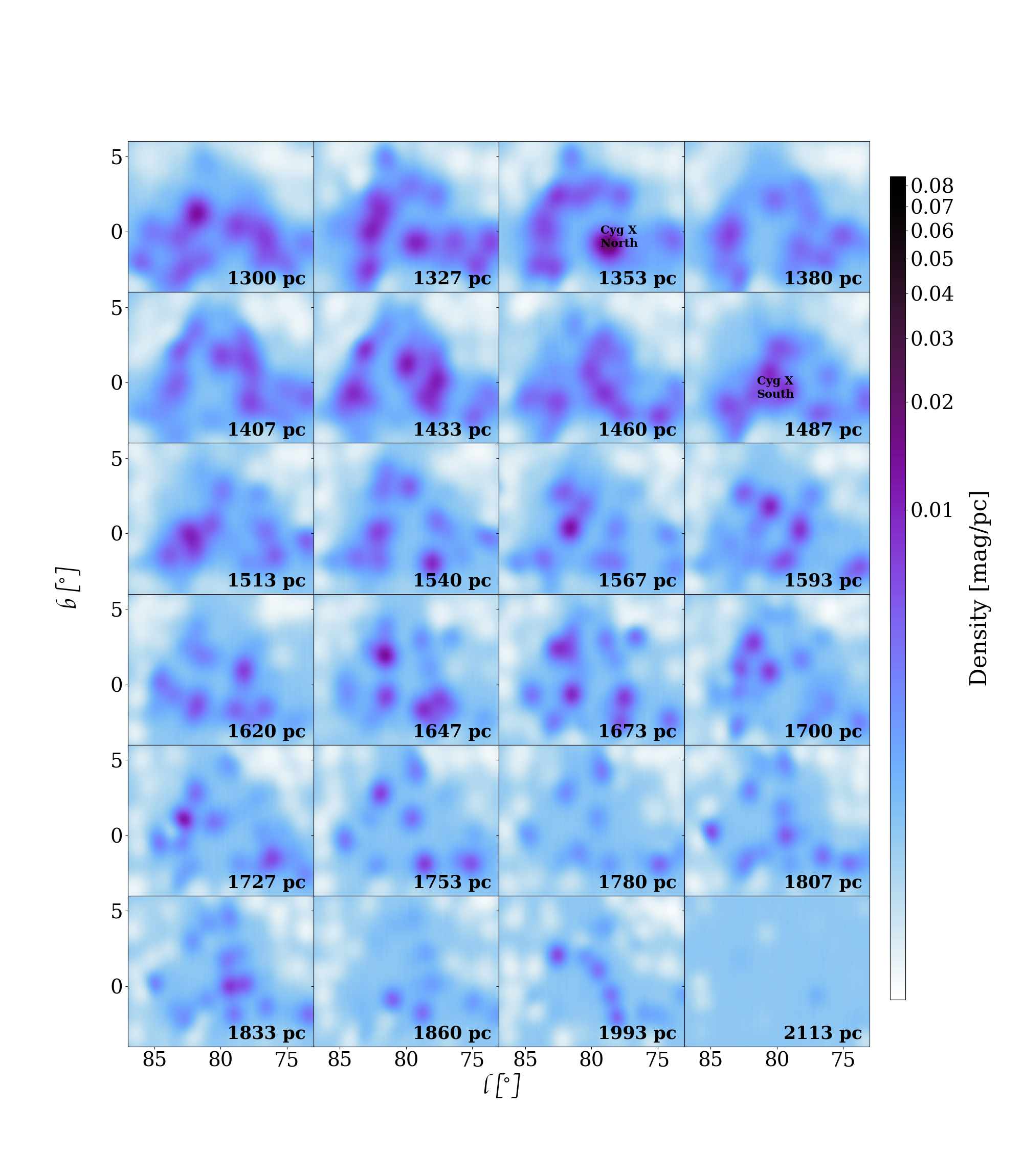}
     \caption{Three-dimensional dust density distribution of Cygnus X sampled at the indicated distances.}
    \label{fig:CygXDensSlices}
\end{figure*}

The distance to  Cygnus X has historically proven difficult to determine, in part because of its location in the Galactic Plane. Distances in the literature span from 1.3~kpc \citep{Rygl2010_CygXDist1} to 2~kpc \citep{Schneider2006_CygX}. With our 3D density mapping algorithm we are able to localise the densest regions of Cygnus X to 1.3--1.5~kpc as shown by fig~\ref{fig:CygXDensSlices}. Our distance estimate is consistent with  the maser parallax distances measured by \citet{Rygl2010_CygXDist1} and \citet{Rygl2012_CygXDist2}.
Using CO data in which adjacent regions appear to be interacting, \citet{Schneider2006_CygX} inferred that Cygnus X is a monolithic structure at 1.7~kpc. 
However, we detect small clumps throughout our distance range, corroborating suggestions in the literature that the Cygnus X region  may contain both the large star-forming region (which we place at 1.3--1.5~kpc) and a number of smaller clouds spread along the line of sight, as first suggested by \citet{Dickel1969_CygXstructure} based on extinction measurements. The clumpy structure and upper distance limit also matches well with \citet{Zucker2020}. While \citet{Zucker2020} recover some clumps closer to us than 1200~pc, we do not because we do not have any stars closer than 1200~pc included in our model.

\begin{figure*}
    \centering
    \includegraphics[width=\textwidth, trim=4cm 4cm 5.5cm 4cm, clip]{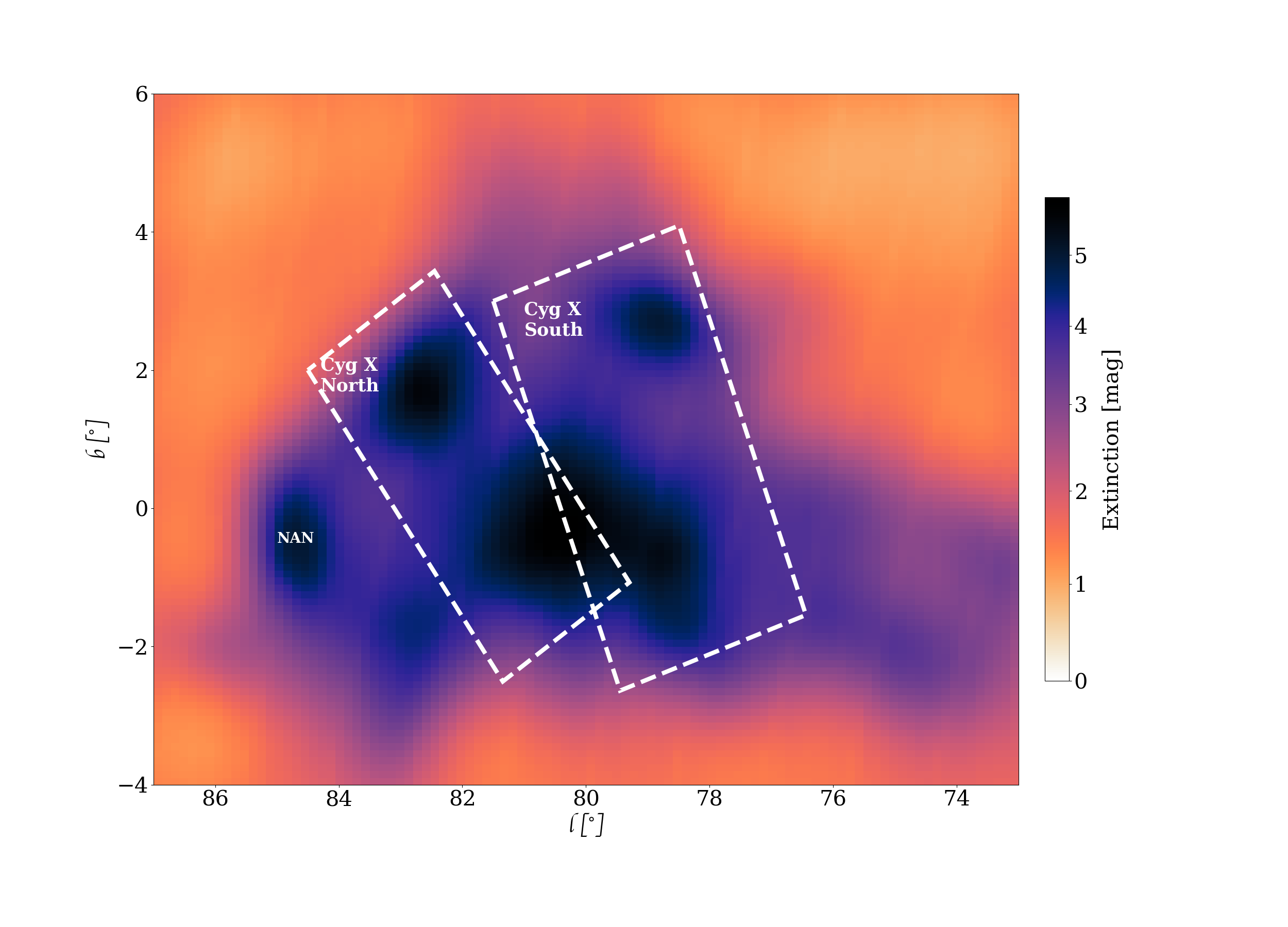}
    \caption{Total Extinction of Cygnus X reconstructed by our model at a distance of 1600~pc.}
    \label{fig:CygXCumExt}
\end{figure*}

We see a separation in the Cygnus X North and South Clouds in figure \ref{fig:CygXDensSlices} and find that Cygnus X South is closer to us, with the dense region in Cygnus X South starting at 1300~pc. The south cloud is densest at 1350~pc, while the north cloud appears to be densest at around 1500~pc. We also see the presence of the North American Nebula (NAN) in the extinction figure \ref{fig:CygXCumExt} at $l = 0^{\circ}, b = 85^{\circ}$. At 600~pc \citep{Schneider2006_CygX} it is much closer to us than the region encompassed by our model and therefore not visible in our 3D density plot. Further, as expected given the high extinction in the region and its placement in the Galactic plane, Cygnus X has the highest mean density of any of the regions we examine. 

\subsubsection{Perseus}

\begin{figure*}
    \centering
    \includegraphics[width=\textwidth, trim=1cm 3.5cm 1cm 7cm, clip]{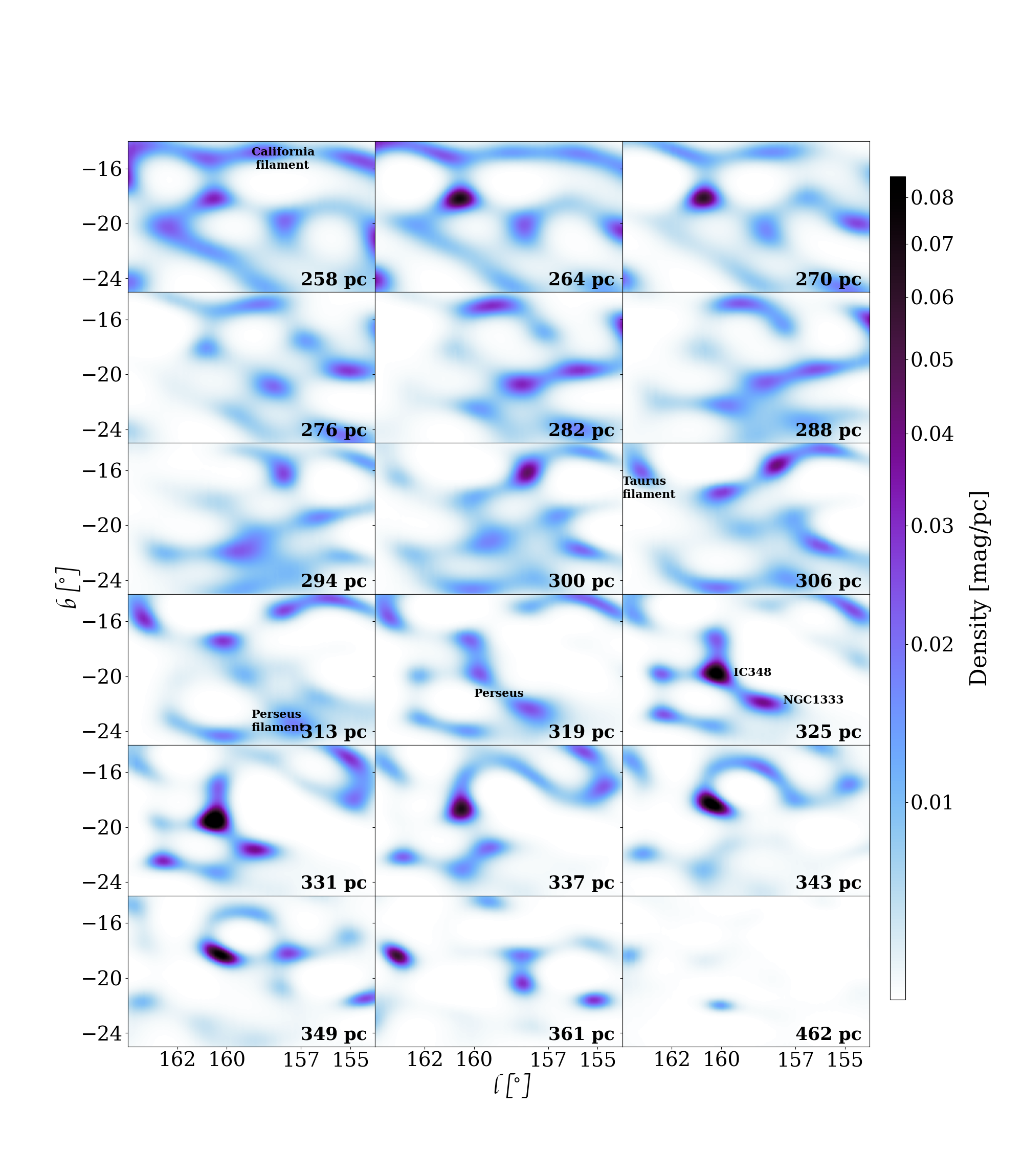}
     \caption{Three-dimensional dust density distribution of Perseus sampled at the indicated distances.}
    \label{fig:PerseusDensSlices}
\end{figure*}

Our density distribution (Fig.~\ref{fig:PerseusDensSlices}) shows that Perseus itself lies between 300 and 350~pc. This is larger than the mean distance of 240~pc determined by \citet{Lombardi2010_Perseus_Taurus} (hereafter \citetalias{Lombardi2010_Perseus_Taurus}) who used foreground stars to arrive at this distance. However it agrees well with the mean distance of 310~pc from \citet{Yan2019_Taurus_Perseus} (using Gaia DR2 extinction and parallaxes) and the range of 294--350~pc found by \citet{Zucker2018_PerseusDist} who used photometry, Gaia DR2 astrometry and $^{12}$CO data. 

\begin{figure*}
    \centering
    \includegraphics[width=\textwidth, trim=4cm 4cm 5.5cm 4cm, clip]{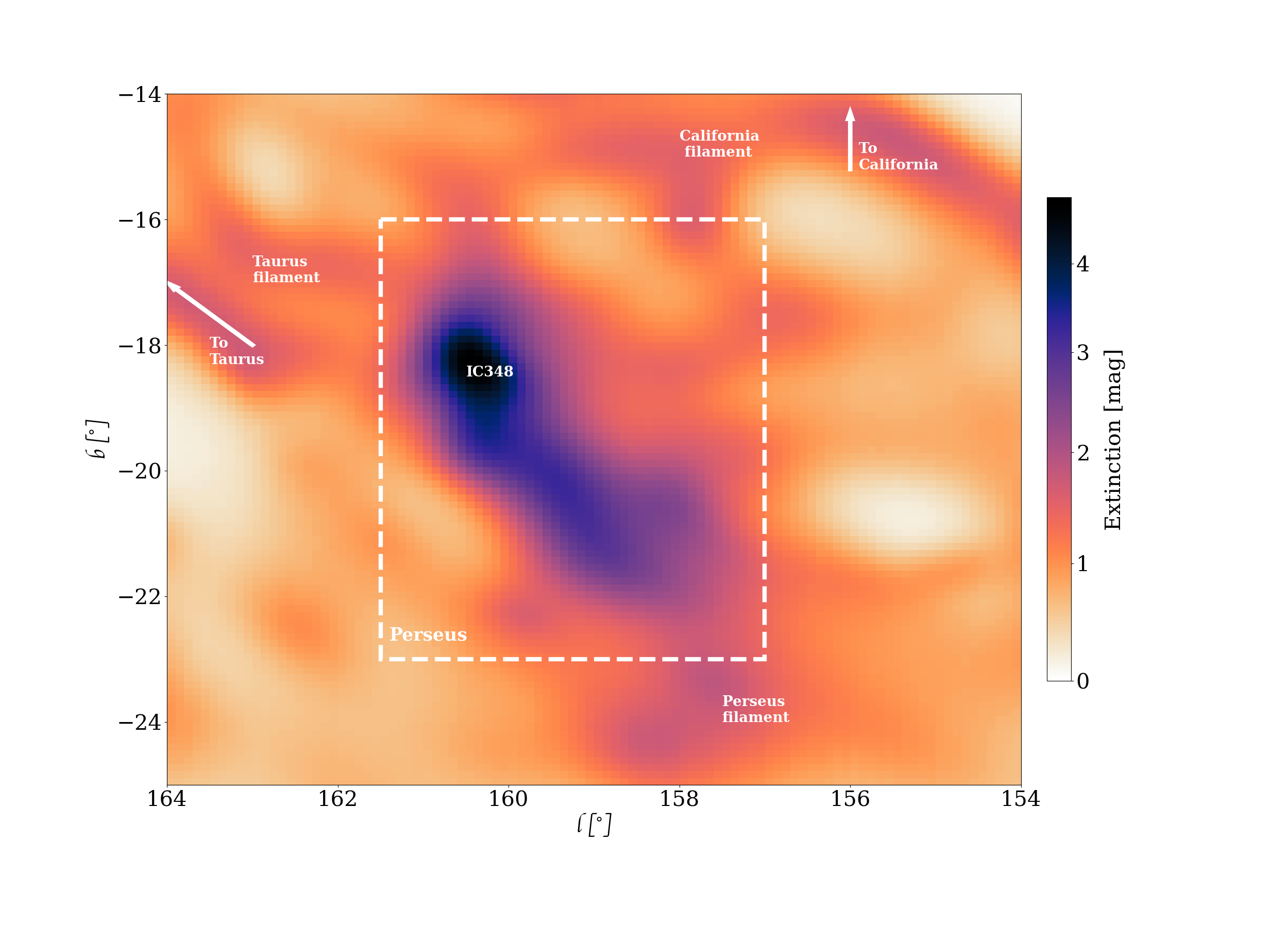}
     \caption{Total Extinction of Perseus reconstructed by our model at a distance of 500~pc.}
    \label{fig:PerseusCumExt}
\end{figure*}

We identify the two largest clusters IC348 ($l=160^{\circ}, b=-18^{\circ}$) and NGC1333 ($l=158^{\circ}, b=-20^{\circ}$) at the eastern and western edges of the cloud complex at similar distances to one another. We also detect a clump just above IC 348 at $l=160^{\circ}, b=-17.5^{\circ}$ located at a distance of $\sim 306-330$ pc. It is densest at a distance of $\sim 325$ pc and is clearly detected in both the Planck dust emission in Fig.~\ref{fig:Planck_Tau_All} as well as \citet{Zucker2020} who place it at a distance of $331\pm16$ pc. We do not localise the smaller clumps associated with the Perseus SFR. Although IC348 is also visible in our total extinction map (Fig.~\ref{fig:PerseusCumExt}) NGC1333 is not as distinct, further demonstrating the importance of 3D mapping to fully understand the structure of molecular cloud regions. 

Interestingly, we also recover the elongated filaments which appear to extend towards Perseus from the Taurus and California molecular clouds when viewed on sky. We also detect filamentary structure extending from the western edge of Perseus away from the Galactic plane. We localise these filaments in 3D for the first time. These filaments are visible in Planck emission (see Figs.\ref{fig:Planck_Tau_All}, \ref{fig:Planck_850micronflux_All}, \ref{fig:Planck_DustFlux_All}) as well as extinction maps presented by \citetalias{Lombardi2010_Perseus_Taurus}, but they have not been localised in 3D before this work. The filament associated with California may in fact be on the edge of Perseus with a distance starting at 250~pc out to nearly 350~pc. The filament between Perseus and Taurus corresponds to a distance of 300~pc extending to $\sim 330$~pc. The filament furthest from the Galactic plane is at 300~pc but appears to be more diffuse than the other filaments and we cannot clearly associate it to Perseus or any other nearby star forming regions. We refer to these filaments as \textit{California filament}, \textit{Taurus filament}, and \textit{Perseus filament} respectively in our figures.

\subsubsection{Taurus}

\begin{figure*}
    \centering
    \includegraphics[width=\textwidth, trim=4cm 4cm 5.5cm 4cm, clip]{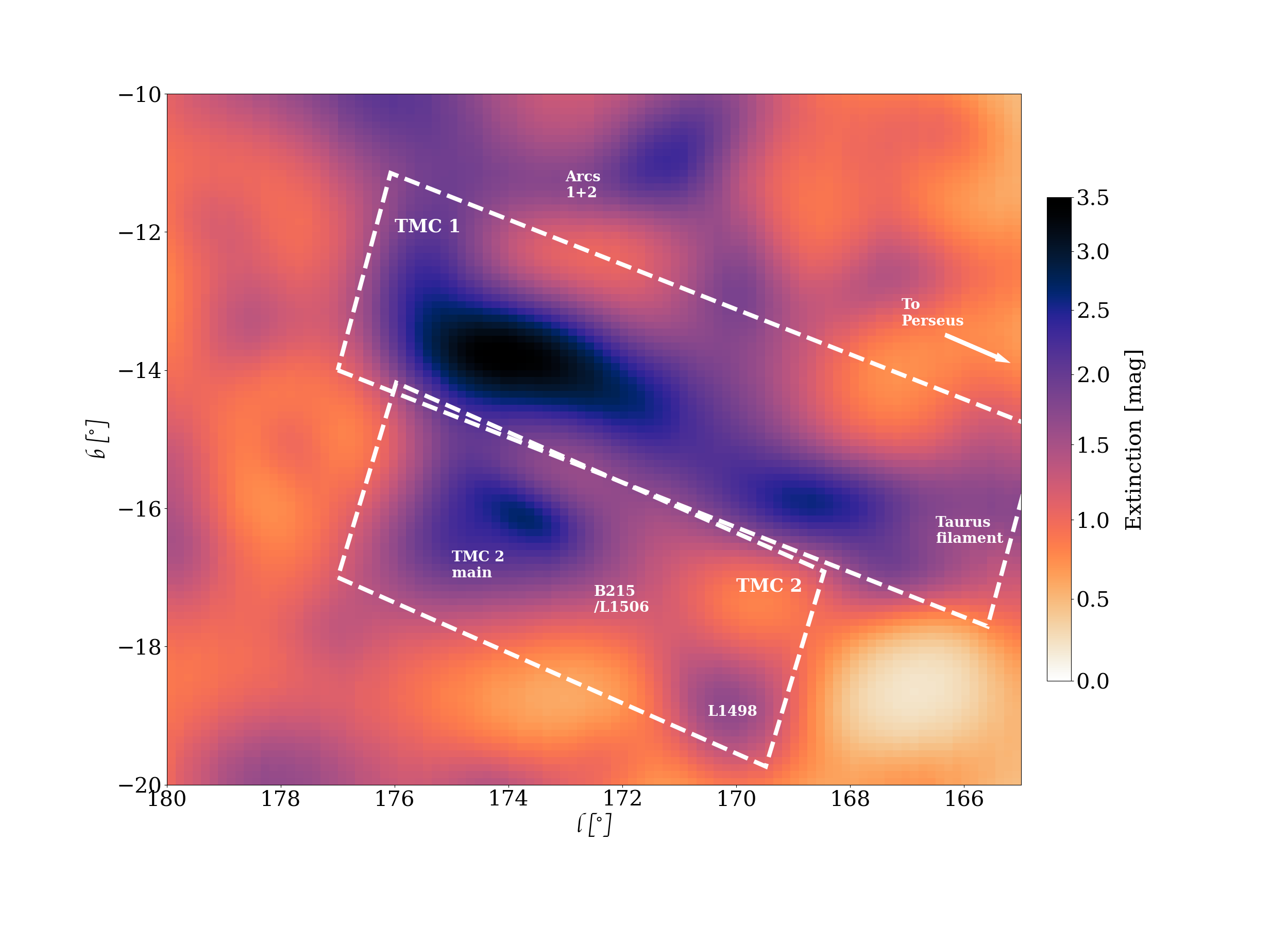}
     \caption{Total Extinction of Taurus reconstructed by our model at a distance of 350~pc.}
    \label{fig:TaurusCumExt}
\end{figure*}

\begin{figure*}
    \centering
    \includegraphics[width=\textwidth, trim=1cm 3.5cm 1cm 7cm, clip]{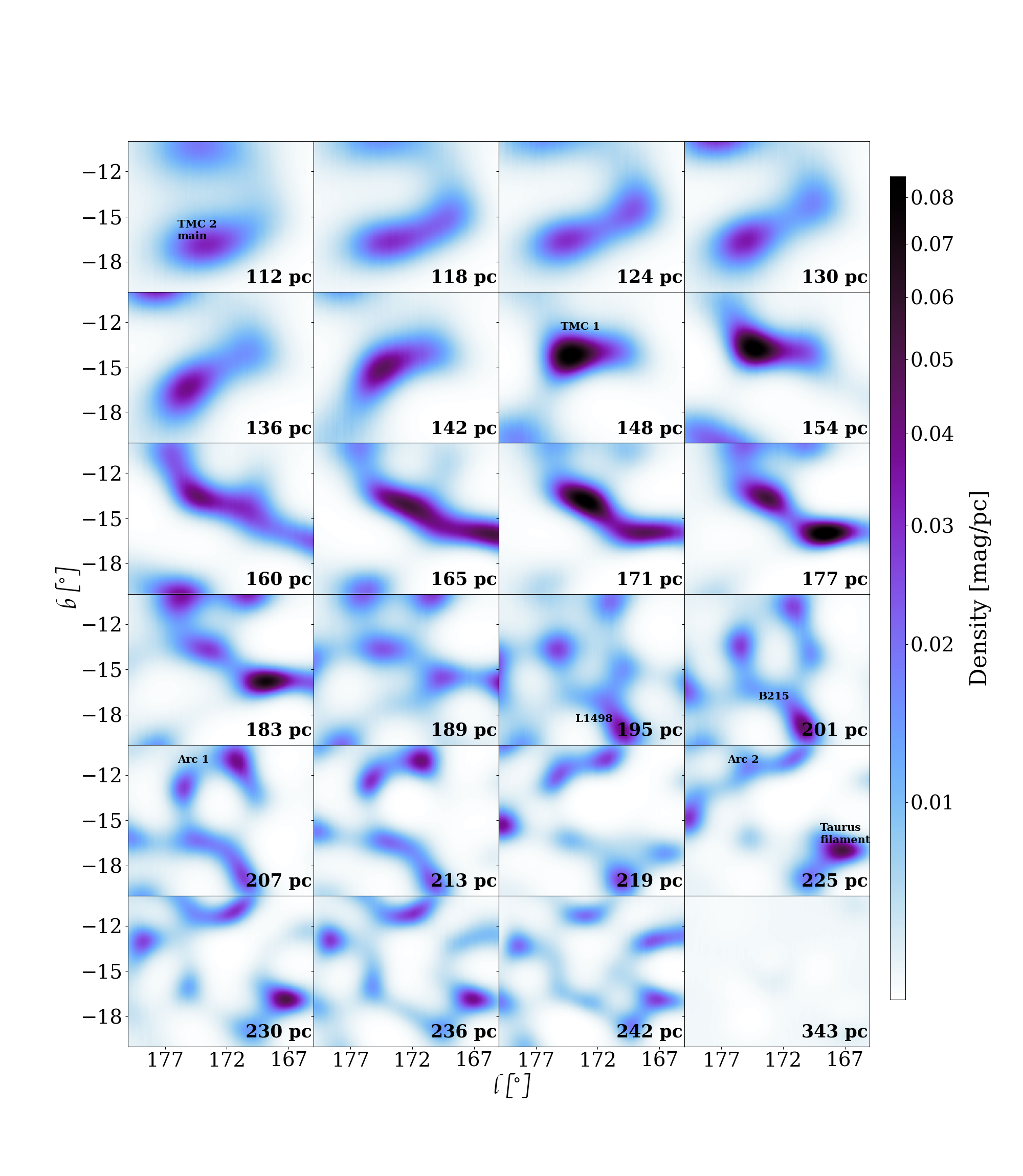}
     \caption{Three-dimensional dust density distribution of Taurus sampled at the indicated distances.}
    \label{fig:TaurusDensSlices}
\end{figure*}

Taurus has been separated into two components, TMC 1 and TMC 2, based on their sky positions \citep{Lombardi2010_Perseus_Taurus}. In our 3D density distribution plot shown in Fig.~\ref{fig:TaurusDensSlices}, the Taurus cloud complex extends from 110--190~pc, with the highest density region at 150--180~pc. This is consistent with \citet{Yan2019_Taurus_Perseus}, who found a mean distance of 145~pc. The dense cloud at $l=174^{\circ}, b=-16^{\circ}$ in Fig.~\ref{fig:TaurusCumExt} corresponds to the main component of TMC 2. This component is elongated along the line-of-sight (Fig.~\ref{fig:TaurusDensSlices}), starting at 110~pc and continuing to 145~pc, therefore TMC 2 is closer to us than TMC 1. While the main component of TMC 2 ("TMC 2 main" in Figs.~\ref{fig:TaurusCumExt} and \ref{fig:TaurusDensSlices}) is home to several smaller clumps we do not separate these clumps in our predictions. 

TMC 1, on the other hand, appears at 145~pc and continues to 190~pc, showing that it is distinct from TMC 2. Similar to TMC 2 main, we cannot separate the smaller components of TMC 1 in our maps. This suggests that TMC 1 is a coherent structure containing the clumps. 

The overdensity at $l=170^{\circ}, b=-19^{\circ}$ and 195~pc can be associated with the L1498 clump, part of TMC 2. Furthermore, the elongated structure at 195--220~pc which appears to connect TMC 2 main and clump L1498 on sky can be associated with the B215/L1506 extended clump, also part of TMC 2 on sky. While the B215/L1506 is clearly visible in 3D density in figure \ref{fig:TaurusDensSlices} it is barely visible in total extinction as seen by figure \ref{fig:TaurusCumExt}; this reiterates the importance of 3D density localisation compared to extinction maps where information is lost in the integration along the line-of-sight. 

We also see evidence for a structure at 220--250~pc at $l=165^{\circ}, b=-17^{\circ}$ (\textit{Taurus filament} in Fig.~\ref{fig:TaurusDensSlices}). This likely corresponds to the other (closer) end of the filament identified in the Perseus data, corroborating this feature. It appears to lie at a distance intermediate to the two star-forming regions, along a trajectory linking them. Whether this is an independent feature or a bridge between the two regions remains to be seen. In our density figure, we also see two arc-like structures at $\sim b<-14^{\circ}$ appearing one after the other extending from 201~pc to the end of our distance range. This could be associated to the filamentary structure adjoining TMC 1, and holds the L1540, L1507 and L1503 clumps. As noted in the Sec.~\ref{sec:SimData} there are uncertainties associated with the localisation of structure in all four regions. These uncertainties include shifts in position by approximately one scale length. 

\section{Total Mass}
\label{sec:Mass}

\input{Tables/DustMasses}

To calculate the dust and total masses of our regions, we take our integrated extinction (calculated up to the upper distance boundary) as described in Sect.~\ref{sec:LosInt} and convert it to a column density using the dust mass-extinction coefficient (opacity) of $\kappa_\mathrm{0} = 26000 \  \mathrm{cm^{2}g^{-1}}$ calculated at 547~nm. To calculate this opacity we use an absorption coefficient of $8.5 \times 10^{3} \mathrm{cm^{2}g^{-1}}$ and albedo of $0.67$ at 547~nm\footnote{data obtained from \url{https://www.astro.princeton.edu/~draine/dust/dustmix.html}} \citep{Draine2003a,Draine2003b}.
We integrate the total extinction along and $l$ and $b$ to derive a dust mass as follows:

\begin{equation}
    M_d = d_{max}^{2} \int_{b_{min}}^{b_{max}} \cos b~\mathrm{d}b \int_{l_{min}}^{l_{max}} \mathrm{d}l~\frac{ A_\mathrm{mod,d_{max}}\left(l, b\right)}{1.086 \kappa_\mathrm{0}},
    \label{eq:totmass}
\end{equation}
where the factor $1.086 = \frac{1}{\ln 2.512}$ converts from magnitudes of extinction to optical depth, $A_\mathrm{mod,d_{max}}$ is the total extinction map predicted by our algorithm up to upper distance boundary and $D$ is the distance assumed for the mass calculation which in our case is assumed to be the upper distance boundary of our model setup presented in table~\ref{tab:HPs}, $d_{max}$.

To derive the total mass we use a gas:dust ratio of 124 \citep{Draine2003a, Draine2003b} and multiply the dust mass by this ratio. We estimate the uncertainties in the masses in Table 3 by sampling over the uncertainties in our extinction maps, which in turn come from the finite spread of the Gaussian posteriors in the dust densities. Table \ref{tab:DustMasses} presents the calculated masses.

Both \citetalias{Lombardi2011_Orion} and \citetalias{Schlafly2015_OrionDust} derived the total masses for Orion using the same $l$ and $b$ range as each other. This range is approximately two degrees smaller on each axis compared to our measured region and excludes the filamentary structure we detect. They also assumed smaller upper distance boundaries. In order to be able to directly compare our masses to \citetalias{Lombardi2011_Orion} and \citetalias{Schlafly2015_OrionDust} we carry out the following steps: 

\begin{enumerate}
    \item Recalculate the mass from this work using the same $l,b$ boundaries used in the literature.
    \item Scale the literature masses up to the $d_{max}$ used in this work. 
\end{enumerate}

Following step 1, our total mass for Orion recalculated to the smaller $l,b$ range used in \citetalias{Lombardi2011_Orion} and \citetalias{Schlafly2015_OrionDust} is $735^{+260}_{-177} \times 10^{3} \ \mathrm{M_{\odot}}$. This estimate is now directly comparable to the literature masses from \citetalias{Lombardi2011_Orion} and \citetalias{Schlafly2015_OrionDust} scaled to our distance of 550~pc (step 2), which are $504 \times 10^{3} \ \mathrm{M_{\odot}}$ and $444 \times 10^{3} \ \mathrm{M_{\odot}}$ respectively.  We find our recalculated mass to be 1.4 times larger than \citetalias{Lombardi2011_Orion} and 1.6 times larger \citetalias{Schlafly2015_OrionDust}. 

Our input $A_{0}$ values (see Sec.~\ref{sec:Data}) are derived using longer wavelengths than either \citetalias{Lombardi2011_Orion} or \citetalias{Schlafly2015_OrionDust}, and hence are sensitive to higher total extinction. This means we are more likely to recover higher column densities and therefore greater masses in the present work. Another reason for the variations between the masses are the two different approaches used. While we use a conversion from extinction to mass using $\kappa_\mathrm{0}$ and a dust:gas ratio, both \citetalias{Lombardi2011_Orion} and \citetalias{Schlafly2015_OrionDust} use an approach that adopts a hydrogen column density to extinction relationship. The typical (systematic) uncertainties on the conversions in each case are approximately a factor of two \citep[e.g.][]{Draine2003a}. 

In the cases of Perseus and Taurus we use the same $l,b$ range as \citetalias{Lombardi2010_Perseus_Taurus}, but not the same $d_{max}$ boundaries. Therefore we can skip step 1 given for Orion and only carry out step 2, where we scale up the literature mass up to our $d_{max}$ in order to be able to directly compare our masses.  Once scaled we derive total masses of $117 \times 10^{3} \ \mathrm{M_{\odot}}$ for the Perseus region and a total mass of $110 \times 10^{3} \ \mathrm{M_{\odot}}$ for the Taurus region for \citetalias{Lombardi2010_Perseus_Taurus}. Our total masses are 1.5 times and 1.3 times larger for Perseus and Taurus respectively similar to the variation observed in Orion. The causes of variations discussed above also applies to these two regions as well. A recent publication by \citet{Zucker2021} predicted total masses for Perseus, Taurus, Orion A and B as well as $\lambda$ Ori and found them to be of the order of $10^{4} \ \mathrm{M_{\odot}}$, which is inconsistent with our work as well as the works we compare to. We have not identified a clear reason for this and will be exploring it in a forthcoming paper. 

For the Cygnus X region \citet{Schneider2016_CygXMass} derived a total mass of $7.9 \times 10^{6} \ \mathrm{M_{\odot}}$ using CO 2--1 and 3--2 data which we have scaled up to our $d_{max}$ of 2200~pc. However the $l,b$ range employed by \citet{Schneider2016_CygXMass} is about a degree shifted from ours; thus recalculating our mass to match this $l,b$ range we derive $8.7^{+0.7}_{-0.6} \times 10^{6} \ \mathrm{M_{\odot}}$, making the two masses now directly comparable. The two masses are consistent with each other with our mass being 1.1 times larger compared to \citet{Schneider2016_CygXMass}. Further the mass of Cygnus X is more than 10 times larger than the masses of all three of our other regions, as expected given that Cygnus X is the most massive SFR within 2~kpc while the other regions are low- or intermediate-mass.


\section{Comparison to the Planck legacy products}
\label{sec:Planck}

In addition to its all-sky flux maps, the Planck sub-mm survey has produced maps of various physical quantities, such as dust optical depth, at a resolution of around 10\,arcmin \citep{Planck2016}. We compare these here to our dust maps.

A natural product to compare to extinction is the sub-mm optical depth of dust at 353~GHz ($\tau_{353}$). Like extinction, this measures the column density, but along the entire line-of-sight instead of just to certain stars. Both of these quantities depend only on the dust column density and the dust properties at their respective wavelengths. Extinction is proportional to the dust cross sections at optical wavelengths, and $\tau_{353}$ is proportional to the dust cross sections at sub-mm wavelengths. This means that taking the ratio the two cancels out the column density, leaving only the dust properties.

\begin{figure*}
\centering
\begin{subfigure}{0.49\textwidth}
  \centering
 \includegraphics[width=\textwidth, trim=4cm 4cm 5.5cm 4cm, clip]{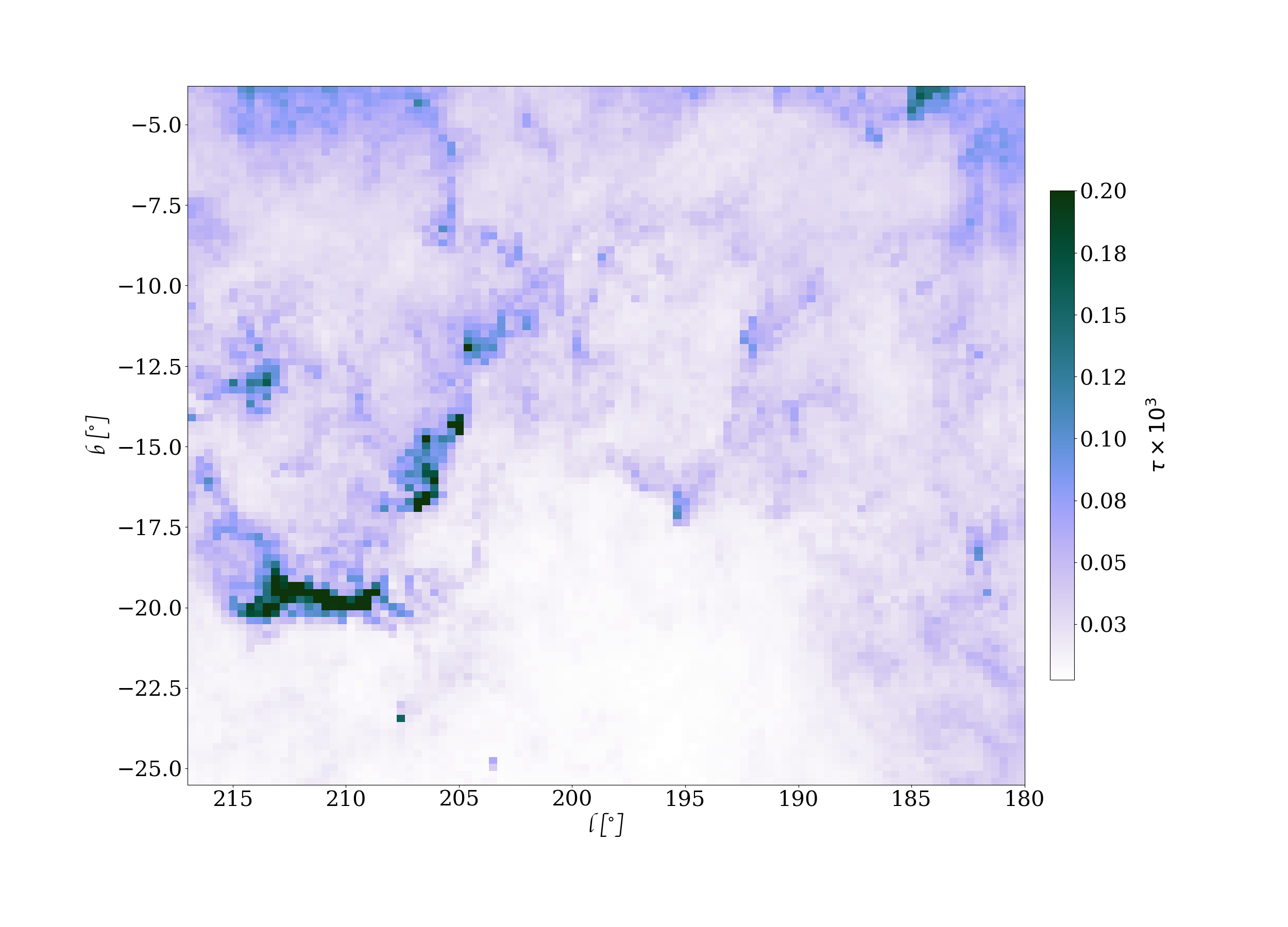}
  \end{subfigure}
\begin{subfigure}{0.49\textwidth}
  \centering
   \includegraphics[width=\textwidth, trim=4cm 4cm 5.5cm 4cm, clip]{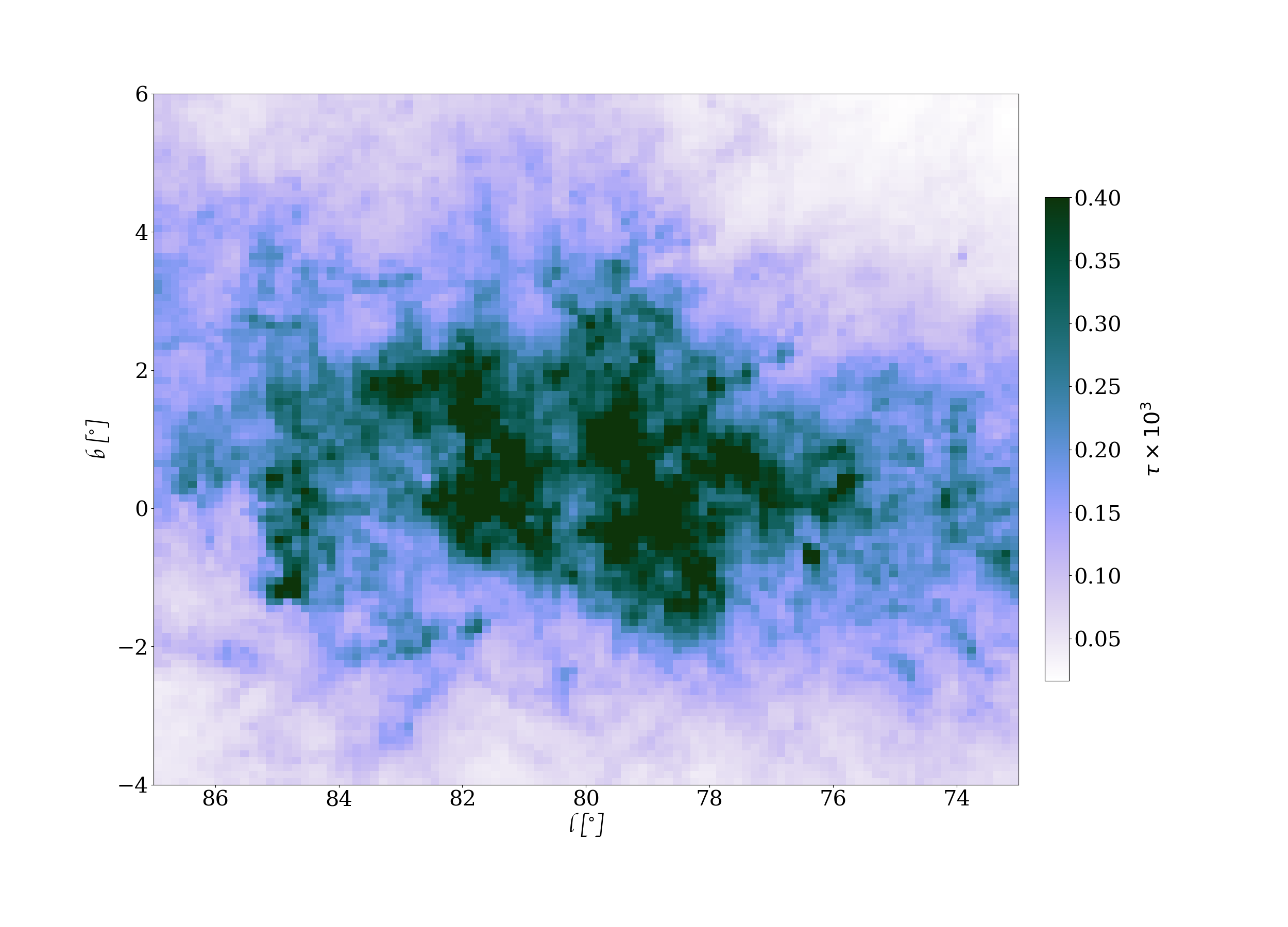}
  \end{subfigure}
 
\begin{subfigure}{0.49\textwidth}
  \centering
 \includegraphics[width=\textwidth, trim=4cm 4cm 5.5cm 4cm, clip]{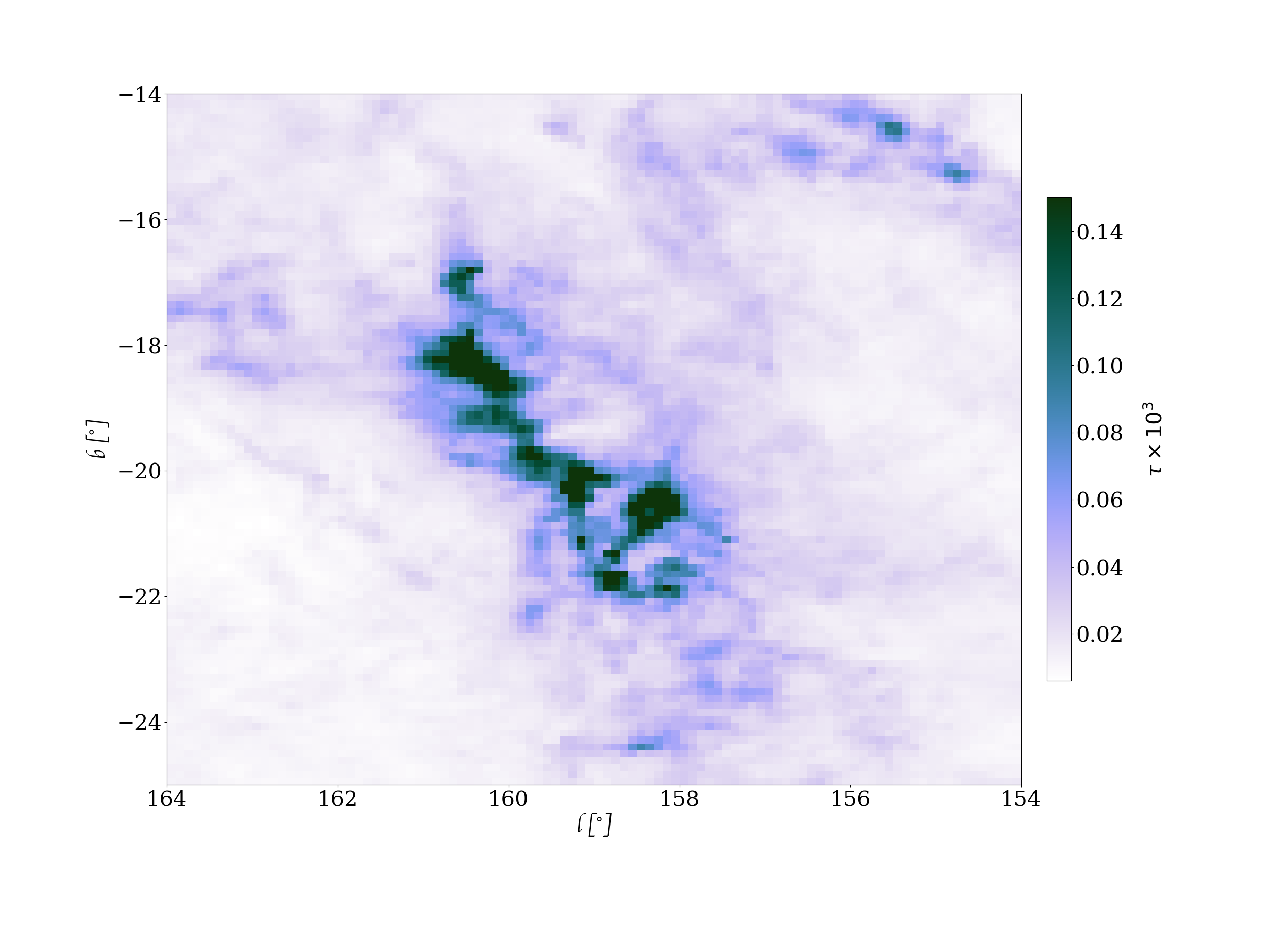}
  \subcaption{}
    \label{fig:PerseusPlanck_Tau}
  \end{subfigure}
\begin{subfigure}{0.49\textwidth}
  \centering
   \includegraphics[width=\textwidth, trim=4cm 4cm 5.5cm 4cm, clip]{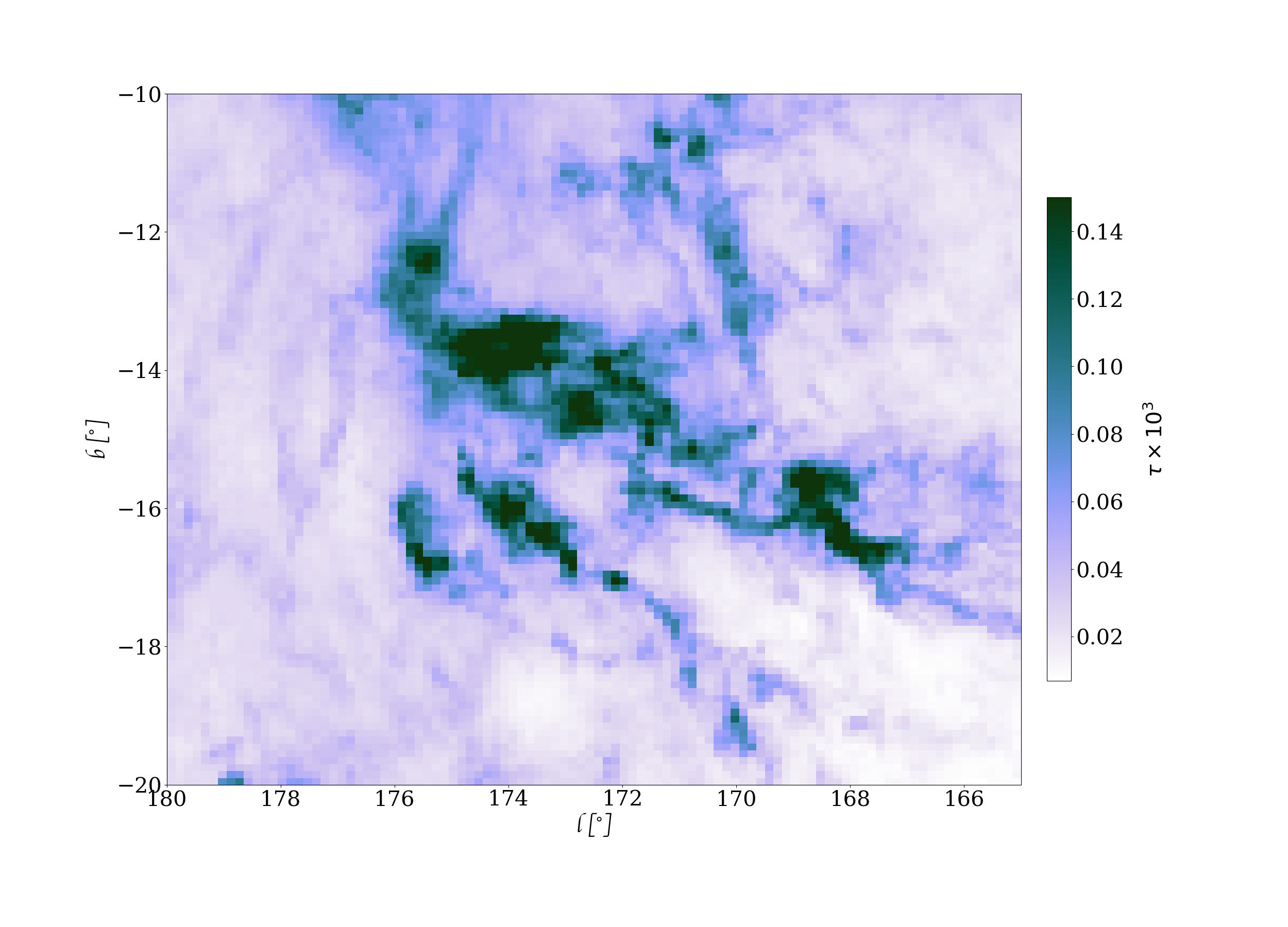}
  \subcaption{}
  \label{fig:TaurusPlanck_Tau}
  \end{subfigure}
  
\begin{subfigure}{0.49\textwidth}
  \centering
 \includegraphics[width=\textwidth, trim=4cm 4cm 5.5cm 4cm, clip]{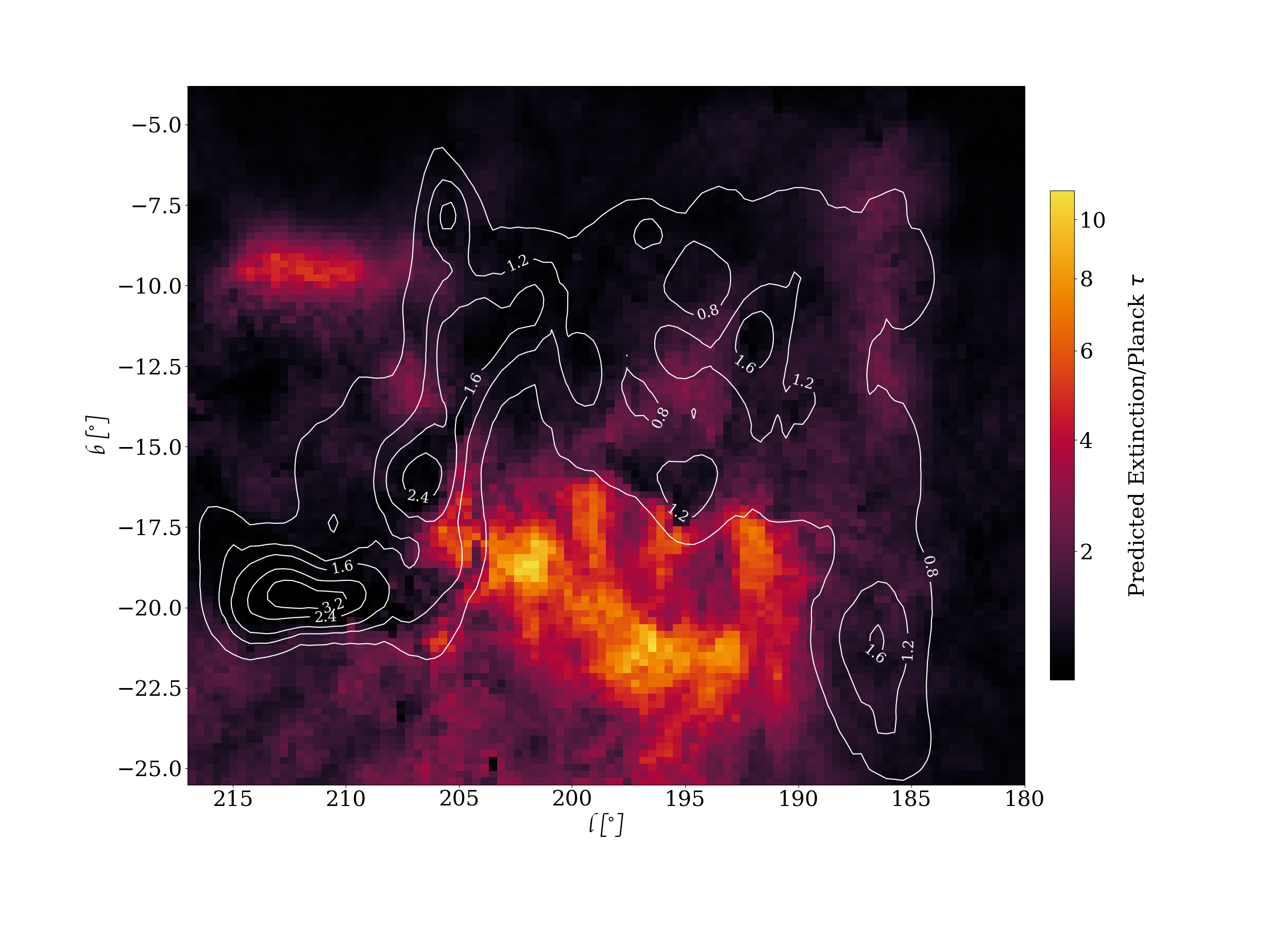}
  \end{subfigure}
\begin{subfigure}{0.49\textwidth}
  \centering
   \includegraphics[width=\textwidth, trim=4cm 4cm 5.5cm 4cm, clip]{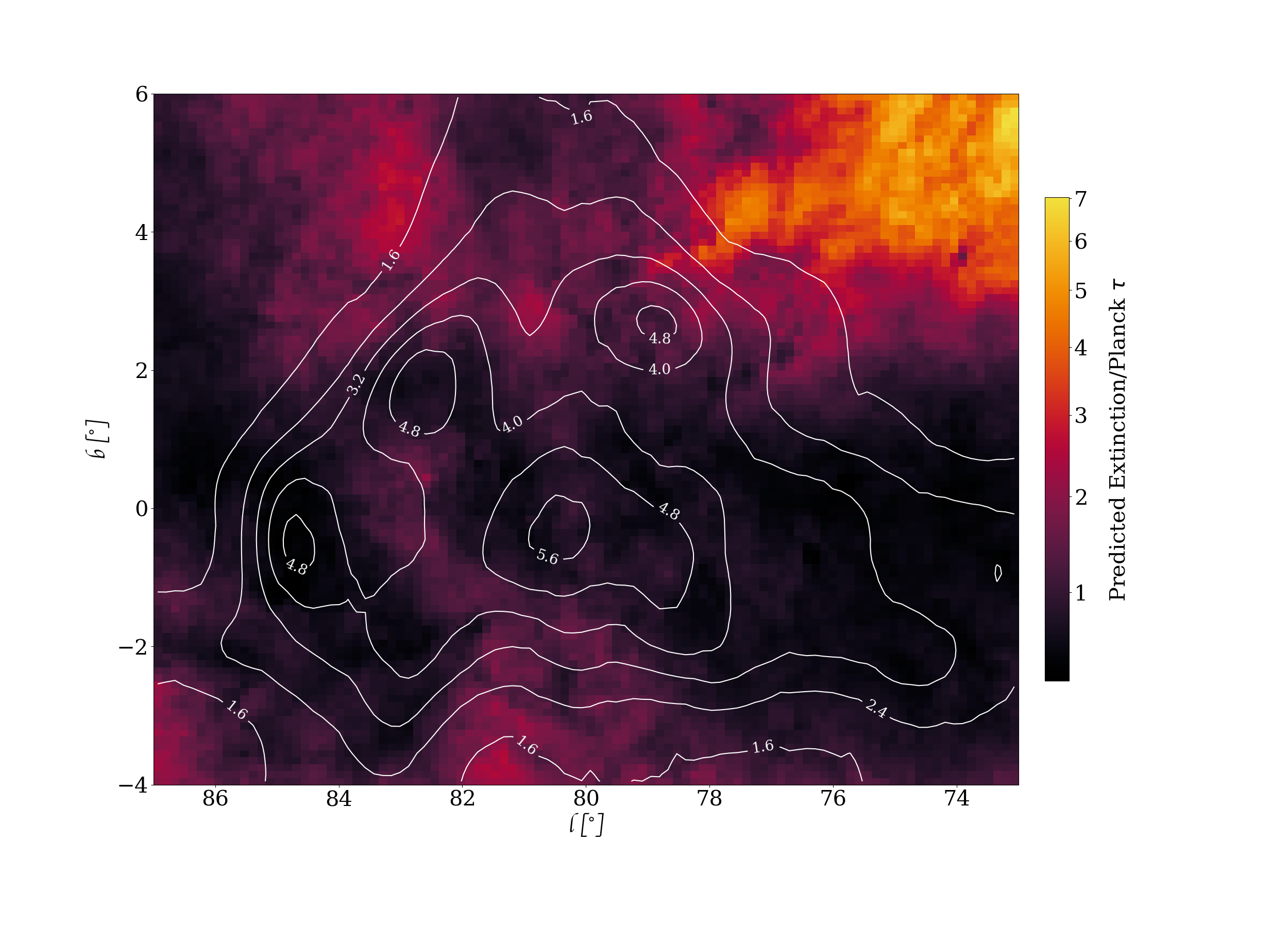}
  \end{subfigure}
 
\begin{subfigure}{0.49\textwidth}
  \centering
 \includegraphics[width=\textwidth, trim=4cm 4cm 5.5cm 4cm, clip]{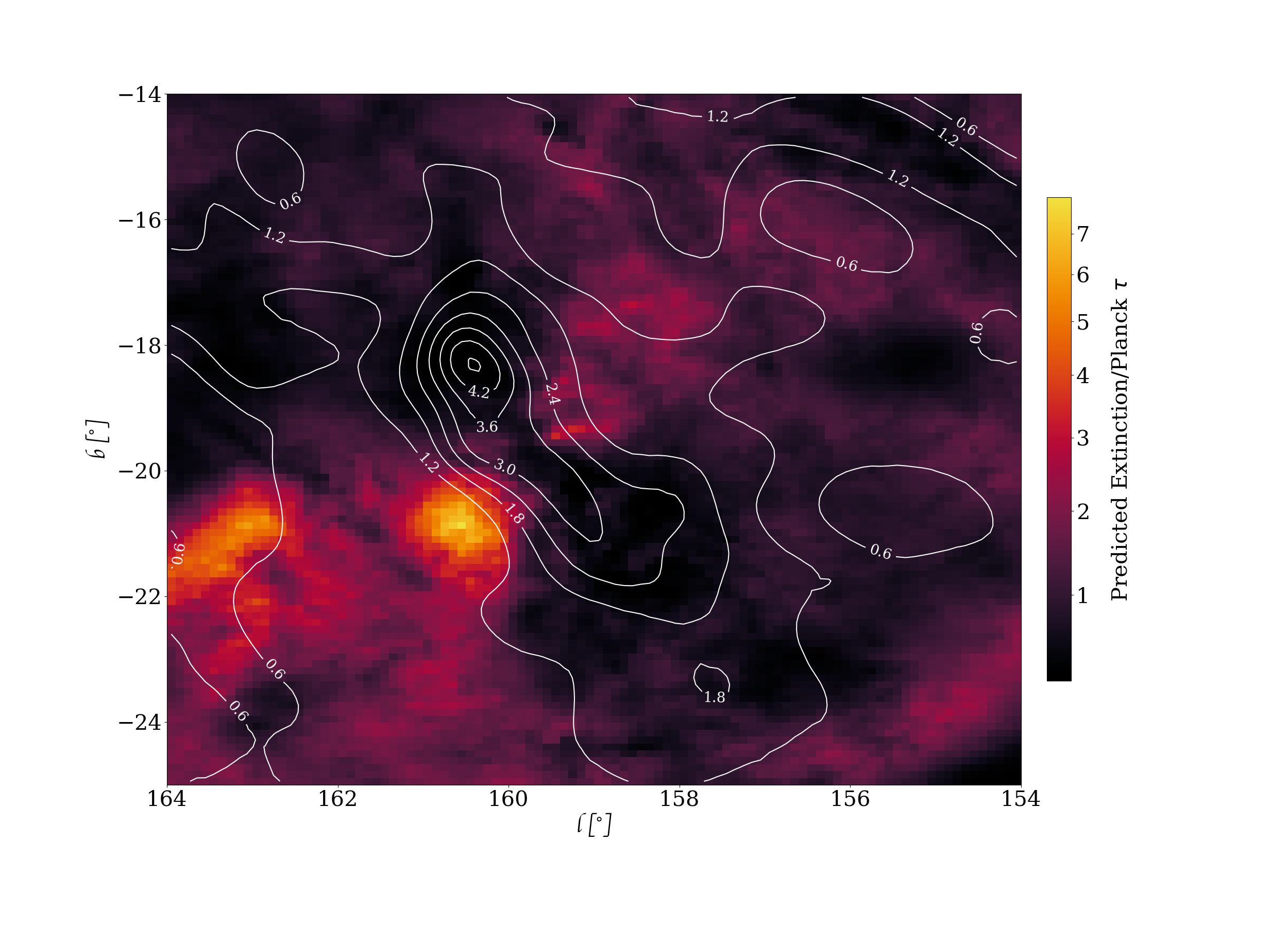}
  \end{subfigure}
\begin{subfigure}{0.49\textwidth}
  \centering
   \includegraphics[width=\textwidth, trim=4cm 4cm 5.5cm 4cm, clip]{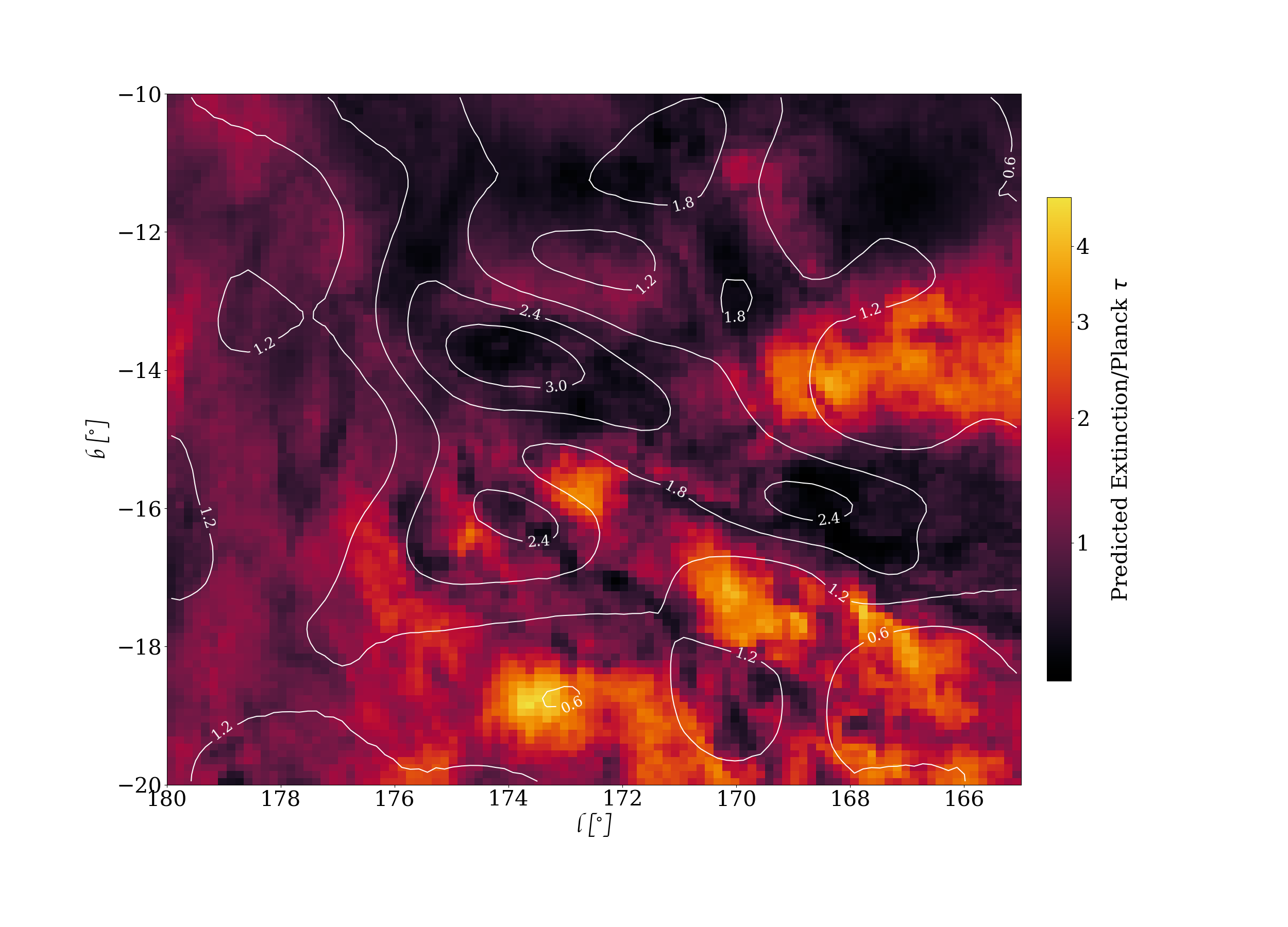}
  \end{subfigure}
 \caption{\emph{Top four panels}: Planck dust optical depth maps for Orion (top left), Cygnus X (top right), Perseus (upper middle left), Taurus (upper middle right); \emph{Bottom four panels}: Ratio of total extinction predicted by our model to Planck optical depth normalised by the median of the ratio. Predicted extinction contours overlaid. Ratio plots are arranged in the same pattern as the Planck emission maps.}
  \label{fig:Planck_Tau_All}
\end{figure*}

We show the Planck $\tau_{353}$ maps of our star-formation regions in Fig.~\ref{fig:Planck_Tau_All}. It also includes ratio maps of our predicted extinction to Planck $\tau_{353}$. This ratio is proportional to the ratio of the optical to the sub-mm dust cross-section. A similar comparison of the predicted extinction to Planck sub-mm flux and predicted dust emission is shown in figs~\ref{fig:Planck_850micronflux_All}, \ref{fig:Planck_DustFlux_All}. In Fig.~\ref{fig:Planck_Tau_All} we can identify
in the Planck optical depth map
the same features we see in our extinction map. 
The global structure of the clouds is reproduced well, but the finer details e.g.\ of filaments in Taurus TMC 2 are not; our model is not able to resolve such small structures on account of the length scale we have adopted, but instead encloses several features that are distinct in the Planck data as a single overdensity. Similar behaviour is seen in the smallest substructure in Perseus as well. 

The ratio of extinction to $\tau_{353}$ changes significantly between high and low extinction lines-of-sight, as seen in figure \ref{fig:Planck_Tau_All}. Lines-of-sight that intersect with the molecular clouds (dense regions) all have rather low ratios of extinction to sub-mm (dust) emission, while the values are much higher, by up to a factor of 10, in the lines-of-sight with low density (diffuse) regions. The most likely physical explanation for the variation in the $\frac{A_0}{\tau_{353}}$ ratio between dense and diffuse regions is a change in the ratio of the cross-sections between the visual and the sub-mm between these lines-of-sight, for example as a result of changes in grain size or composition. As grains grow, the ratio of the optical to sub-mm cross sections tends to shrink, because the optical properties change from Rayleigh to Mie scattering and eventually to the geometric-optics regime, yet the sub-mm opacities stay in the Rayleigh regime. This relationship between opacity and extinction (and hence changes in grain properties) has been described in the near-infrared by, for example, \citet{Lombardi2014} in an unresolved sense when looking only within the cores of molecular clouds. We can extend this relationship to the wider environments of our star formation regions and their surroundings. If the dust properties are forced to stay the same, the observed trends could also be explained if there is additional material at larger distances not traced in our extinction maps, but that is detected by Planck in emission; this would, however, require an unlikely preference for additional dust clouds only along lines-of-sight that already have dust clouds near to them.  

Following the argument above, the variation in the ratio of optical extinction to sub-mm emission allows us to map changes in grain properties in detail on large scales in the ISM and molecular clouds. This probes the dust processing (probably grain growth) that occurs as density increases. As the grains get larger, the opacities change, and the range of variation tells us something about how much they change. However as the plots are normalised they probe only the changes in size and not the absolute sizes. 

Models of star formation show that grains start off in the diffuse ISM (smaller grains and standard composition, higher ratios) and as the cloud collapses the density increases, allowing them to grow larger or change in composition (lower ratios) \citep[e.g.,][]{Guillet2007_GrainGrowth, Hirashita2011_GrainGrowth}. What we see in the plot is that the grain size changes relatively little in the outer layers of the clouds (but still enough to differentiate them from the diffuse material around it) and as you go deeper into the cloud the processing accelerates at higher densities, so that the ratio changes by a much larger factor moving from the outer layers of the cloud into their cores than it does when moving from the diffuse medium to the outer parts of the cloud.

In effect, we are mapping the trend between $A\left(\lambda\right)/\tau_{353}$ and $R_{\rm V}$ identified by \citet{Zelko2020_DustProp}, showing that this trend can be used to trace where and when grain processing occurs in detail. The trend persists on much smaller scales than those identified by \citet{Schlafly2016_DustProp}, more similar to the assumption of star-forming regions as the typical scale.

\section{Conclusions}


We have introduced an algorithm that reconstructs the 3D dust density distribution of our Galaxy from the extinction to stars sources along multiple lines-of-sight. Based on latent variable GPs and variational inference techniques, our algorithm enforces positive density and thus monotonically non-decreasing extinction along any given line-of-sight. Our approach scales favourably in computation time and memory up to large samples and distances. It is built on publicly-available software for GPs, machine learning, and Bayesian inference, thereby ensuring portability and reproducibility while maximising stability of the code.

We applied our algorithm to four well known star forming regions in the Milky Way: Orion, Cygnus X, Perseus, and Taurus. Our predicted 3D dust densities improve on previous work by recovering features in 3D density, 
for example filaments in Orion and condensations in Taurus, that are otherwise lost in 2D extinction maps. Our maps shed light on debates over the structure of Cygnus X, suggesting that the main star-forming regions are at 1300--1500~pc and that there is a collection of more distant clumps which likely contributes to its on sky projection. We find that the filament seen in extinction between Taurus and Perseus is a unified structure that lies upon a line joining the two regions and may connect the two clouds. We suggest that several clusters in the Orion region outside the main star-forming regions may be components of a large filament at around 300~pc from us; distance measurements to these clusters are required to confirm this.

Based on the predicted extinction maps, we estimate masses for the clouds. Our estimates are comparable to those in the literature, but higher by a factor of 1.1 -- 1.5, with a typical uncertainty of $10\%$ but $35\%$ in the case of Orion. This may be a result of the inclusion of WISE photometry in the determination of the input extinctions, improving the recovery of the highest column densities.

Comparing our results with the Planck maps shows that lines-of-sight that intersect dense clouds are systematically different from those that do not, most likely as a result of differences in grain size or composition; dense regions have much smaller ratios between optical and sub-mm cross-sections, as expected if the average grain size is larger or the composition is different in dense regions. This shows that the previously observed trend toward a lower ratio of optical extinction to sub-mm optical depth when passing from diffuse regions to more dense regions also holds in a resolved sense.

The flexibility of our method means that it can easily be extended. For example, future work could use multiple kernels to account for different structures in the interstellar medium, enabling 
structures of variable sizes to be recovered more effectively, something that is particularly relevant when mapping a large fraction of the Galaxy in one go. Alternatively, the model could take into account the wavelength dependence of extinction (R$_\mathrm{0}$) to map changes in the grain-size distribution in three dimensions.

\begin{acknowledgements}

This project is partially funded by the Sonderforschungsbereich SFB 881 “The Milky Way System” of the German Research Foundation (DFG).

\end{acknowledgements}

\bibliographystyle{aa} 
\bibliography{Bib_MWExt}



\begin{appendix}

\section{The approximate posterior}
\label{sec:Appendix:ApproxPosterior}

We use two approximations to compute our posterior. The first is that the posterior distribution over $\log_{10}(\rho)=\phi$ can be approximated as a Gaussian. The second is to compute the covariances of the GP prior only using a subset of points at which the prior is defined.

The first approximation arises as follows. We adopt a Gaussian likelihood for the observed extinctions, $A_\mathrm{obs}$, which have uncertainties $\sigma_\mathrm{obs}$, i.e.\
\begin{equation}
    \ln P(A_\mathrm{obs}\,\vert\,A_\mathrm{mod}) = -\frac{1}{2}\sum\left(\frac{A_\mathrm{obs} - A_\mathrm{mod}}{\sigma_\mathrm{obs}}\right)^2 + {\rm const.}
\end{equation}
where $A_\mathrm{mod}$ are the corresponding extinctions predicted by the model, and the sum is taken over the measured extinctions for different lines-of-sight.

Our goal is to infer the densities, $\rho(s)$, at various distances $s$ along the different lines-of-sight. For one line-of-sight, the integral of these dust densities gives the extinction to a star at distance $d$,
\begin{equation}
    A_\mathrm{mod} = \int_{0}^{d} \rho_\mathrm{mod}(s)\,\mathrm{d}s .
\end{equation}
Thus the likelihood can be written
\begin{equation}
    \ln P(A_\mathrm{obs}\,\vert\,\rho_\mathrm{mod}) = -\frac{1}{2}\sum\left(\frac{A_\mathrm{obs} - \int_{0}^{d} \rho_\mathrm{mod}\,\mathrm{d}s}{\sigma_\mathrm{obs}}\right)^2  + {\rm const.} 
\end{equation}
In principle we could use this to solve for  $\rho_\mathrm{mod}$, for which we require some regularisation of these model densities because they are otherwise degenerate. Specifically, if we adopted a Gaussian prior on $\rho_\mathrm{mod}$ we have a Gaussian posterior in $\rho_\mathrm{mod}$ and so a straight forward closed form solution (see sections 2.2 and 2.3 of~\citep{Rezaei2017}).

However, we also want to impose non-negativity on $\rho$, and so we choose to infer $\phi = \log_{10}\rho$ instead of $\rho$. The likelihood is
\begin{equation}
    \ln P(A_\mathrm{obs}\,\vert\,\rho_\mathrm{mod}) =  -\frac{1}{2}\sum\left(\frac{A_\mathrm{obs} - \int_{0}^{d} 10^{\phi}\,\mathrm{d}s_i}{\sigma_\mathrm{obs}}\right)^2  + {\rm const.}
\end{equation}
which makes it explicit that the posterior will not be Gaussian in $\phi$ when we have a Gaussian prior in $\phi$.We therefore have two options for computing the posterior: we could solve for the distribution of $\phi$ at all points in space by sampling the full posterior (e.g.\ using MCMC), or we could approximate the posterior with something more tractable. We choose the second option, and approximate the posterior as a Gaussian using variational inference. That is, when we compute the likelihood from equation A.4 and combine it with the prior, we assume that the resulting posterior has a Gaussian distribution. This is our first approximation.

Ideally we would calculate the covariances of the GP using all the grid points, but for computational time reasons we do this instead only at a subset of points, known as the inducing points. Using an iterative procedure we then find the set of inducing points, as well as the mean and covariance of the distribution at these points that best approximates the exact posterior over all the data. This is our second approximation. This procedure is implicit to \emph{Gpytorch}. 

Once these two approximations have been made, the Gaussian approximation of the posterior can be written 
\begin{equation}
    P(\phi\,\vert\,\vec{x}, \vec{x}_i, \phi_i) = \mathcal{N}(\phi\,\vert\,\vec{\mu}_\phi, \Sigma_\phi),
\end{equation}
where $\phi_i$ is the set of log densities $\log_{10} \rho$ at the inducing points whose means and covariances are those of the variational distribution (see Sect.~\ref{sec:VI}), $\vec{x}_i$ is the set of positions of the inducing points, and $\vec{x}$ is the set of other positions at which the posterior is computed (e.g. the grid points). The mean of the posterior is 
\begin{equation}
    \vec{\mu}_\phi = K^{T}_x\ K^{-1}_{x_i} \phi_i
\end{equation}
where $K_{x_i} = K(x_i , x_i)$ (as defined in eqn.~\ref{eqn:rbfKernel}) is an $M \times M$ matrix where $M$ is the number of inducing points and $K_{x} = K(x_i , x)$ is an $M \times N$ matrix where $N = n_l \times n_b \times n_d$ is the total number of points in the grid. The covariance of the posterior is
\begin{equation}
    \Sigma_\phi = K_{xx} - K^{T}_x\ K^{-1}_{x_i}\ K_x,
\end{equation}
where $K_{xx} = K(x, x)$ is an $N \times N$ matrix. 

\section{Additional figures showing the behaviour of the model with simulated data introduced in Sec~\ref{sec:SimData}} 
\label{sec:Appendix:SimAlongLBDFigs}

In this appendix we present the additional figures which demonstrate the behaviour of our model using the simulated data and their model predictions introduced in Sec.~\ref{sec:SimData}. First we show the simulated vs. model predicted extinctions and densities as a function of distance for a selected set of lines-of-sight. This is followed by simulated and model predicted densities as a function of $l$ and $b$ for the same set of distance points as Fig~\ref{fig:SimDenseAll}. 

\subsection{Extinction and density along selected lines-of-sight for the simulated and model predicted data introduced in Sec~\ref{sec:SimData}}

In this subsection we present the extinctions and densities along a set of selected lines-of-sight for our simulated data in comparison to our model predicted data of the simulations. From these figures we infer a typical uncertainty of $\lesssim 20\%$ in density. 

\begin{figure*}
\centering
\begin{subfigure}{0.48\textwidth}
  \centering
  \includegraphics[width=\textwidth]{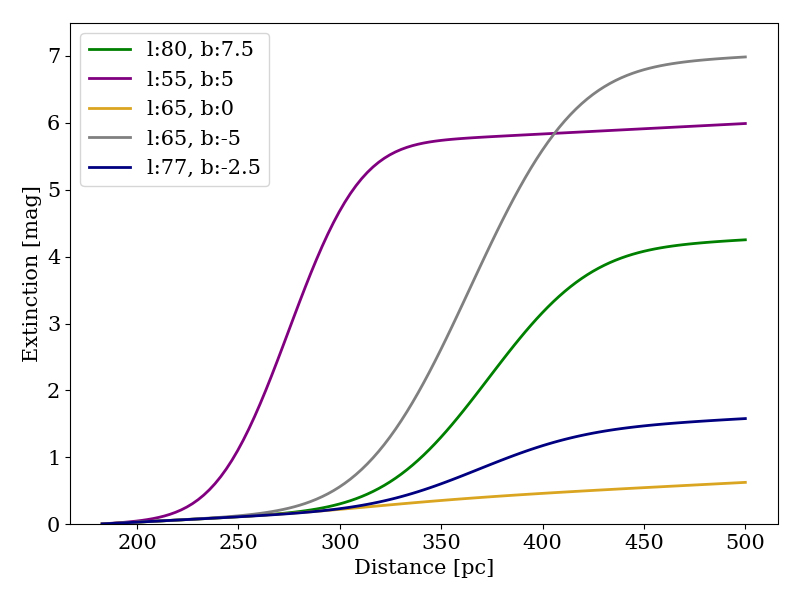}
  \end{subfigure}
\begin{subfigure}{0.48\textwidth}
  \centering
  \includegraphics[width=\textwidth]{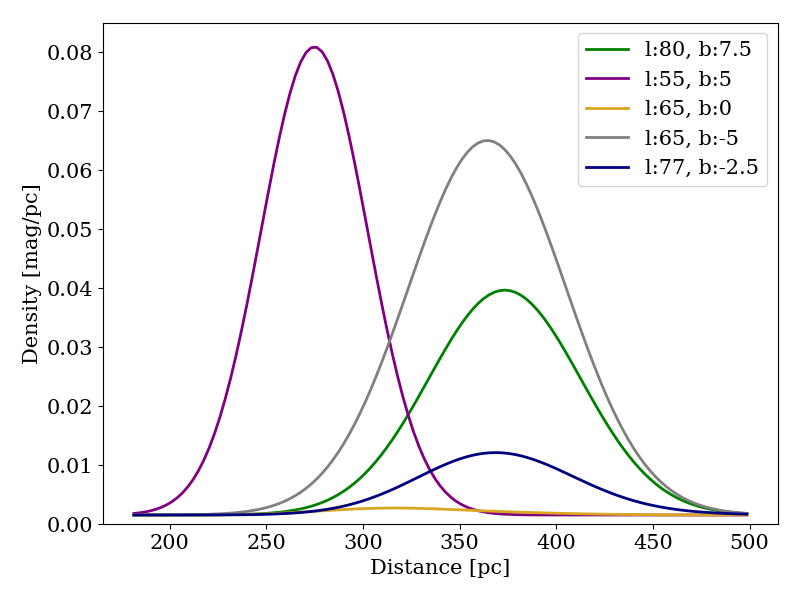}
  \end{subfigure}
  
 \begin{subfigure}{0.48\textwidth}
  \centering
  \includegraphics[width=\textwidth]{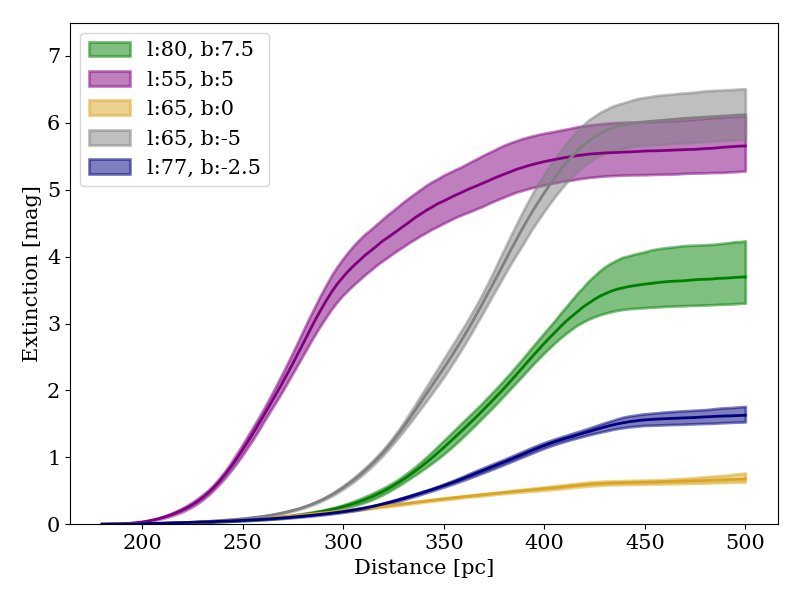}
  \end{subfigure}
\begin{subfigure}{0.48\textwidth}
  \centering
  \includegraphics[width=\textwidth]{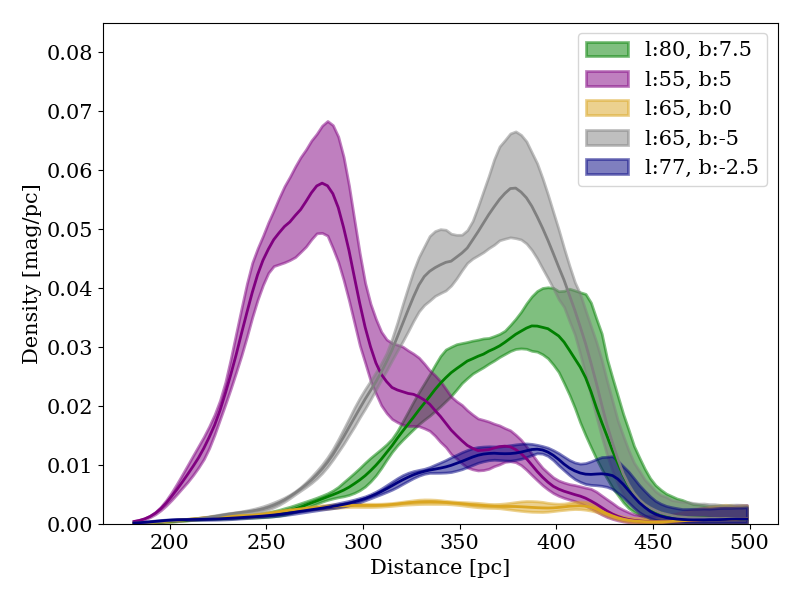}
  \end{subfigure}
  
 \begin{subfigure}{0.48\textwidth}
  \centering
  \includegraphics[width=\textwidth]{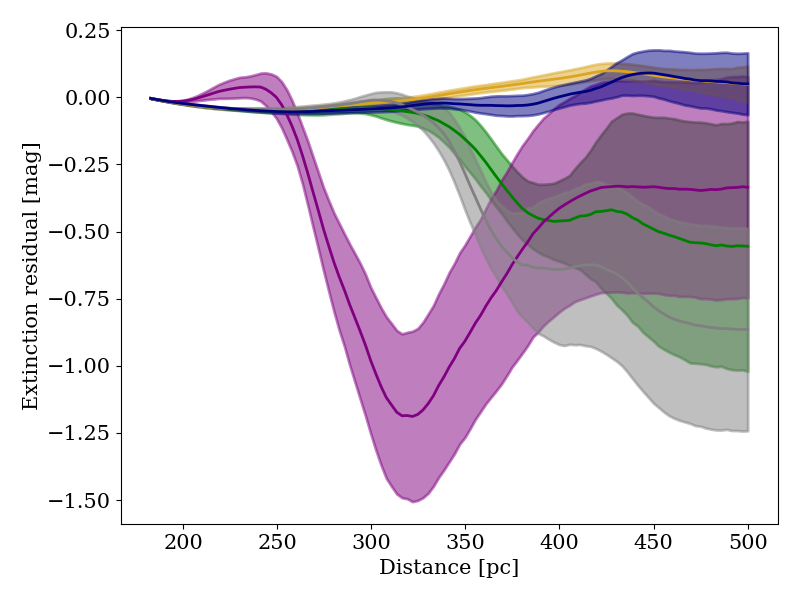}
  \end{subfigure}
\begin{subfigure}{0.48\textwidth}
  \centering
  \includegraphics[width=\textwidth]{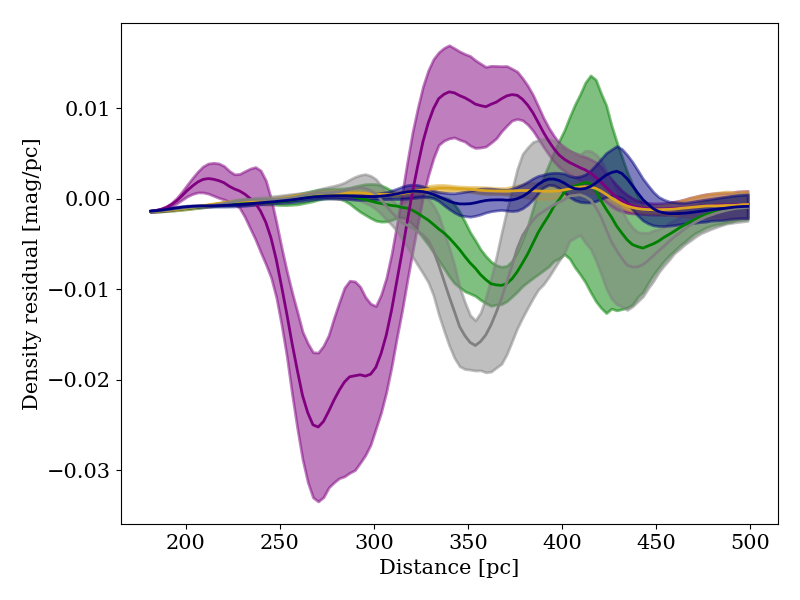}
  \end{subfigure}
  \caption{\emph{Top}: Density and extinction sampled along selected lines-of-sight for the same simulated data presented in Sec~\ref{sec:SimData}; \emph{Middle}: Model reconstructed densities and extinctions sampled along the same lines-of-sight; \emph{Bottom}:Residual of model predicted density - simulated density.}
    \label{fig:SimDense_Dist}
\end{figure*}

\subsection{Simulated and model predicted densities along the l and b axes for the same distance points as Fig~\ref{fig:SimDenseAll} in Sec~\ref{sec:SimData}}

Similar to the previous subsection here we present the simulated and model predicted densities, however this time along l and b axes for specific distances. The combination of these two subsections allow us to infer a typical uncertainty of $\lesssim 20\%$ in  predicted density. 

\begin{figure*}
\centering
\begin{subfigure}{\textwidth}
  \centering
  \includegraphics[width=\textwidth]{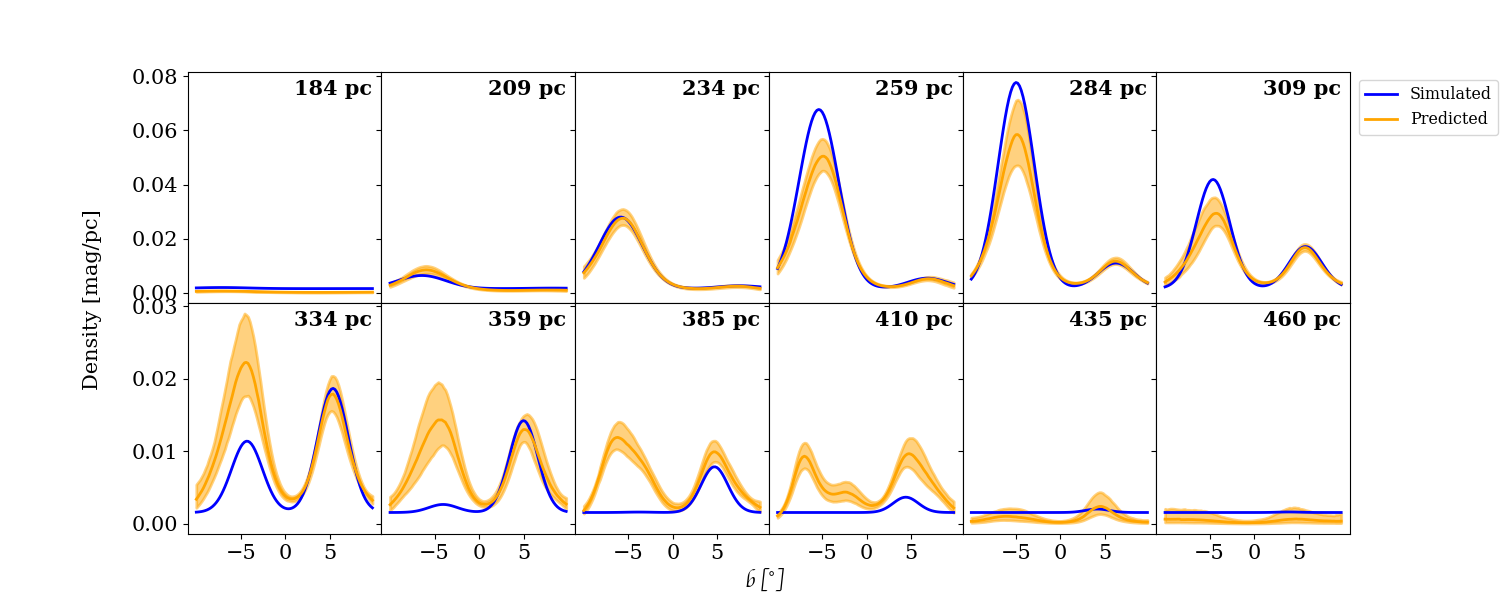}
  \end{subfigure}
\begin{subfigure}{\textwidth}
  \centering
  \includegraphics[width=\textwidth]{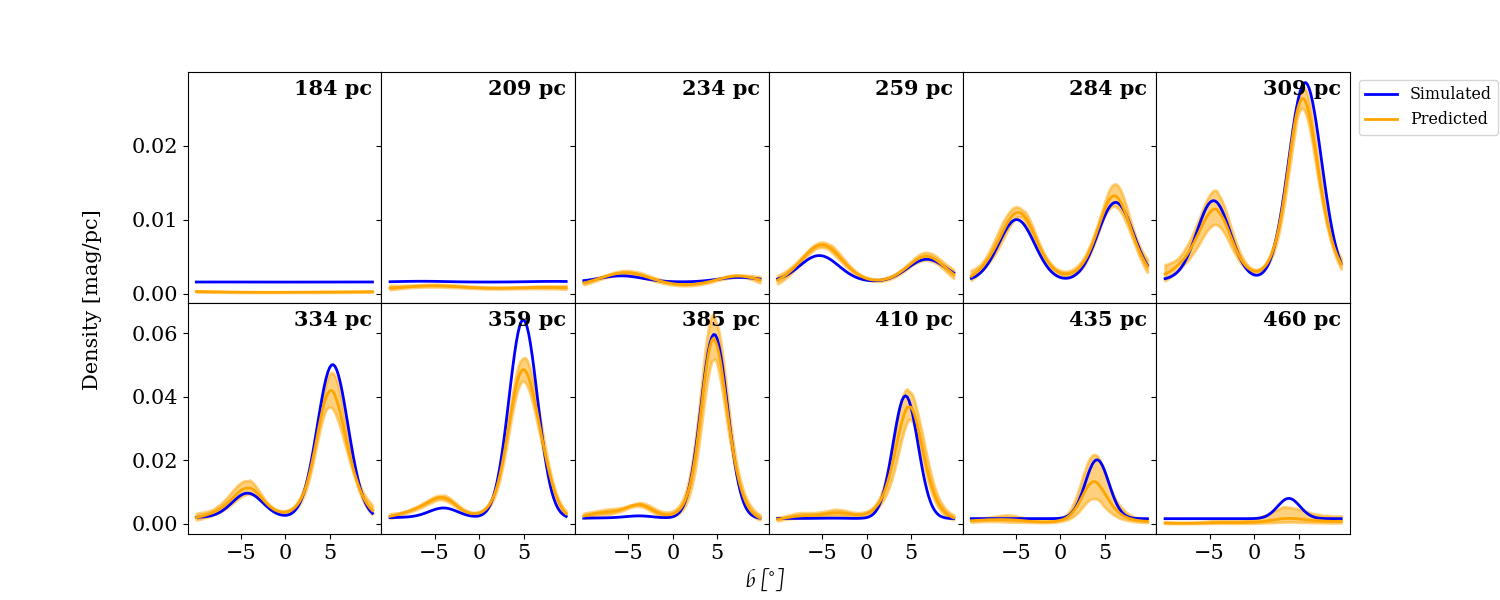}
  \end{subfigure}
   \begin{subfigure}{\textwidth}
  \centering
  \includegraphics[width=\textwidth]{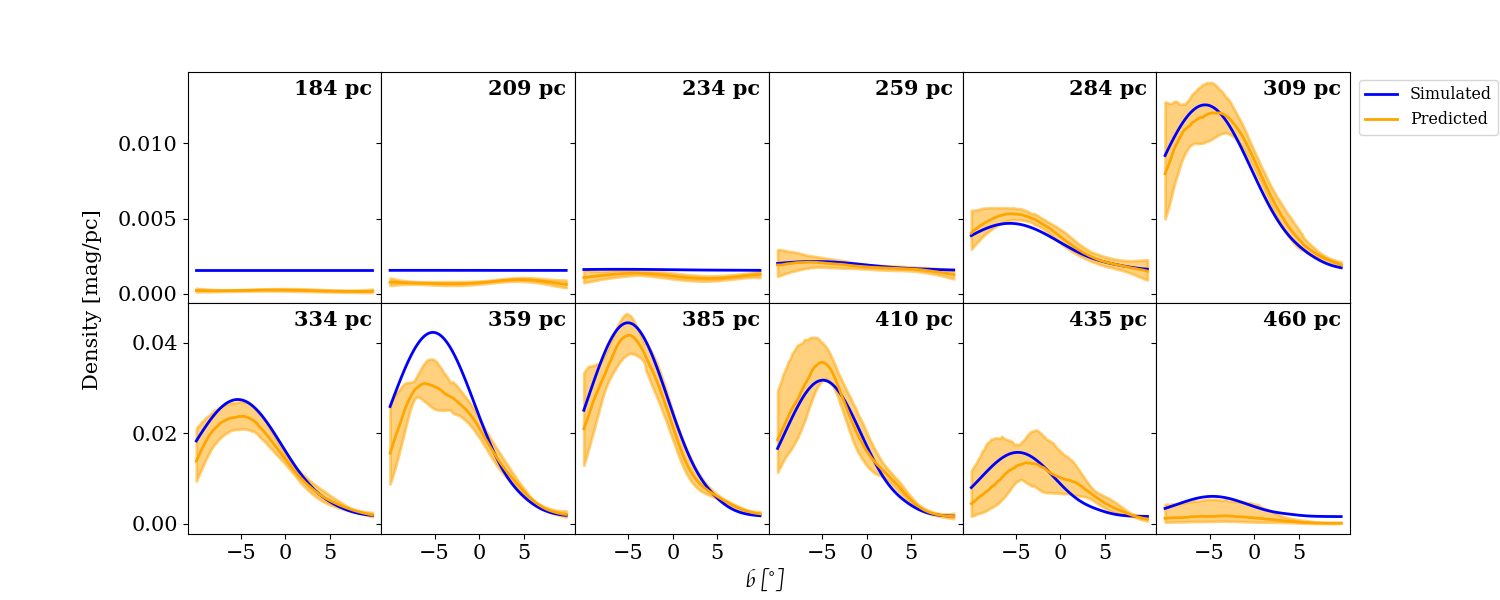}
  \end{subfigure}
  \caption{Profiles of simulated and model predicted densities along $b$ for the same distance points as Fig~\ref{fig:SimDenseAll} in Sec~\ref{sec:SimData}. \emph{Top}: $l = 55^{\circ}$; \emph{Middle}: $l = 65^{\circ}$; \emph{Bottom}: $l = 80^{\circ}$.}
    \label{fig:SimDense_l}
\end{figure*}

\begin{figure*}
\centering
\begin{subfigure}{\textwidth}
  \centering
  \includegraphics[width=\textwidth]{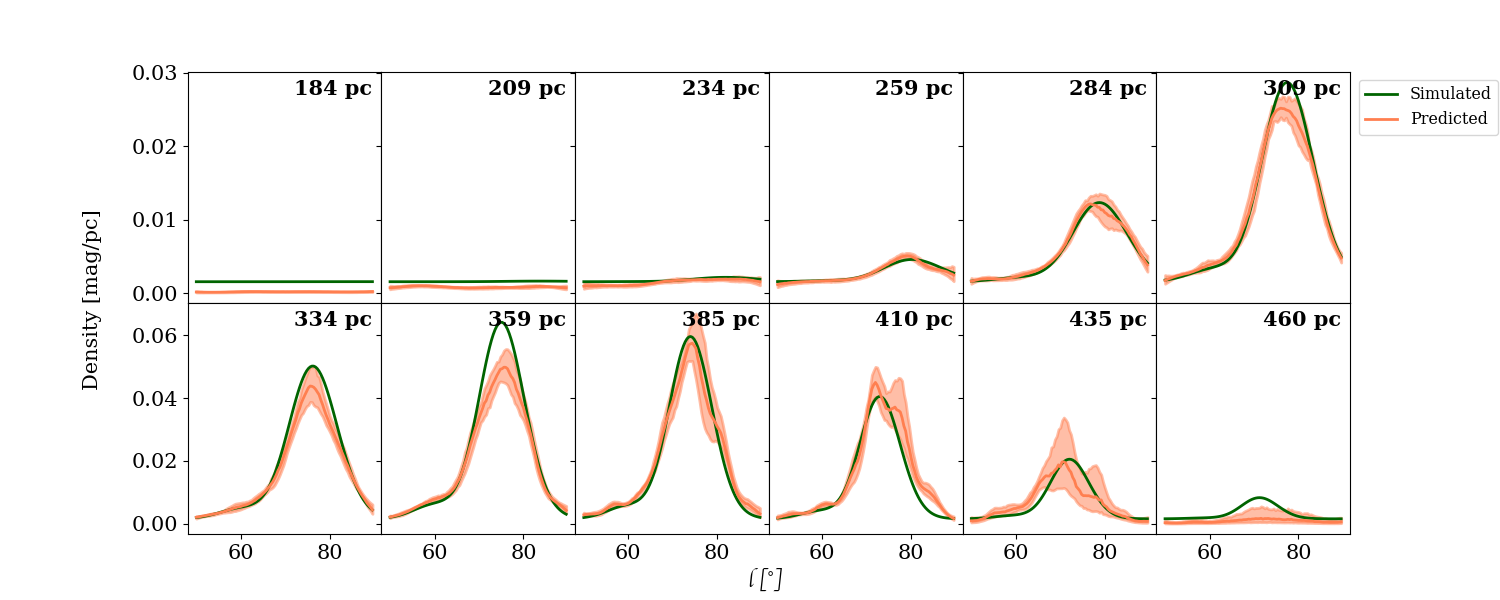}
  \end{subfigure}
\begin{subfigure}{\textwidth}
  \centering
  \includegraphics[width=\textwidth]{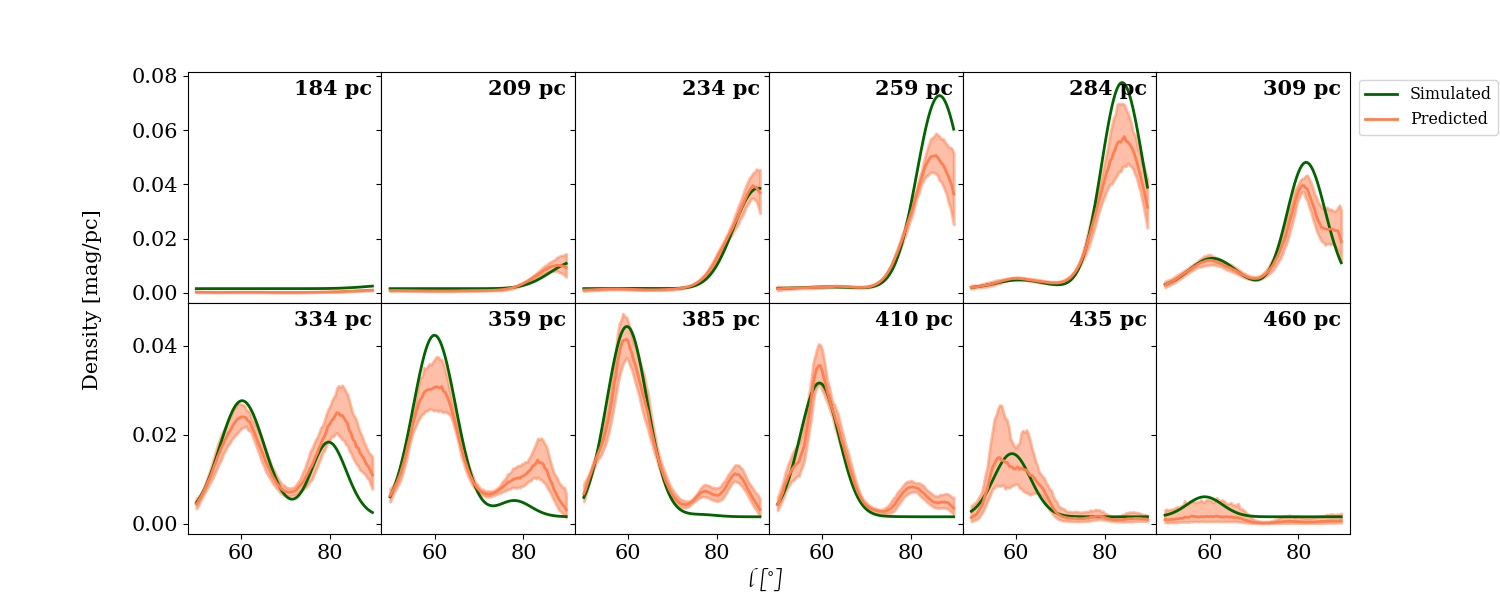}
  \end{subfigure}
  \caption{Profiles of simulated and model predicted densities along $l$ for the same distance points as Fig~\ref{fig:SimDenseAll} in Sec~\ref{sec:SimData}. \emph{Top}: $b = -5^{\circ}$; \emph{Bottom}: $b = 5^{\circ}$.}
    \label{fig:SimDense_b}
\end{figure*}

\section{Three-dimensional extinction distributions predicted by our algorithm}

Here we present the integrated extinction to the given distance points predicted by our model for the SFRs Orion (Sec.~\ref{sec:Orion}), Cygnus X, Perseus and Taurus (Sec.~\ref{sec:Cyx_Per_Taurus}). 

\begin{figure*}
    \centering
    \includegraphics[width=\textwidth, trim=1cm 3.5cm 1cm 7cm, clip]{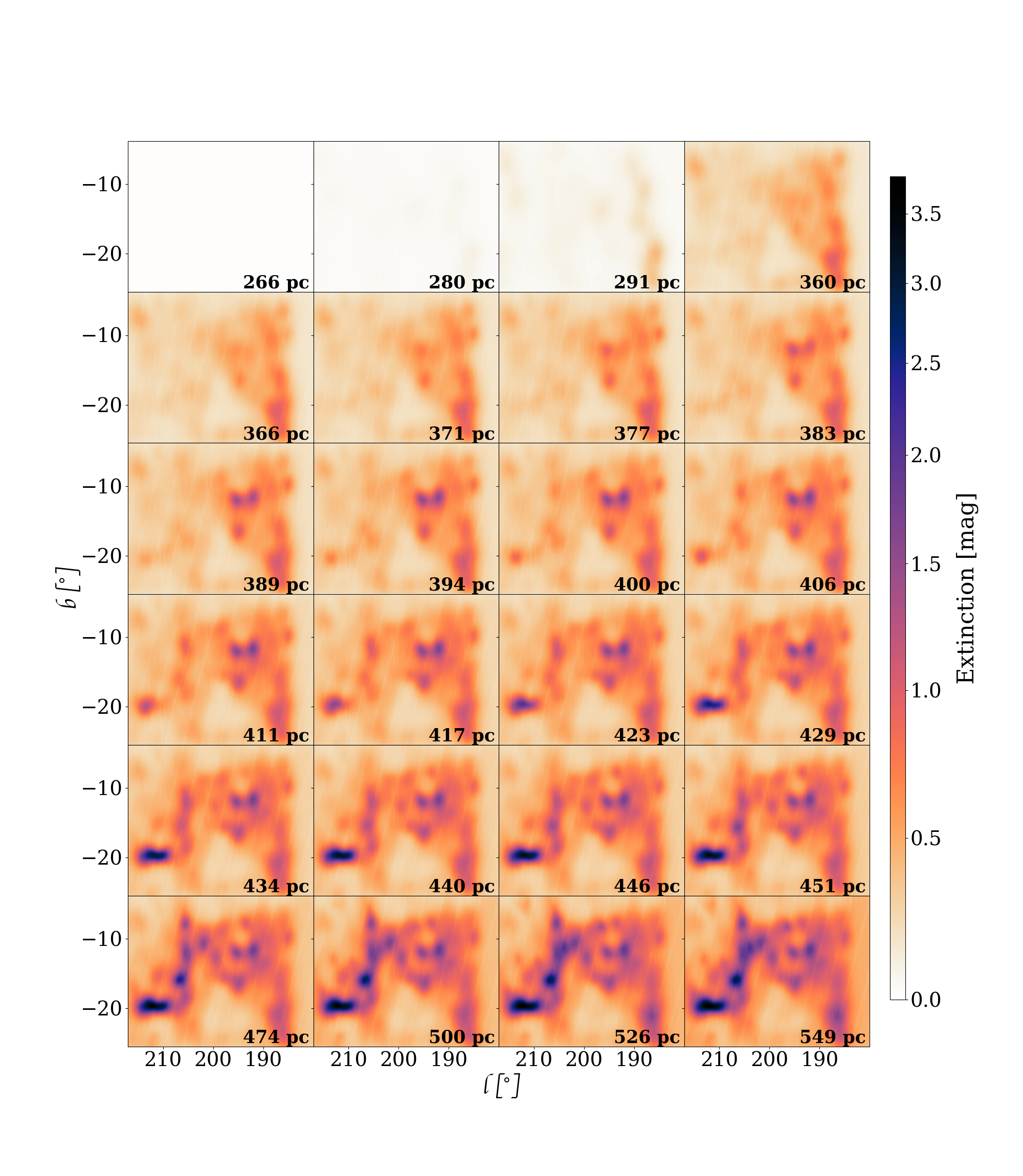}
    \caption{Three-dimensional extinction distribution of Orion up to the indicated distances.}
    \label{fig:OrionExtSlices}
\end{figure*}

\begin{figure*}
    \centering
    \includegraphics[width=\textwidth, trim=1cm 3.5cm 1cm 7cm, clip]{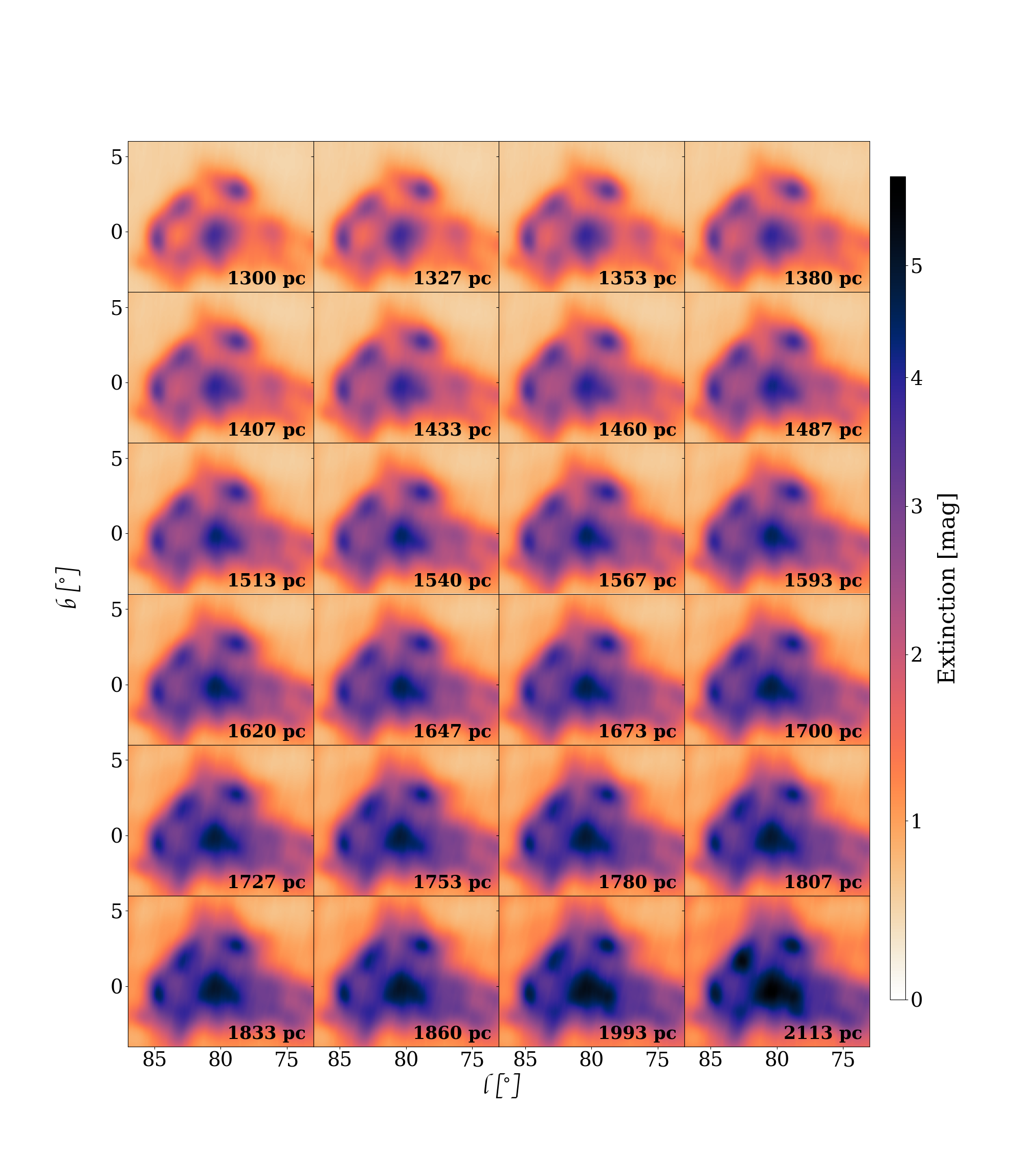}
    \caption{Three-dimensional extinction distribution of Cygnus X up to the indicated distances.}
    \label{fig:CygXExtSlices}
\end{figure*}

\begin{figure*}
    \centering
    \includegraphics[width=\textwidth, trim=1cm 3.5cm 1cm 7cm, clip]{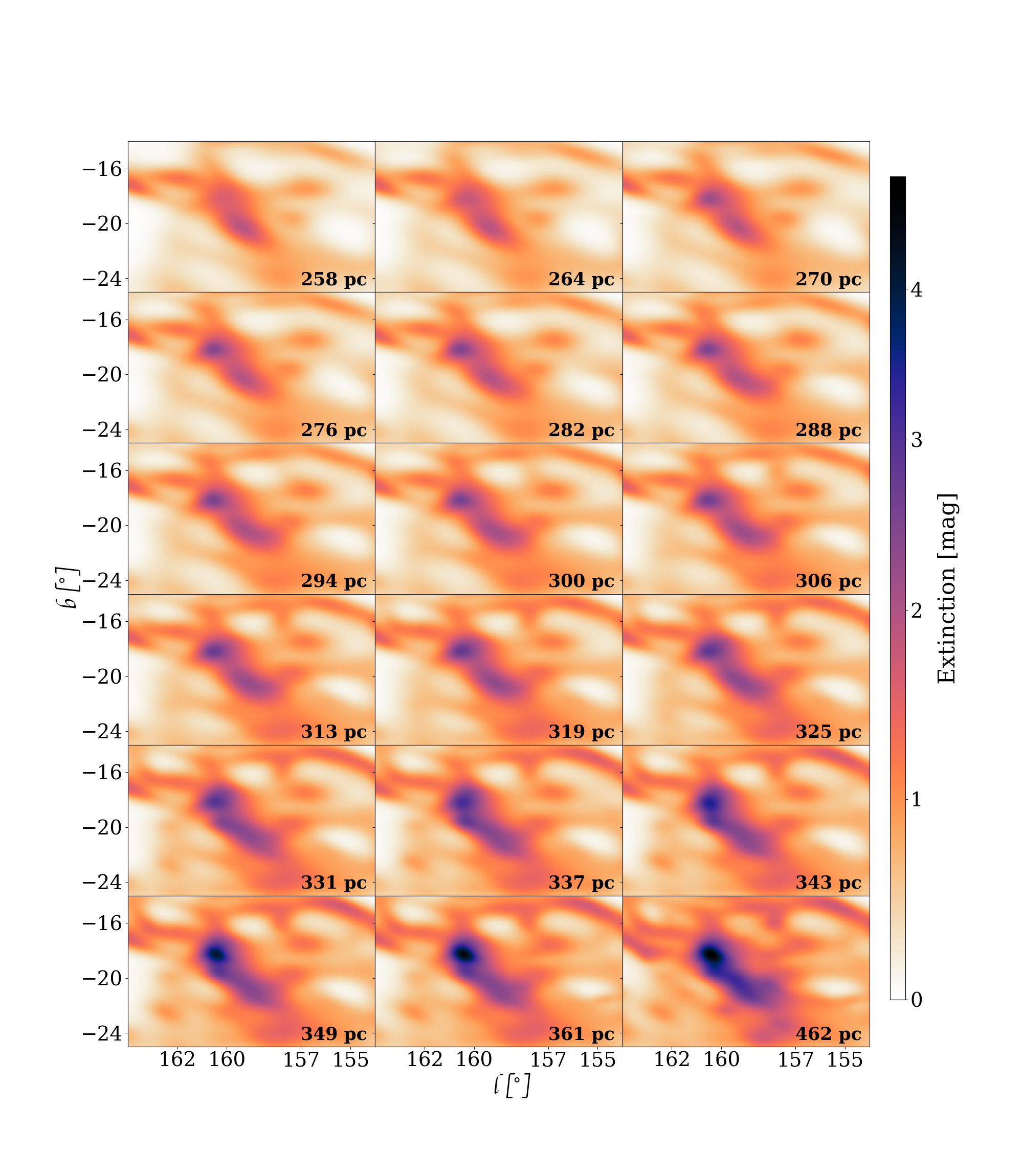}
   \caption{Three-dimensional extinction distribution of Perseus up to the indicated distances.}
    \label{fig:PerseusExtSlices}
\end{figure*}

\begin{figure*}
    \centering
    \includegraphics[width=\textwidth, trim=1cm 3.5cm 1cm 7cm, clip]{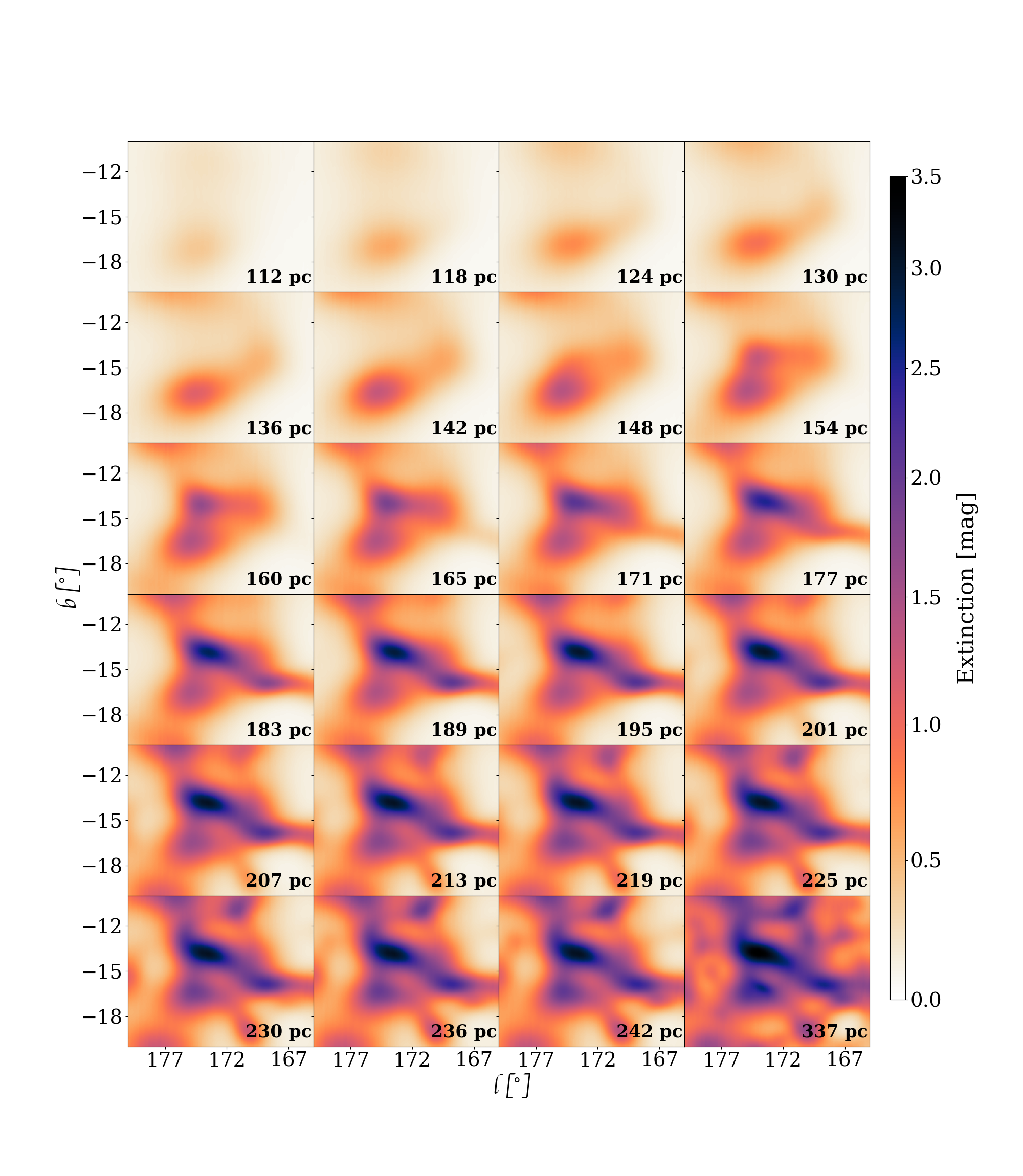}
    \caption{Three-dimensional extinction distribution of Taurus up to the indicated distances.}
    \label{fig:TaurusExtSlices}
\end{figure*}

\section{Planck $850~\mathrm{\mu m}$ flux and $550~\mathrm{\mu m}$ dust intensity maps} 

Following from Sec.~\ref{sec:Planck}, in this appendix we show the Planck $850~\mathrm{\mu m}$ flux and $550~\mathrm{\mu m}$ dust intensity maps for our four SFRs. We also show the normalised ratio of Planck data over model predicted total integrated extinction. The overlaid contours are our total model predicted extinctions up to the upper distance boundary introduced by figures~\ref{fig:OrionCumExt}, \ref{fig:CygXCumExt}, \ref{fig:PerseusCumExt}, \ref{fig:TaurusCumExt}. 

\begin{figure*}
\centering
\begin{subfigure}{0.49\textwidth}
  \centering
 \includegraphics[width=\textwidth, trim=4cm 4cm 5.5cm 4cm, clip]{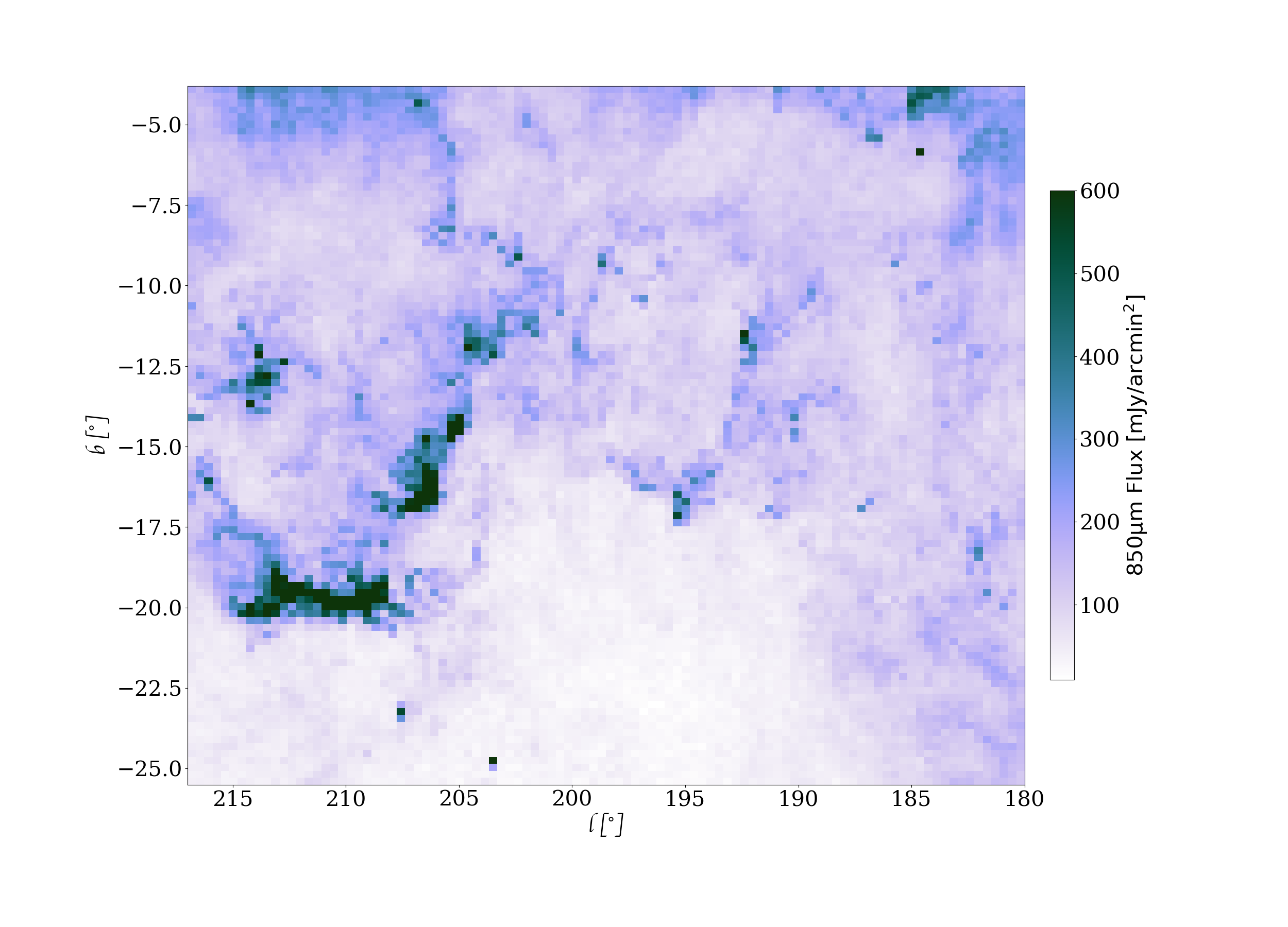}
  \end{subfigure}
\begin{subfigure}{0.49\textwidth}
  \centering
   \includegraphics[width=\textwidth, trim=4cm 4cm 5.5cm 4cm, clip]{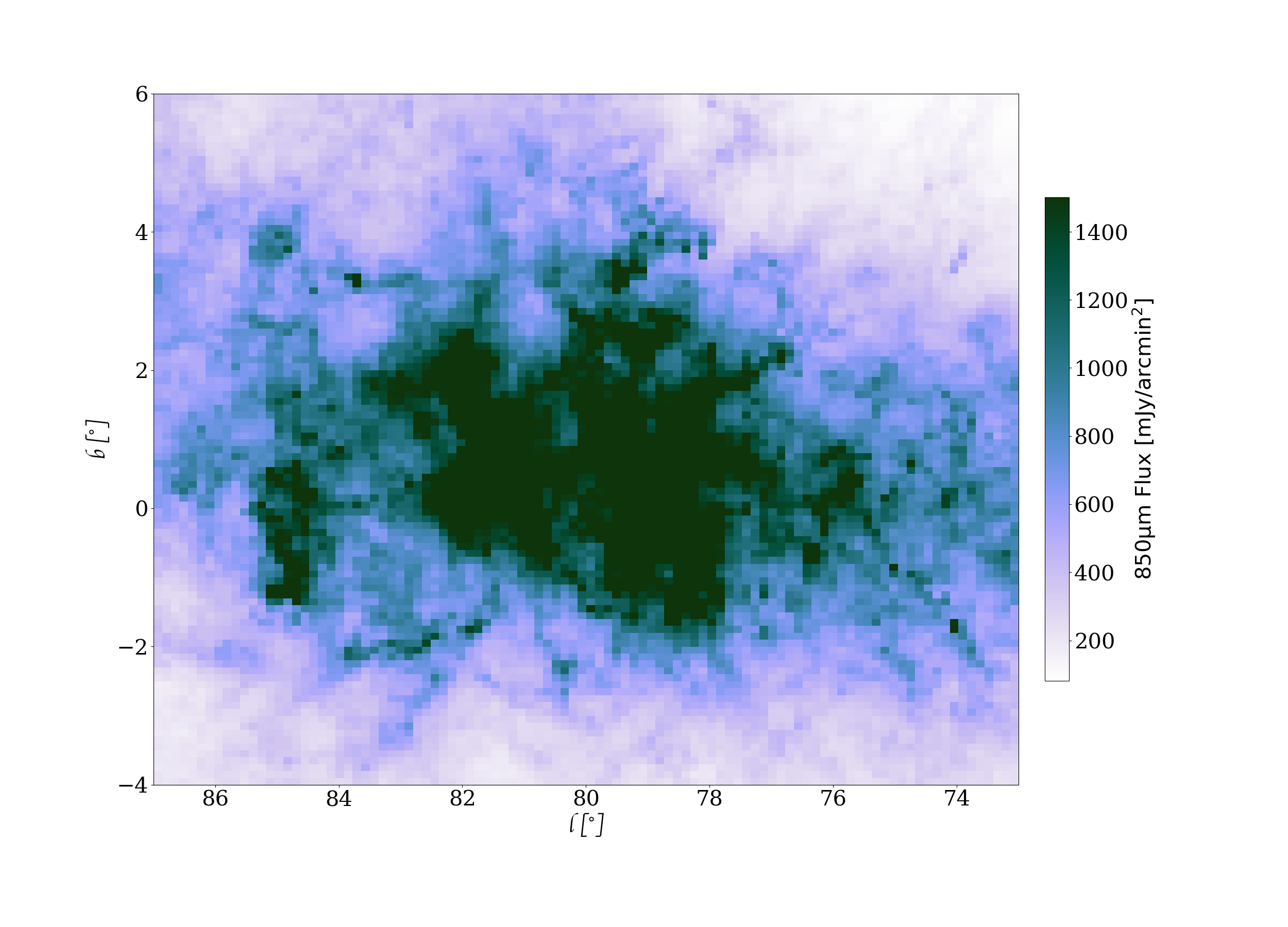}
  \end{subfigure}
 
\begin{subfigure}{0.49\textwidth}
  \centering
 \includegraphics[width=\textwidth, trim=4cm 4cm 5.5cm 4cm, clip]{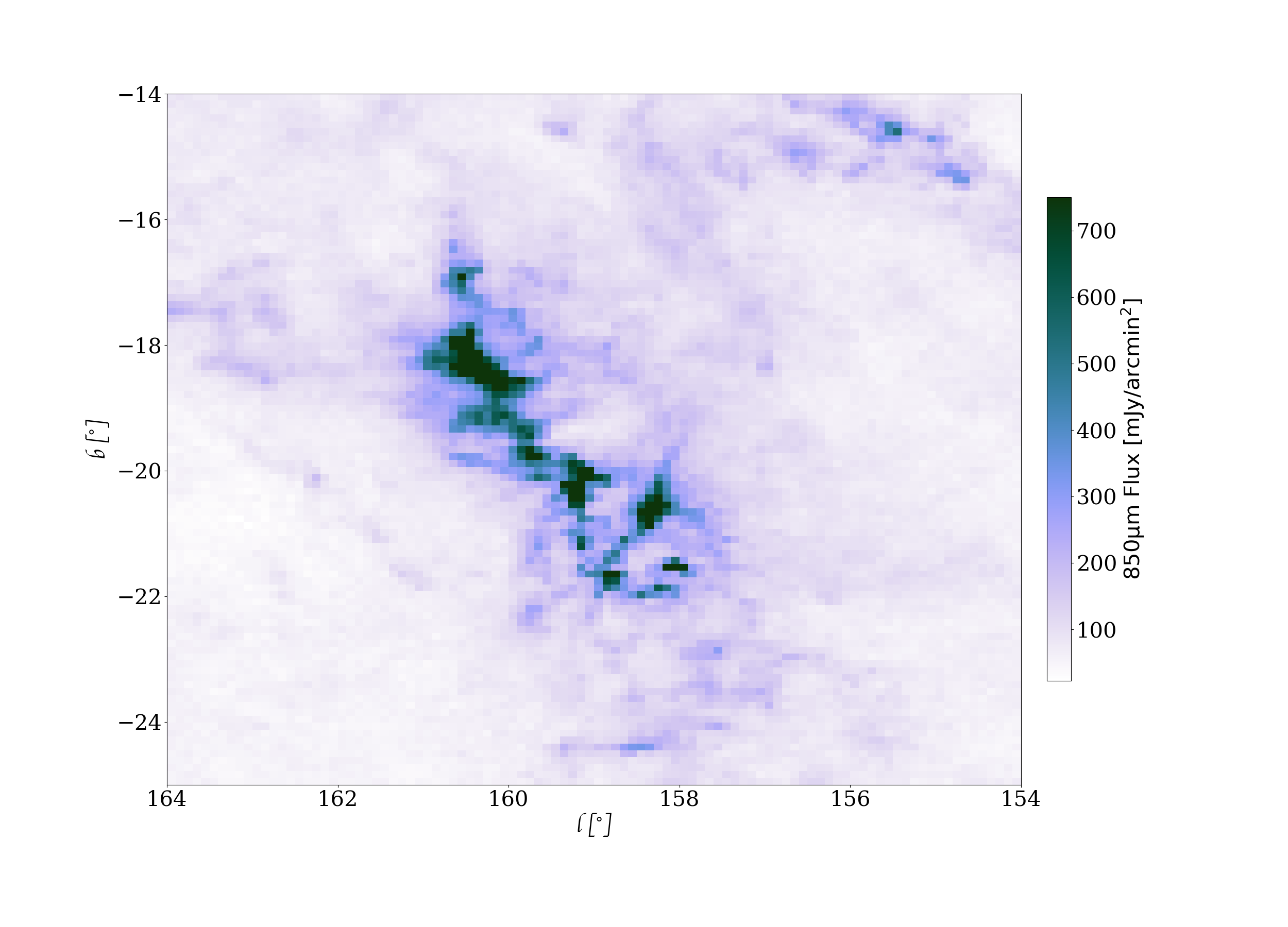}
  \end{subfigure}
\begin{subfigure}{0.49\textwidth}
  \centering
   \includegraphics[width=\textwidth, trim=4cm 4cm 5.5cm 4cm, clip]{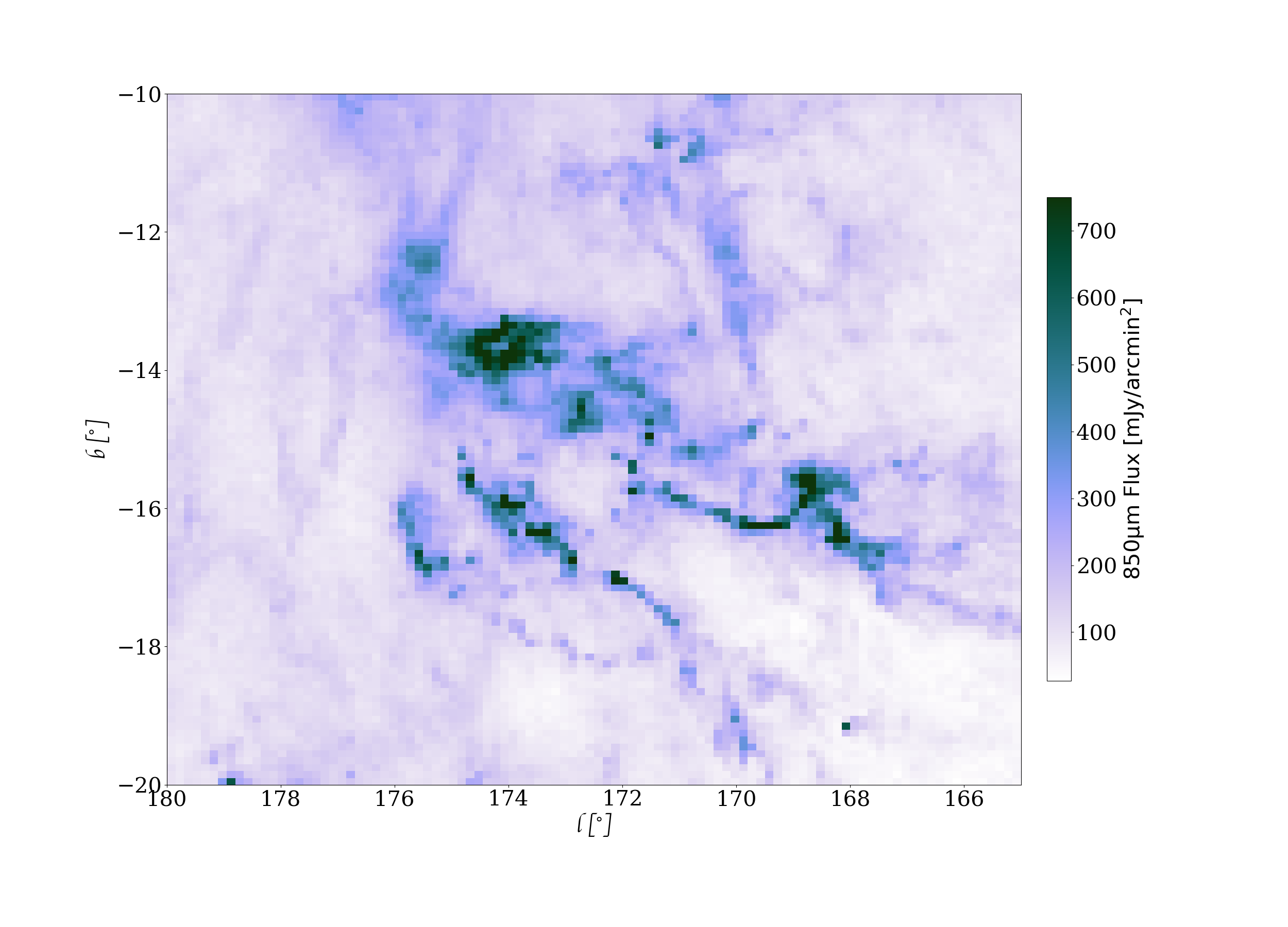}
  \end{subfigure}

\begin{subfigure}{0.49\textwidth}
  \centering
 \includegraphics[width=\textwidth, trim=4cm 4cm 5.5cm 4cm, clip]{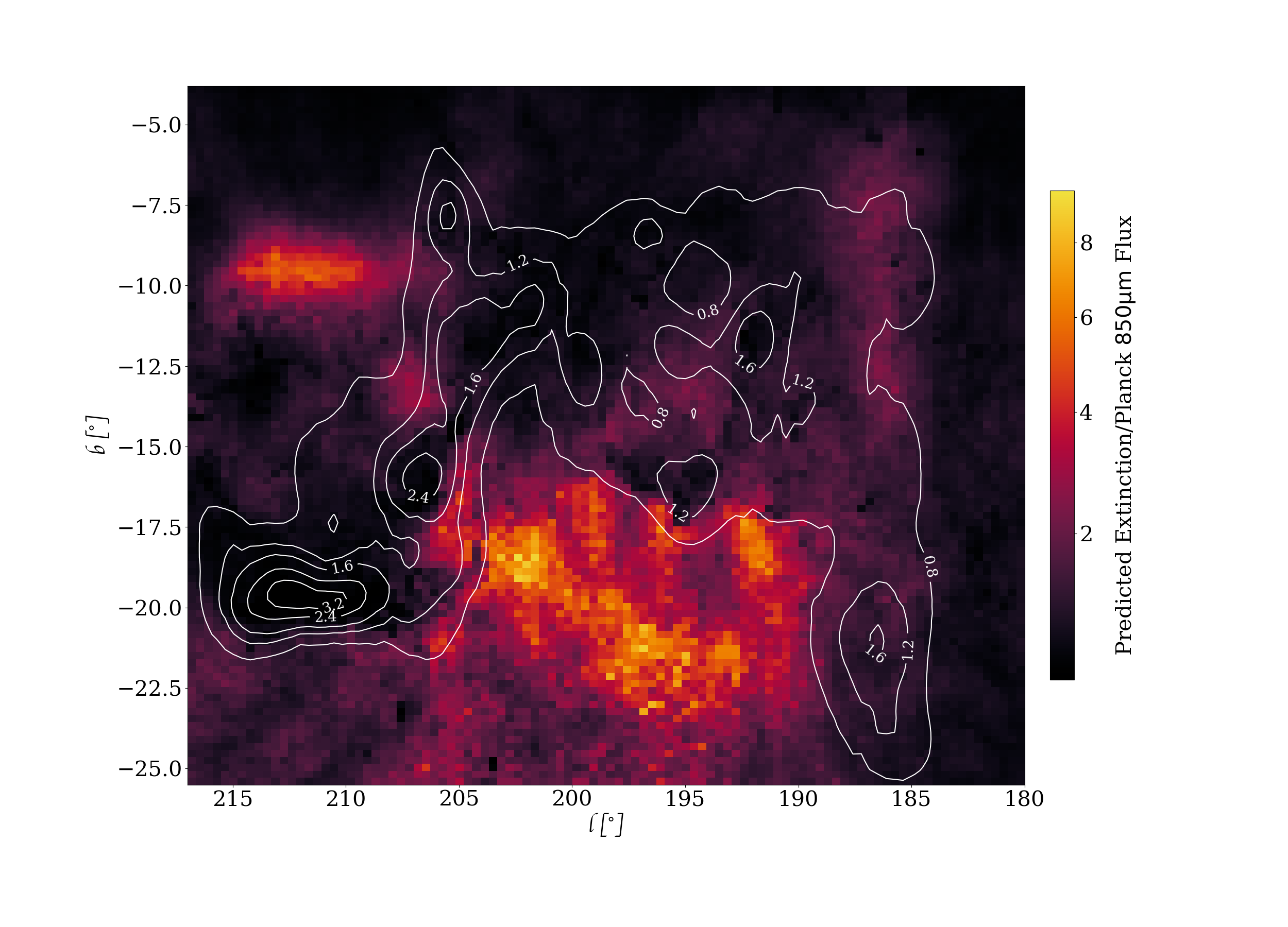}
  \end{subfigure}
\begin{subfigure}{0.49\textwidth}
  \centering
   \includegraphics[width=\textwidth, trim=4cm 4cm 5.5cm 4cm, clip]{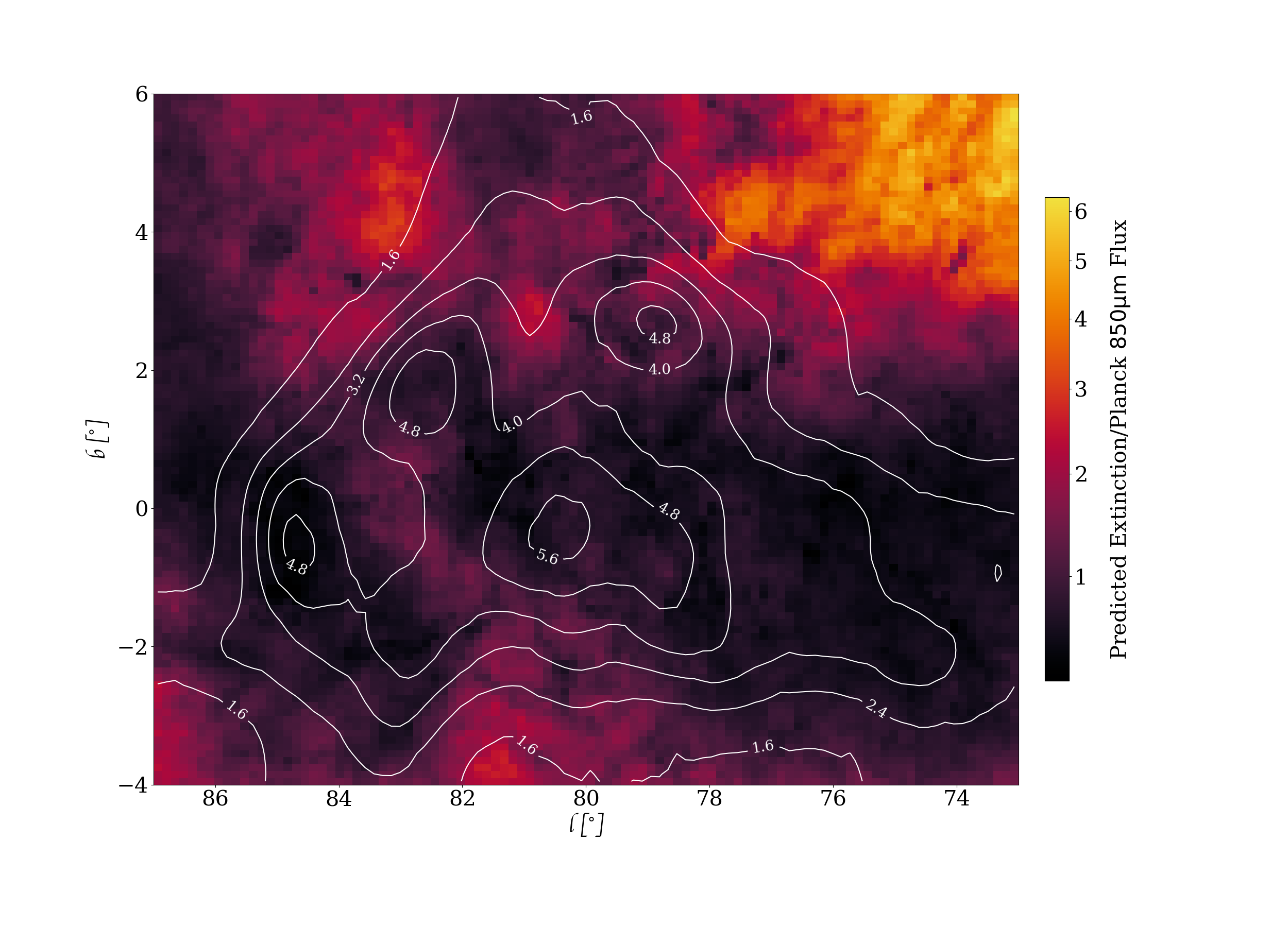}
  \end{subfigure}
 
\begin{subfigure}{0.49\textwidth}
  \centering
 \includegraphics[width=\textwidth, trim=4cm 4cm 5.5cm 4cm, clip]{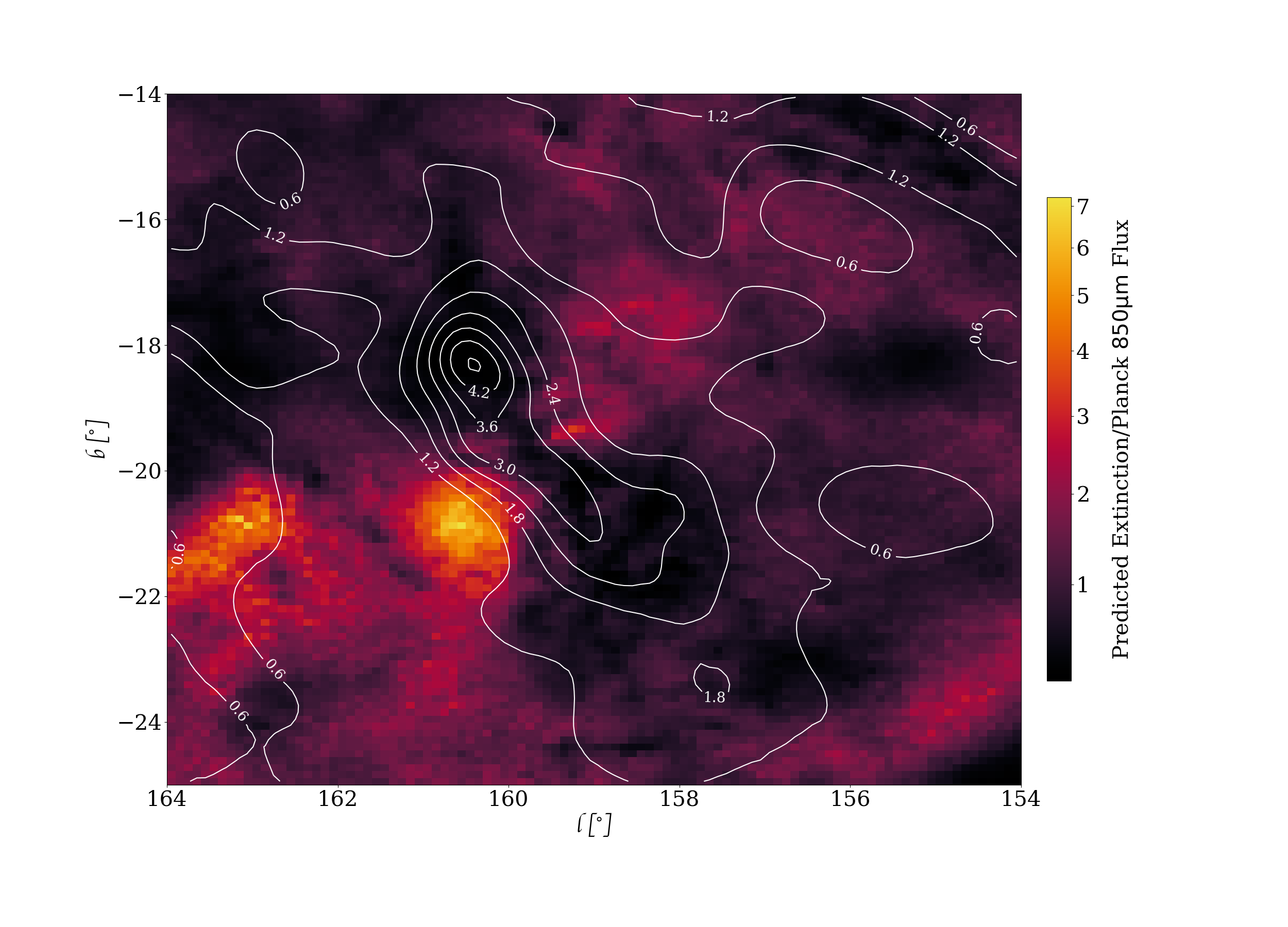}
  \end{subfigure}
\begin{subfigure}{0.49\textwidth}
  \centering
   \includegraphics[width=\textwidth, trim=4cm 4cm 5.5cm 4cm, clip]{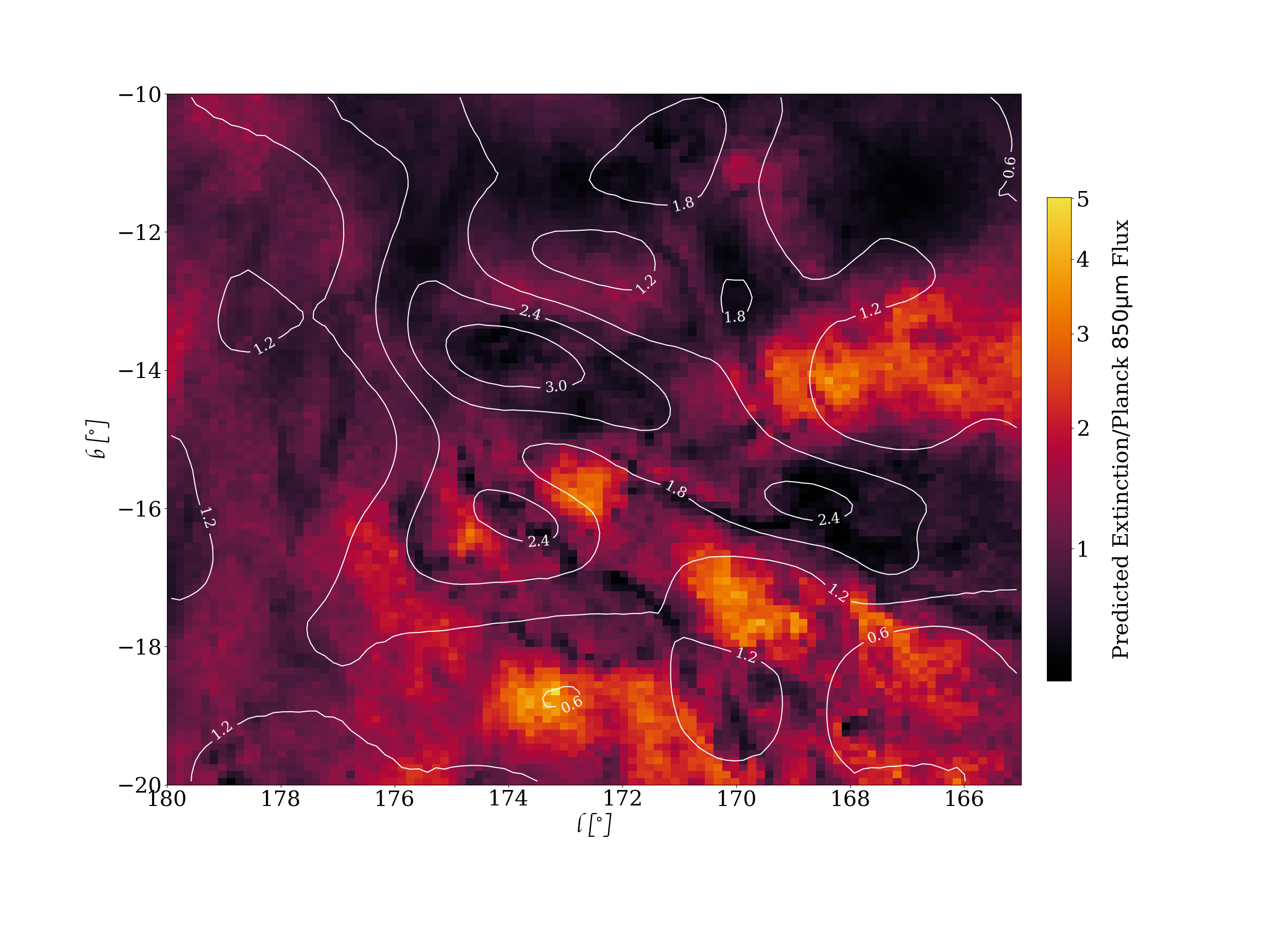}
  \end{subfigure}
 
  \caption{\emph{Top four panels}: Planck $850 \ \mathrm{\mu m}$ maps for Orion (top left), Cygnus X (top right), Perseus (upper middle left), Taurus (upper middle right); \emph{Bottom four panels}: Ratio of total extinction predicted by our model to Planck $850 \ \mathrm{\mu m}$ normalised by the median of the ratio. Predicted extinction contours overlaid. Ratio plots are arranged in the same pattern as the Planck emission maps.}
  \label{fig:Planck_850micronflux_All}
\end{figure*}

 \begin{figure*}
\centering
\begin{subfigure}{0.49\textwidth}
  \centering
 \includegraphics[width=\textwidth, trim=4cm 4cm 5.5cm 4cm, clip]{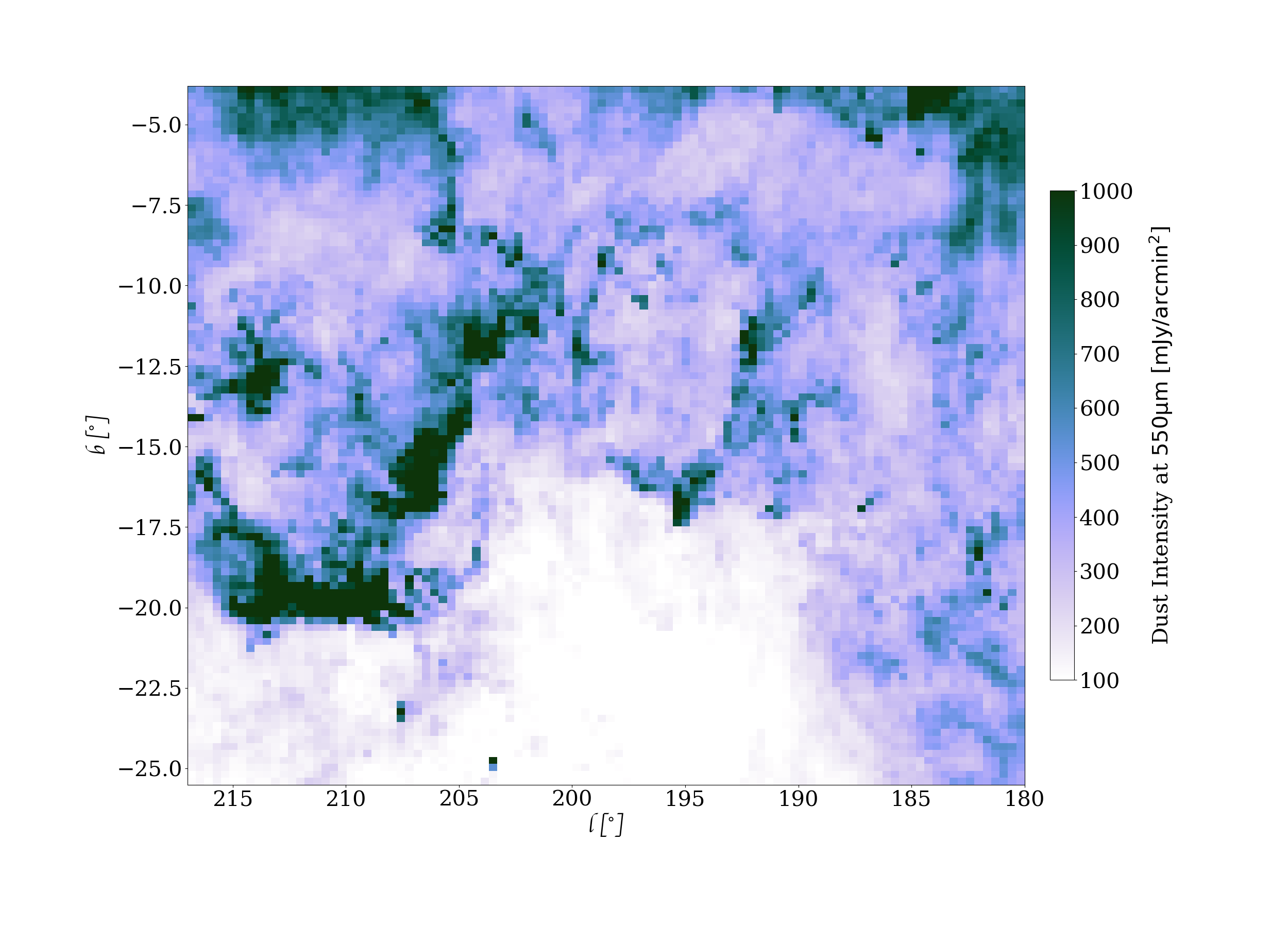}
  \end{subfigure}
\begin{subfigure}{0.49\textwidth}
  \centering
   \includegraphics[width=\textwidth, trim=4cm 4cm 5.5cm 4cm, clip]{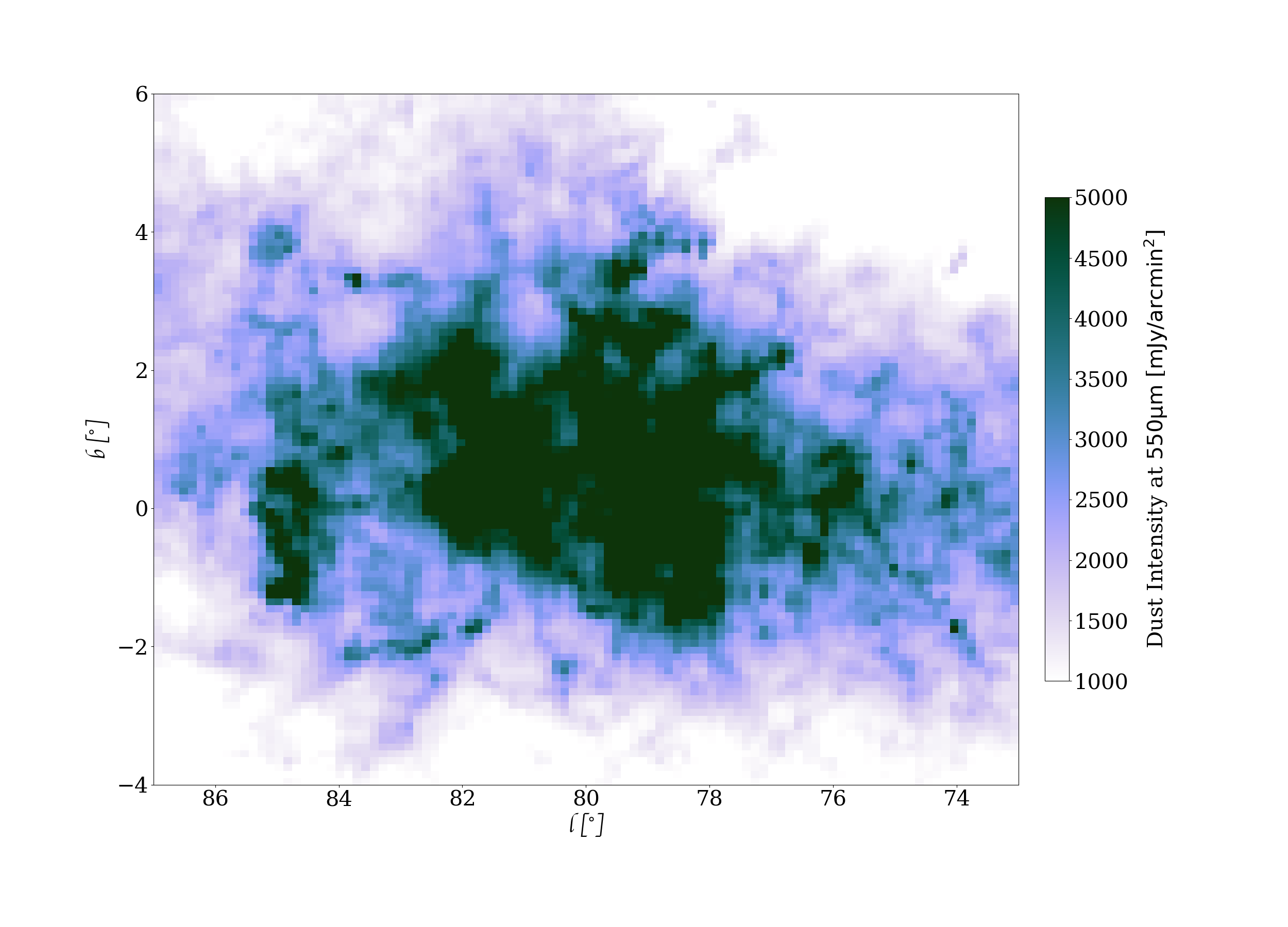}
  \end{subfigure}
 
\begin{subfigure}{0.49\textwidth}
  \centering
 \includegraphics[width=\textwidth, trim=4cm 4cm 5.5cm 4cm, clip]{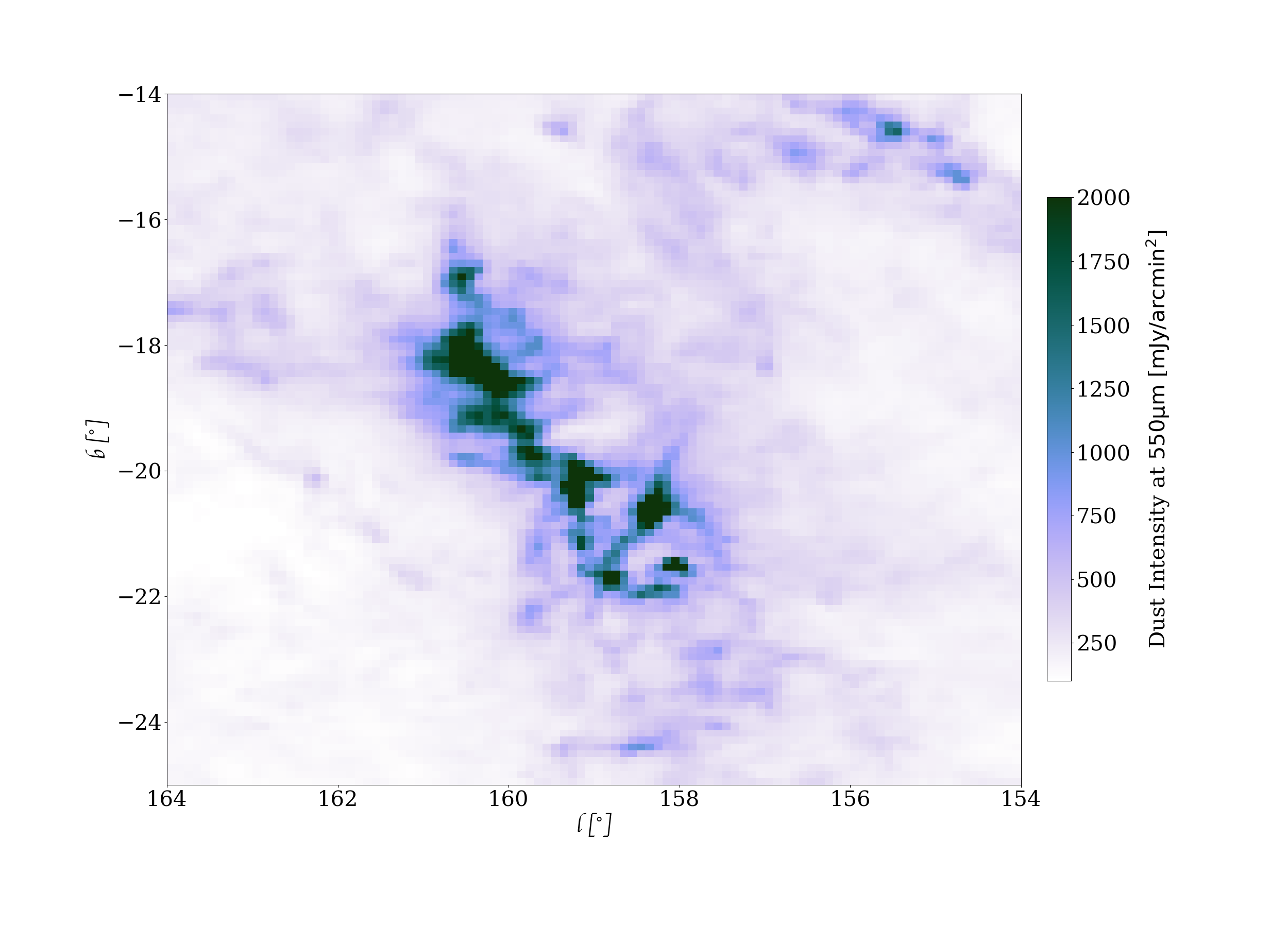}
  \end{subfigure}
\begin{subfigure}{0.49\textwidth}
  \centering
   \includegraphics[width=\textwidth, trim=4cm 4cm 5.5cm 4cm, clip]{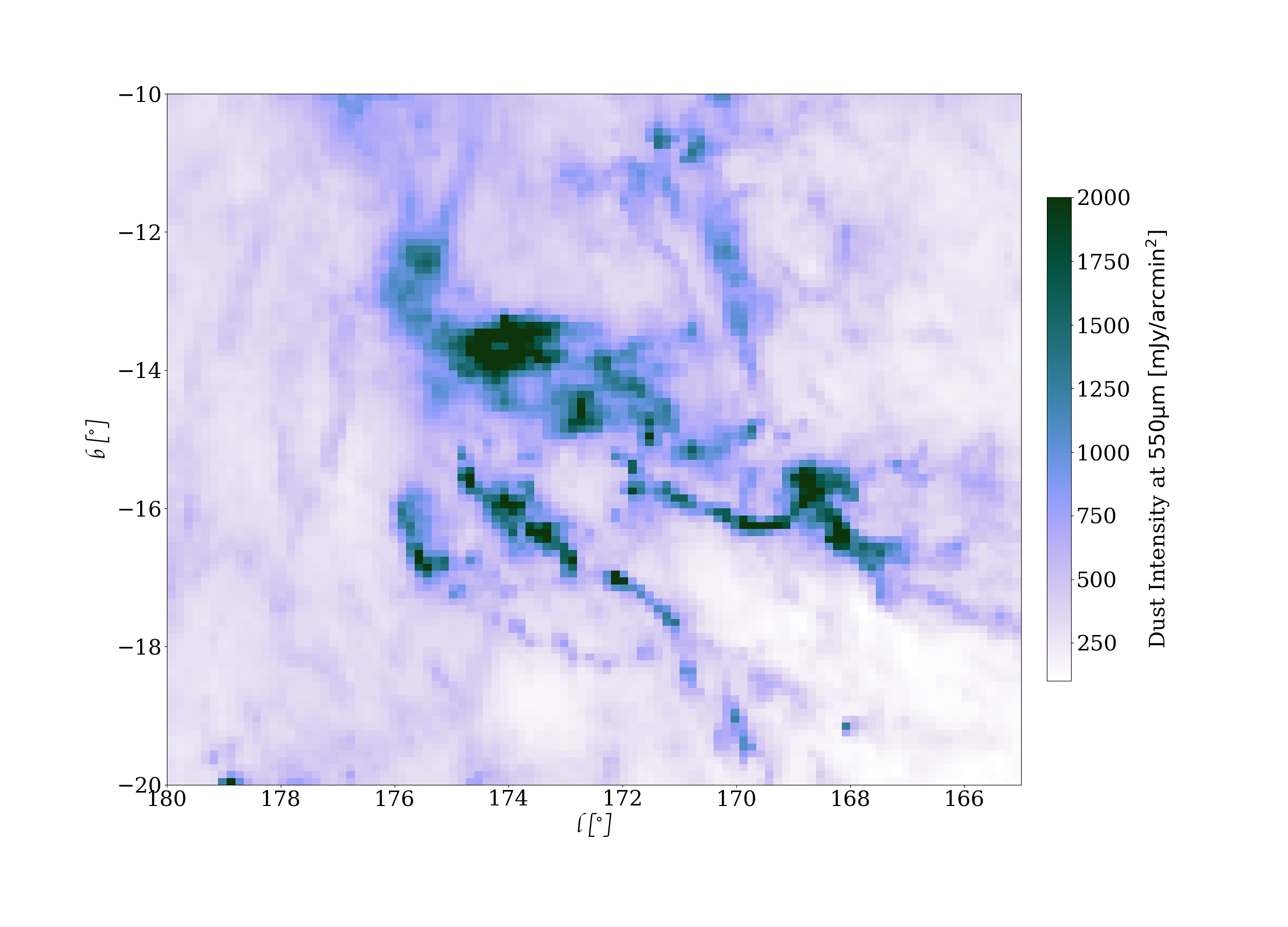}
  \end{subfigure}
  
  \begin{subfigure}{0.49\textwidth}
  \centering
 \includegraphics[width=\textwidth, trim=4cm 4cm 5.5cm 4cm, clip]{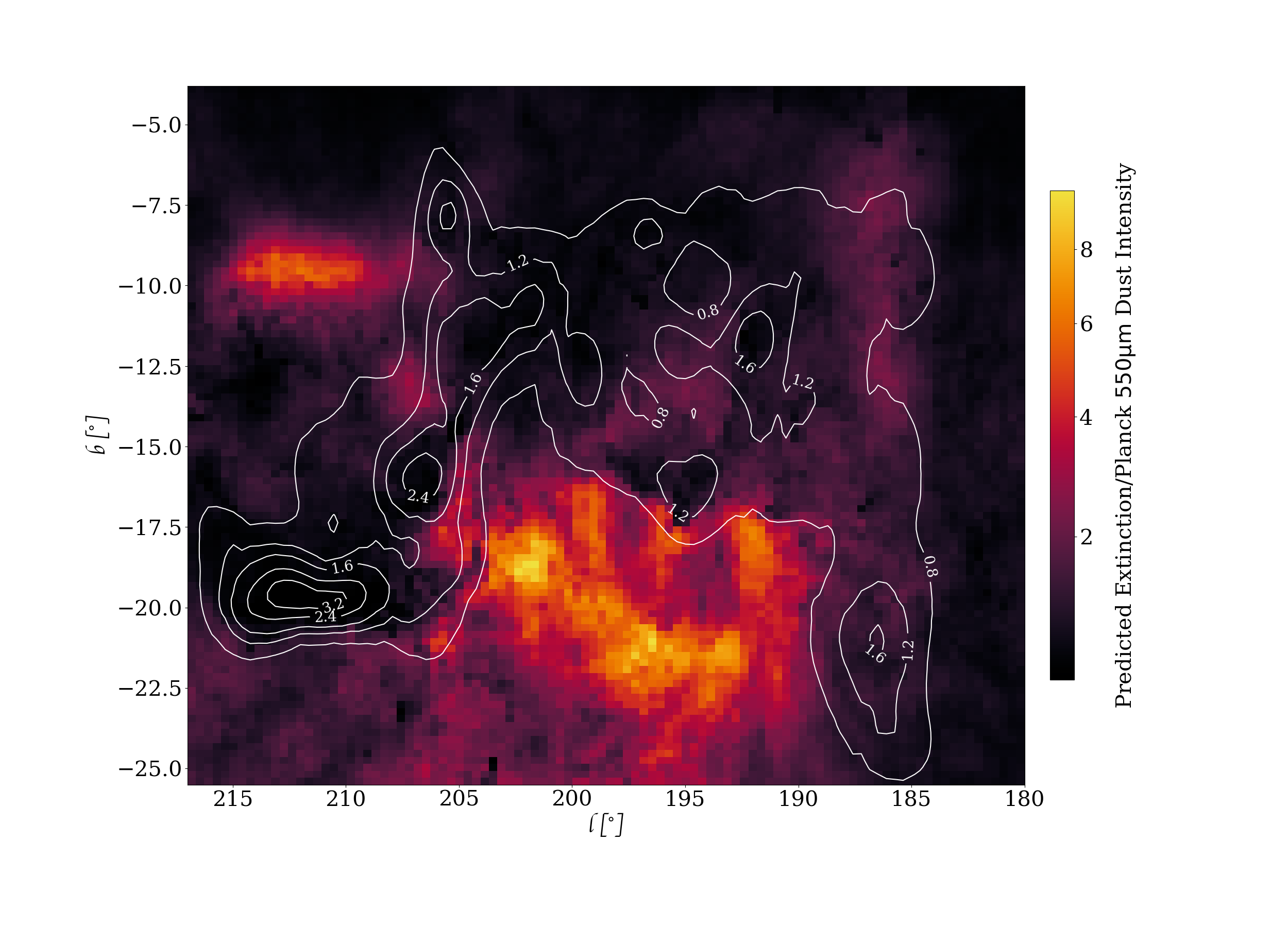}
  \end{subfigure}
\begin{subfigure}{0.49\textwidth}
  \centering
   \includegraphics[width=\textwidth, trim=4cm 4cm 5.5cm 4cm, clip]{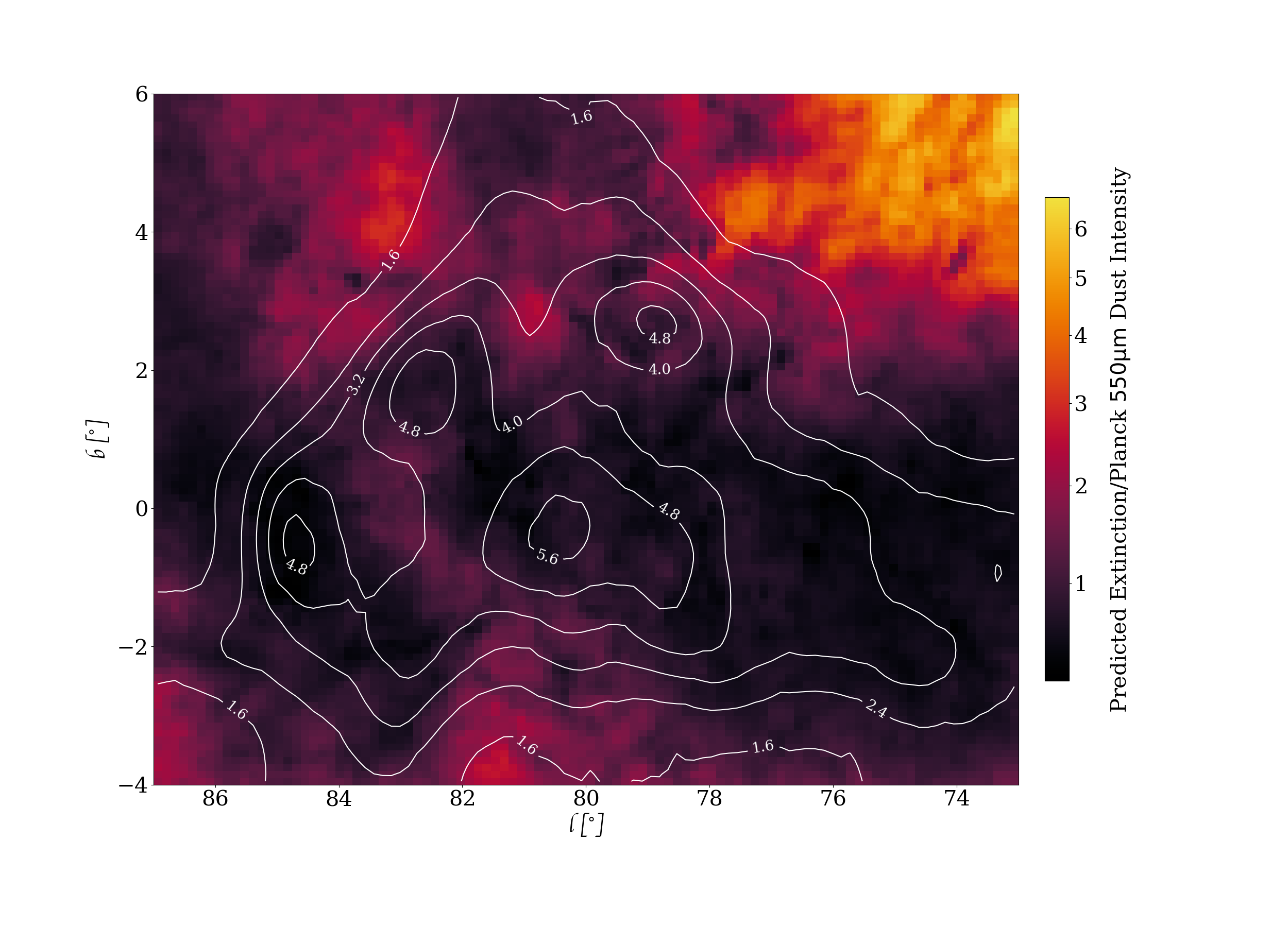}
  \end{subfigure}
 
\begin{subfigure}{0.49\textwidth}
  \centering
 \includegraphics[width=\textwidth, trim=4cm 4cm 5.5cm 4cm, clip]{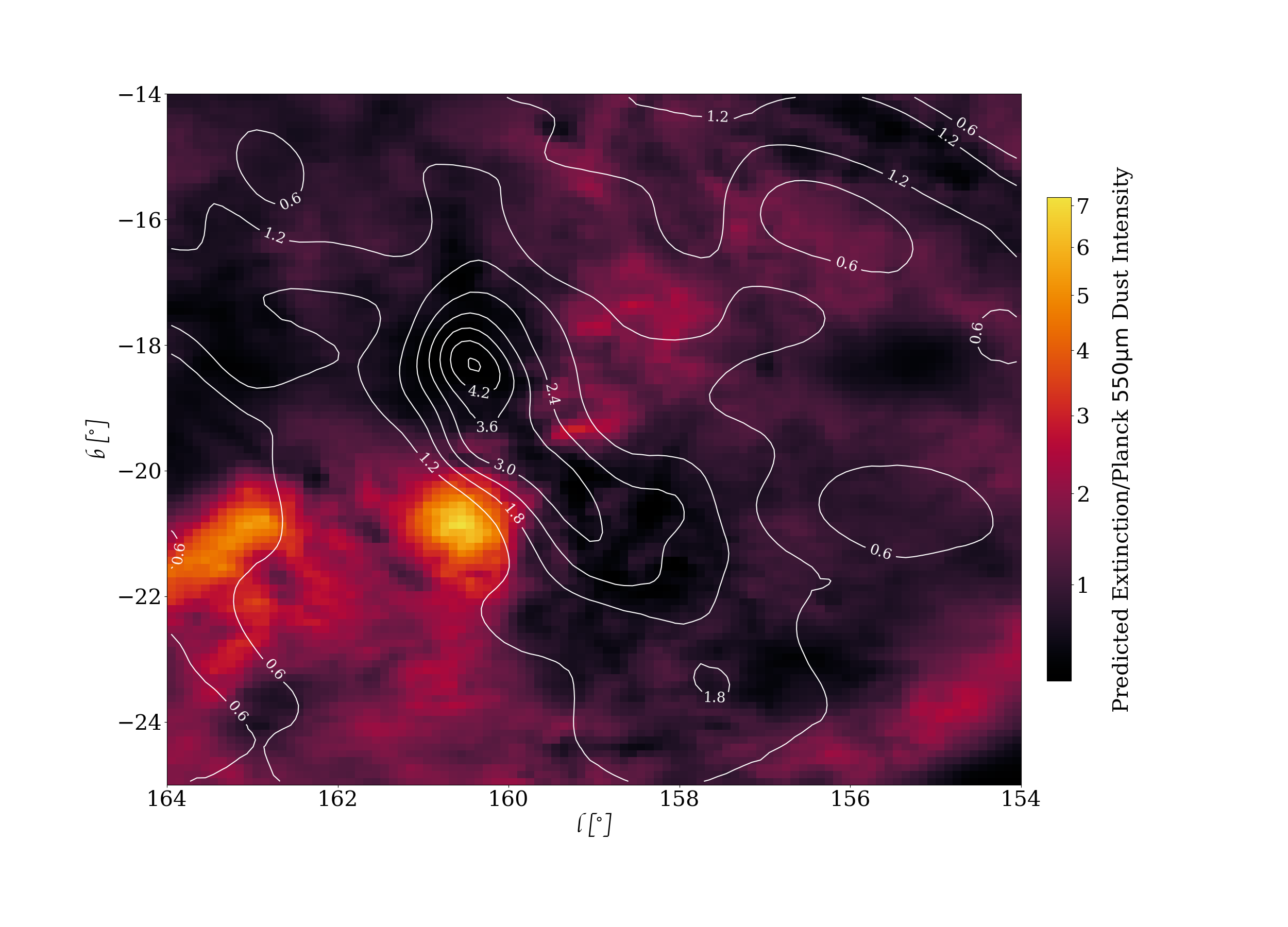}
  \end{subfigure}
\begin{subfigure}{0.49\textwidth}
  \centering
   \includegraphics[width=\textwidth, trim=4cm 4cm 5.5cm 4cm, clip]{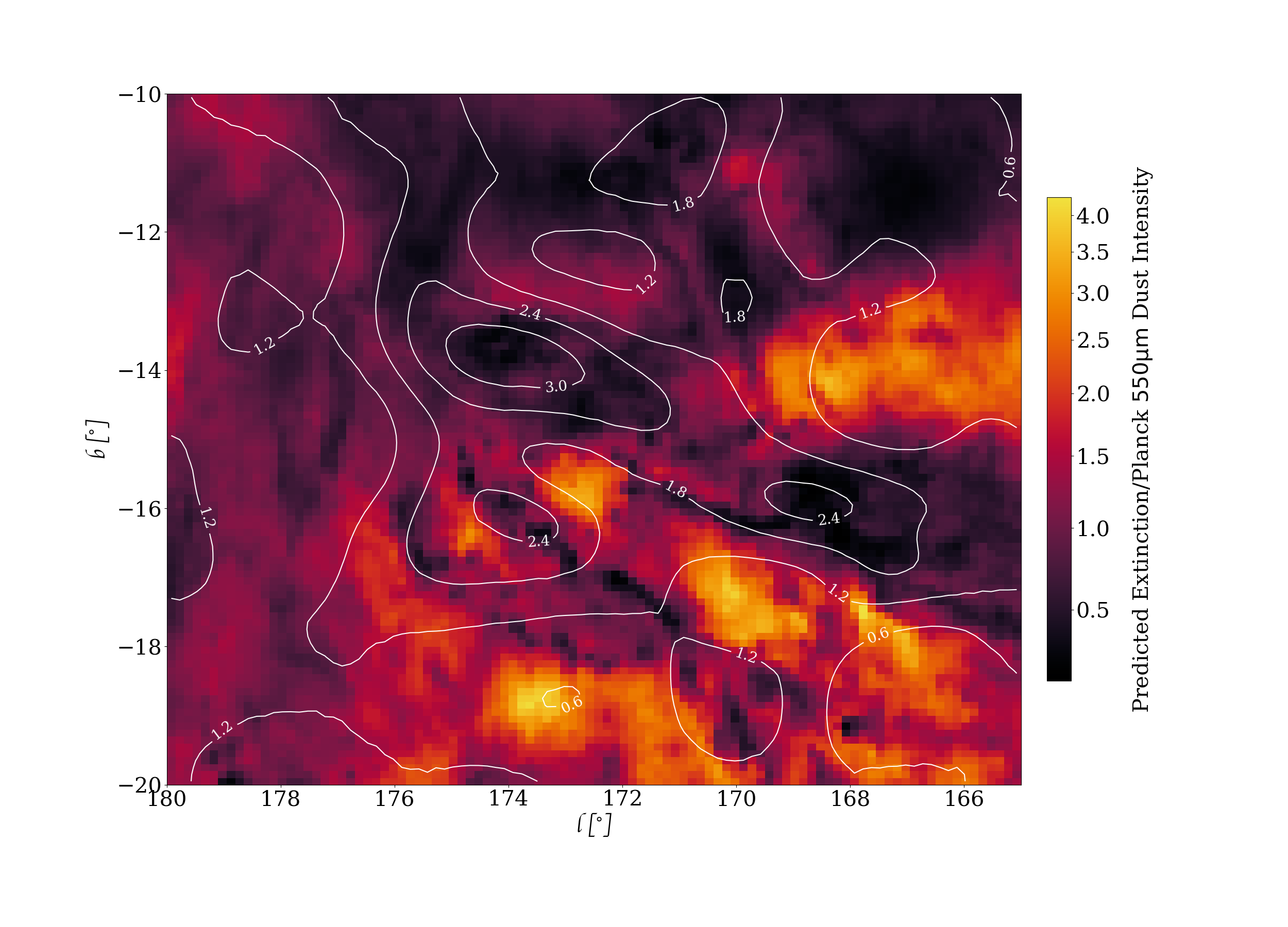}
  \end{subfigure}
 
  \caption{As Fig.~\ref{fig:Planck_850micronflux_All}, showing Planck Dust Intensity.}
  \label{fig:Planck_DustFlux_All}
\end{figure*}

\end{appendix}

\end{document}

%% file: Tables/HPs.tex
\begin{table*}
    \centering
\begin{threeparttable}[]
    \caption{Summary of model setup and behaviour for the selected star formation regions}
    \label{tab:HPs}
    \begin{tabular}{lcccc}
        \hline
        \\
         & Orion & Cygnus X & Perseus & Taurus \\
        \hline \hline
        \\

	    $l$ bounds ($^{\circ}$) & $180 \leq l \leq 217$ &  $73 \leq l \leq 87$ &  $154 \leq l \leq 164$ & $165\leq l \leq 180$\\
        
        $b$ bounds  ($^{\circ}$) & $-25.5 \leq b \leq -3.8$  & $-4 \leq b \leq 6$ & $-25 \leq b \leq -14$ & $-20 \leq b\leq -10$\\

        $d$ bounds (pc) & $250 \leq d \leq 550$ & $800 \leq d \leq 2200$ & $180 \leq d \leq 500$ & $40 \leq d \leq 350$ \\
        
        stars incorporated from (pc) & 270 & 1200 & 200 & 90 \\

        training $n_l,n_b,n_d$ (cells) & 40, 40, 55 & 40, 40, 65 & 30, 30, 35 & 30, 30, 35  \\

        training resolution (pc) & 2.3 & 3.5 & 1.0 & 0.23 \\

        number of sources & 87283 & 703291 & 92829 & 107051 \\

        initial scale length ($x,y,z$; pc) & 10 &  15 & 10 & 5\\

        final scale length ($x,y,z$; pc) & 9.5, 9.3, 9.5 & 19.1, 19.1, 19.2 & 12.8, 5.8, 6.4 & 9.2, 6.7, 4.9\\

        initial mean density ($\mathrm{mag \ pc}^{-1}$) & $4.6\times10^{-4}$ & $3.0\times10^{-3}$ & $4.6\times10^{-4}$ & $1.5\times 10^{-3}$\\

        final mean density ($\mathrm{mag \ pc}^{-1}$) & $1.06\times10^{-3}$ & $1.2\times10^{-3}$  &  $6.0\times10^{-4}$ & $1.1\times 10^{-3}$\\ 

        initial scale factor (dex) & 0.297 & 0.060 & 0.297 & 0.291\\

        final scale factor (dex) & 0.231 & 0.057 & 0.293 &0.136\\ 
        
        learning rate & 0.01 & 0.01 & 0.01 & 0.005\\ 
        
        number of iterations & 500 & 500 & 1000 & 1500\\ 
        
        number of inducing points & 1000 & 1000 & 1000 & 500\\ 

        predicting $nl,nb,nd$ (cells) & 100, 100, 105 & 100, 100, 105 & 100, 100, 105 &  100, 100, 105 \\

        predicting resolution (pc) & 0.9 &  1.4 & 0.3 & 0.07 \\

        run time (hours)\tnote{1} & 132 & 287 & 39 & 59\\

         \hline
    \end{tabular}
    
    \begin{tablenotes}
\item[1] The run times quoted here are the wall clock times for a machine setup of a single node with 2$\times$Intel(R) Xeon(R) CPU E5-2680 v3 @ 2.50GHz, no hyperthreading, 24 cores, 180GB RAM. 
\end{tablenotes}
    
\end{threeparttable}
\end{table*}

%% file: Tables/ClumpSizes.tex
\begin{table*}
    \centering
    \caption{{Locations and approximate sizes of the main components identified in the star formation regions}}
    \label{tab:ClumpSizes}
    \begin{tabular}{lccc}
        \hline
         & central $l,b$  &  & \\
        Component & coordinates ($^{o}$) & los extent (pc) & Notes\\
        \hline \hline
        
    \textit{Orion}: &&&\\
        Orion complex & 200,-15 & 280 - 500 & \\
        Orion A & 210,-20 & 340-450 & \\
        Orion A bubble & 215,-20 & 380-430 & \\
        Orion B comp 1 & 206,-16 & 380-400, 430-475 & separated into two segments along los \\
        Orion B comp 2 & 204,-11 & 380-420 & \\
        $\lambda$ Ori & 195,-12 & 360-470 & \\
        $\lambda$ Ori center & 195,-12 & 360-390 & \\
        AB arc & 210,-15 & 400-450 & \\
        Filament $\lambda$ & $l=$186 & 270-300 & could hold LDN 1558 and TGU L156 clumps \\
    \\
    \textit{Cygnus X}: &&&\\
        Cygnus X complex & 80,0 & 1300 - 1500 & \\
        Cyg X North & 82,0 & 1500 & densest region\\
        Cyg X South & 79,-1 & 1350 & densest region\\
        NAN & 0,85 & <800 & located closer than Cygnus X \\

    \\
    \textit{Perseus}: &&&\\
        Perseus complex & 160,-20 & 300 - 350 & \\
        IC348 & 160,-18 & 325 & \\
        NGC 1333 & 158,-20 & 325 & \\
        California filament & 158,-16 & 250-350 & \\
        Taurus filament & 163,-17 & 300-330 & \\
        Perseus filament & 157,-24 & 300-320 & connecting Perseus and Taurus \\
        
    \\
    \textit{Taurus}: &&\\
        Taurus complex & 172,-15 & 110-190 & densest at 150-180 pc \\
        TMC 1 & 174, -14 & 145-190 & \\
        TMC 2 main & 174,-16 & 110-145 & \\
        L1498 clump & 170,-19 & 195 & part of TMC 2 \\
        B215/L1506  extended  clump & 172,-17.5 & 195-220 \\
        Taurus filament & 165,-17 & 220-250 & connecting Perseus and Taurus\\
        Arcs 1+2 & $b<-14^{\circ}$ & 190-250 &  could hold L1540, L1507 and L1503 \\

         \hline
    \end{tabular}
\end{table*}

%% file: Tables/DustMasses.tex
\begin{table}[]
    \centering
    \caption{Inferred Masses}
    \label{tab:DustMasses}
    \begin{tabular}{ccc}
        \hline
        \\
        Region & Dust Mass ($10^{3} \mathrm{M_{\odot}}$) & Total Mass ($10^{3} \mathrm{M_{\odot}}$)  \\
        \\
        \hline \hline
        \\
        \vspace{0.3cm}  
        Orion & $9.1^{+3.2}_{-2.2}$ & $1130^{+400}_{-270}$ \\ \vspace{0.3cm} 
         
        Cygnus & $88.2^{+7.0}_{-6.2}$ & $10900^{+900}_{-800}$ \\ \vspace{0.3cm} 

        Perseus & $1.5^{+0.1}_{-0.1}$ & $187^{+17}_{-14}$ \\ \vspace{0.3cm} 
         
        Taurus & $1.2^{+0.1}_{-0.1}$ & $149^{+17}_{-14}$  \\ 
         
         \hline
    \end{tabular}
\end{table}